\documentclass[conference,compsoc]{IEEEtran}

\ifCLASSOPTIONcompsoc
  \usepackage[nocompress]{cite}
  \usepackage{amsmath,amsfonts}
\usepackage{array}
\usepackage{textcomp}
\usepackage{stfloats}
\usepackage{url}
\usepackage{verbatim}
\usepackage{graphicx}
\usepackage{cite}

\usepackage{wasysym}
\usepackage{multirow}
\usepackage{pifont}
\usepackage[lined,ruled,commentsnumbered]{algorithm2e}
\usepackage{algpseudocode}
\usepackage{pdfpages}
\usepackage{tabularx}
\usepackage{booktabs}
\usepackage{amsmath}
\usepackage{amssymb}
\usepackage{diagbox}
\usepackage{subfigure}
\usepackage{color}
\usepackage{url}
\usepackage[misc]{ifsym} 
\usepackage{cite}
\usepackage{siunitx}

\usepackage[bottom]{footmisc}

\renewcommand{\footnoterule}{%
  \kern -3pt 
  \hrule width 0.4\textwidth height 0.5pt 
  \kern 2.6pt 
}

\fi
\ifCLASSINFOpdf
\else
\fi
\begin{document}
%
\title{FlowMur: A Stealthy and Practical Audio Backdoor Attack with Limited Knowledge}




%
\author{\IEEEauthorblockN{Jiahe Lan\IEEEauthorrefmark{1},
Jie Wang\IEEEauthorrefmark{1},
Baochen Yan\IEEEauthorrefmark{1},
Zheng Yan\IEEEauthorrefmark{1}\IEEEauthorrefmark{2}$^{\textrm{\Letter}}$, and
Elisa Bertino\IEEEauthorrefmark{3}}
\IEEEauthorblockA{\IEEEauthorrefmark{1}The State Key Laboratory on Integrated Services Networks,\\ School of Cyber Engineering, Xidian University, China}
\IEEEauthorblockA{\IEEEauthorrefmark{2}Hangzhou Institute of Technology, Xidian University, China}
\IEEEauthorblockA{\IEEEauthorrefmark{3}Department of Computer Science, Purdue University, USA\\
\{jhlan16, jwang1997, bcyan\}@stu.xidian.edu.cn, zyan@xidian.edu.cn, bertino@purdue.edu}}


\maketitle

\begin{abstract}
\footnotetext{\textit{Accepted for publication at the 45th IEEE Symposium on Security \& Privacy (S\&P 2024). Please cite accordingly.}}Speech recognition systems driven by Deep Neural Networks (DNNs) have revolutionized human-computer interaction through voice interfaces, which significantly facilitate our daily lives. However, the growing popularity of these systems also raises special concerns on their security, particularly regarding backdoor attacks. A backdoor attack inserts one or more hidden backdoors into a DNN model during its training process, such that it does not affect the model's performance on benign inputs, but forces the model to produce an adversary-desired output if a specific trigger is present in the model input. Despite the initial success of current audio backdoor attacks, they suffer from the following limitations: (i) Most of them require sufficient knowledge, which limits their widespread adoption. (ii) They are not stealthy enough, thus easy to be detected by humans. (iii) Most of them cannot attack live speech, reducing their practicality. To address these problems, in this paper, we propose FlowMur, a stealthy and practical audio backdoor attack that can be launched with limited knowledge. FlowMur constructs an auxiliary dataset and a surrogate model to augment adversary knowledge. To achieve dynamicity, it formulates trigger generation as an optimization problem and optimizes the trigger over different attachment positions. To enhance stealthiness, we propose an adaptive data poisoning method according to Signal-to-Noise Ratio (SNR). Furthermore, ambient noise is incorporated into the process of trigger generation and data poisoning to make FlowMur robust to ambient noise and improve its practicality. Extensive experiments conducted on two datasets demonstrate that FlowMur achieves high attack performance in both digital and physical settings while remaining resilient to state-of-the-art defenses. In particular, a human study confirms that triggers generated by FlowMur are not easily detected by participants. The source code of FlowMur is publicly available at \url{https://github.com/cristinalan/FlowMur}.
\end{abstract}


%
\IEEEpeerreviewmaketitle

\section{Introduction}

The advance of Deep Neural Networks (DNNs) has propelled machines to achieve near-human performance in speech recognition~\cite{chorowski2015attention}, which introduces a new and convenient mode of human-machine interaction, known as voice interfaces. Such interfaces not only facilitate our daily lives~\cite{ammari2019music} but also improve digital device accessibility for diverse groups including the elderly and the visually impaired~\cite{pradhan2018accessibility,pradhan2020use}. For example, prominent voice assistants, such as Amazon Alexa~\cite{amazonalexa} and Apple Siri~\cite{applesiri}, empower users to effortlessly access information, control smart devices, and perform diverse tasks via voice interfaces.

While speech recognition technology offers numerous advantages, achieving high accuracy in speech recognition is nontrivial. One main challenge is the substantial demand for computational resources and training data, which makes it difficult for users with limited resources to design and train speech recognition models~\cite{li2022backdoor}. To leverage the convenience of DNNs, many users turn to third-party services, including publicly available datasets, models, and third-party training platforms, to train or acquire speech recognition models. Unfortunately, these practices may introduce potential attack surfaces, resulting in the possibility of backdoor attacks. 
Such an attack inserts one or more hidden backdoors into a DNN model during training. The backdoor does not affect the model's performance on benign inputs, but forces the model to produce an adversary-desired output if a specific trigger is present in the model input~\cite{yao2019latent}. Backdoor attacks are characterized by their harmfulness and stealthiness. For example, they can be used to attack DNNs deployed in safety-critical applications like traffic sign recognition and speech recognition~\cite{tang2020embarrassingly,shi2022audio}. Moreover, it is challenging to detect the presence of backdoors through standard testing procedures, as they do not compromise the performance of infected models on benign samples. These factors have prompted serious research into backdoor attacks.

Backdoor attacks were first identified in the image domain~\cite{gu2019badnets} and have gained increasing attention in the audio domain in recent years. Audio backdoor attacks can be classified into two categories depending on the type of used triggers: audio-specific and audio-agnostic. Audio-specific backdoor attacks assign a unique trigger to each audio sample, which enhances the stealthiness of the attack. But such an attack requires the adversary to obtain a model input beforehand to generate a corresponding trigger. This makes it infeasible in attacking live human speech for real-time execution.
On the other hand, audio-agnostic backdoor attacks utilize a universal trigger for all audio samples, saving time and eliminating the requirement to access model inputs. This makes it possible to launch such an attack in real-time, especially for attacking live human speech. Therefore, we focus on audio-agnostic backdoor attacks in this paper due to their practical significance.

We review existing audio-agnostic backdoor attacks and compare them in terms of trigger design, adversary knowledge, stealthiness, practicality, and defense resistance, as shown in Table \ref{related work}. From the table, we have the following observations: 
(i) Many attacks~\cite{koffas2022can,xin2023natural,liu2022opportunistic} utilize designated triggers, such as birdsong and whistles. They are simple and easy to deploy, but less effective than those with optimized triggers. This is because designated triggers are randomly selected and model-agnostic, while optimized triggers are tailored for an infected model.
(ii) Most attacks~\cite{koffas2022can,shi2022audio,liu2022opportunistic} assume that the adversary has sufficient knowledge about a target model or a victim dataset. This assumption, however, limits the widespread adoption of backdoor attacks and increases the difficulty of their deployment.
(iii) There is a lack of in-depth investigation on the stealthiness of audio backdoor attacks. Few studies meet label consistency (i.e., the content of a sample is consistent with its label) and trigger inaudibility, which are key requirements of a stealthy backdoor attack. Moreover, trigger imperceptibility, reflecting the stealthiness from human perception, has not yet been deeply investigated by past research~\cite{xin2023natural,shi2022audio,koffas2022can,liu2022opportunistic}.
(iv) While designers of attacks have tried to deploy audio backdoor attacks in practice, these deployments~\cite{koffas2022can,xin2023natural,liu2022opportunistic} do not meet dynamicity well. A dynamic backdoor attack can be launched regardless of the attachment position of a trigger to a benign sample, which cannot be achieved by a static backdoor attack.
(v) As for defense resistance, only fine-tuning~\cite{tajbakhsh2016convolutional} and fine-pruning~\cite{liu2018fine} are tested in~\cite{liu2022opportunistic} and~\cite{shi2022audio}, respectively. \textcolor{black}{Other advanced defense methods for audio backdoor attacks, such as STRong Intentional Perturbation (STRIP)~\cite{gao2021design,gao2019strip} and Beatrix~\cite{ma2023beatrix}, as well as traditional audio defense methods like filters~\cite{carlini2016hidden}, are not considered in any of the existing attacks.}

\begin{table*}[tbp]
    \centering 
    \scriptsize
    \caption{A comparison of audio-agnostic backdoor attacks.}
    \vspace{-1mm}
    \label{related work}
    {\vspace{-2mm}\CIRCLE: satisfy a criterion; \LEFTcircle: partially satisfy a criterion; \Circle: do not satisfy a criterion; ``-'': not available.}
    \vspace{-1mm}
\\[2mm]

    \begin{tabular}{@{}c|c|cc|ccc|cc|c@{}}  
        \toprule[1.5pt]
        
    \multirow{3.5}{*}{\makebox[0.04\textwidth][c]{Ref}} &
    \multirow{3.5}{*}{\makebox[0.1\textwidth][c]{Trigger Design}}&
    \multicolumn{2}{c|}{Adversary Knowledge} &
    \multicolumn{3}{c|}{Stealthiness} &
    \multicolumn{2}{c|}{Practicality}&
    \multicolumn{1}{c}{\multirow{3.5}{*}{\makebox[0.06\textwidth]{\begin{tabular}[c]{@{}c@{}}Defense\\ Resistance \end{tabular}}}}\\
    \cmidrule{3-9}
        & &\multicolumn{1}{c}{\begin{tabular}[c]{@{}c@{}}No Need\\ Target Model\end{tabular}}&\multicolumn{1}{c|}{\begin{tabular}[c]{@{}c@{}}Only Require\\Target-Class Data \end{tabular}}&\multicolumn{1}{c}{\begin{tabular}[c]{@{}c@{}}Label\\ Consistency \end{tabular}} &\multicolumn{1}{c}{\begin{tabular}[c]{@{}c@{}}Trigger\\ Inaudibility \end{tabular}} &\multicolumn{1}{c|}{\begin{tabular}[c]{@{}c@{}}Trigger\\ Imperceptibility \end{tabular}} &\multicolumn{1}{c}{\begin{tabular}[c]{@{}c@{}}Dynamicity \end{tabular}}  & \multicolumn{1}{c|}{\begin{tabular}[c]{@{}c@{}} Physical\\ Attack \end{tabular}}  \\
      \midrule
~\cite{koffas2022can}            & designated     & \CIRCLE   & \Circle   & \Circle  &  \CIRCLE      &  - & \Circle  &  \CIRCLE  &  -\\
~\cite{xin2023natural}           & designated     & \CIRCLE   & \CIRCLE   & \CIRCLE  &  \Circle      &  - & \Circle  &  \CIRCLE  &  -\\
~\cite{liu2022opportunistic}     & designated     & \Circle   & \Circle   & \Circle  &  \Circle      &  - & \Circle &  \CIRCLE  &  \LEFTcircle\\
~\cite{shi2022audio}        & optimized  & \Circle   & \Circle   & \Circle  &  \Circle      &  - &\CIRCLE  &  \CIRCLE  &  \LEFTcircle\\ \midrule
FlowMur                    & optimized  & \CIRCLE   & \CIRCLE   & \CIRCLE  &  \LEFTcircle      &  \CIRCLE  &\CIRCLE &  \CIRCLE  &  \CIRCLE\\
        
    \bottomrule[1.5pt]
    \end{tabular}
    \vspace{-2mm}
\end{table*}

In this paper, we propose a stealthy and practical audio backdoor attack named FlowMur, which requires limited adversary knowledge (i.e., access to target class samples in a dataset without the need for gaining other samples or the target model) to be launched and achieves high attack performance. To accomplish these goals, we address the following challenges:

\emph{C1: How to generate an effective trigger with limited knowledge?}

Backdoor attacks that rely on designated triggers ~\cite{koffas2022can,xin2023natural,liu2022opportunistic} demonstrate a high potential for widespread adoption due to their limited need for adversary knowledge. However, they suffer from subpar attack performance, especially when the poisoning rate is low, since the association between a designated trigger and a hidden backdoor is loose. In contrast, backdoor attacks based on optimized triggers~\cite{shi2022audio} can achieve high attack performance since the optimized trigger is tailored for an infected model, facilitating a strong connection between the trigger and the backdoor. Nevertheless, generating the optimized trigger requires adequate adversary knowledge, thereby increasing the difficulty of its deployment. Therefore, to address the conflict between high attack performance and practical deployment, FlowMur utilizes an auxiliary dataset and a surrogate model to complement the limited adversary knowledge so that the adversary is capable of generating optimized triggers. 

\emph{C2: How to improve the stealthiness of a backdoor attack?}

A stealthy backdoor attack can evade human inspection, facilitating successful execution of the attack. Conversely, an easily exposed backdoor attack may lead to attack failure. To enhance the stealthiness of FlowMur, we adopt several strategies. First, we only poison samples that belong to the target class to ensure label consistency. Second, we incorporate constraints on the trigger amplitude into trigger optimization to generate an unnoticeable trigger. Third, we propose an adaptive data poisoning method to further improve trigger inaudibility. This method generates triggers of varying scales for different audio samples according to an adversary-desired Signal-to-Noise Ratio (SNR). 

\emph{C3: How to make the attack practical?}

Most of the previous audio backdoor attacks~\cite{koffas2022can,xin2023natural,liu2022opportunistic} are static, in that the trigger is attached at a fixed position for each audio sample. However, in practice, speech recognition systems usually engage with live humans, where an adversary does not know when the human commences speaking. Therefore, attaching the trigger to a fixed position of the live human speech seems infeasible. To address this challenge, FlowMur should be a dynamic audio backdoor attack, whose trigger position is variable. 
To achieve this objective, we introduce a constraint on trigger duration and take into account the variability of trigger position during trigger generation and data poisoning.
This enables the target model to learn the intrinsic features of the trigger regardless of its position. In addition, we incorporate ambient noise into the process of trigger generation and data poisoning to mitigate the impact of ambient noise on FlowMur and further improve its practicality.

We conduct extensive experiments to evaluate FlowMur considering four aspects: effectiveness, practicality, stealthiness, and defense resistance. We test FlowMur's effectiveness on four different models across two datasets. The experimental results show that, in most cases, FlowMur achieves an attack success rate of over 95\% without compromising the benign accuracy of the models. We also examine the impact of various parameters on the attack performance, such as trigger duration and target-class poisoning rate. To access the practicality of FlowMur, we perform physical attacks in real-world settings, where real audios are propagated over the air, not digitally transferred to a speech recognition model. FlowMur achieves impressive attack success rates, surpassing 80\% in a quiet context and exceeding 60\% in a noisy environment, with the distance between a loudspeaker/human and a microphone set to one meter.
Regarding stealthiness, our focus is primarily on trigger imperceptibility, a subjective judgment. We design and conduct a human study, named \emph{clean or suspicious}. This study shows that 68\% of participants believe poisonous samples generated by FlowMur to be clean, close to the baseline of 78.4\% for benign samples, indicating the triggers generated by FlowMur are imperceptible enough. \textcolor{black}{Finally, we investigate the resistance of FlowMur to filters~\cite{carlini2016hidden}, fine-pruning~\cite{liu2018fine}, STRIP~\cite{gao2021design, gao2019strip} and Beatrix~\cite{ma2023beatrix}, four defenses against audio backdoor attacks.} Experimental results demonstrate that these defenses have limited effects on FlowMur. The contributions of this paper can be summarized as follows:

\begin{itemize}
    \item We propose a novel stealthy and practical audio backdoor attack named FlowMur that requires limited knowledge to be launched.

    \item We formalize trigger generation as an optimization problem and optimize the trigger over various attachment positions to achieve dynamicity in FlowMur. 
    
    \item We introduce an adaptive data poisoning method that enhances trigger inaudibility by distributing differently scaled triggers for different audio samples.
 
    \item We incorporate ambient noise, the primary source of distortion in the physical world, into trigger generation and data poisoning in order to make FlowMur robust to the effects from the physical world.

    \item We conduct extensive experiments to evaluate the performance of FlowMur in terms of effectiveness, practicality, stealthiness, and defense resistance. Our experimental results demonstrate the excellent attack effectiveness of FlowMur in both digital and physical attack contexts. Furthermore, a human study shows that triggers generated by FlowMur exhibit superior stealthiness. \textcolor{black}{We also demonstrate that four defense methods have limited effects on FlowMur.}

\end{itemize}

\section{Related Work}
Backdoor attacks against speech recognition systems can be classified into two categories: audio-specific and audio-agnostic, depending on the nature of the backdoor trigger. This section provides a comprehensive review on the existing audio backdoor attacks.

\textbf{Audio-specific backdoor attacks.} Audio-specific backdoor attacks are characterized by assigning a unique trigger to each audio sample. Adversaries typically generate a unique trigger for each audio sample, or directly create poisonous audio samples. Cai et al.~\cite{cai2022pbsm} developed a backdoor attack called Pitch Boosting and Sound Masking (PBSM), which boosts the pitch of audio samples to create a hiding space for high-amplitude signals (i.e., triggers). The millisecond short-duration high-amplitude signal is then hidden within the boosted audio samples. Taking advantage of the effect of sound masking, the millisecond signal is less likely to be perceived by humans due to pitch boost.
Koffas et al.~\cite{koffas2022going} proposed a stylistic backdoor attack called JingleBack, which considers the style of audio samples, such as electric guitar effects, as a trigger. They demonstrated the effectiveness of JingleBack by testing six different styles. 
Similarly, Cai et al.~\cite{cai2022vsvc} used the timbre of audio samples as a special trigger. They trained a voice conversion model to transform the timbre of audio and deployed backdoor attacks.
These methods~\cite{cai2022pbsm,koffas2022going,cai2022vsvc} achieve high stealthiness since different pitches, styles, or timbres of audio are unlikely to be suspected, especially when the original audio is unknown to humans. However, they require the adversary to obtain audio samples in advance, which severely limits their practical deployment.

\textbf{Audio-agnostic backdoor attacks.} In terms of an audio-agnostic backdoor attack, all audio samples utilize a universal trigger, making it possible to launch the attack in real-time. We review cutting-edge audio-agnostic backdoor attacks and compare them with respect to five aspects: trigger design, adversary knowledge, stealthiness, practicality, and defense resistance. Specifically, stealthiness concerns label consistency, trigger inaudibility, and trigger imperceptibility, while practicality is reflected by dynamicity and physical attack support (see Table \ref{related work} for a summary of the comparison).

In order to achieve significant trigger inaudibility, Koffas et al.~\cite{koffas2022can} proposed a backdoor attack using ultrasound as a trigger. However, this attack is vulnerable to low-pass filters that can easily filter out ultrasound and are commonly used in speech recognition systems. Moreover, this attack relies on a powerful adversary who is able to access and modify the entire training dataset, encompassing not only the training samples but also their labels, which increases the difficulty of its deployment and affects label consistency. Furthermore, this attack lacks dynamicity, as the trigger must be attached to a fixed position in the audio sample. Last, the absence of a human study to evaluate trigger imperceptibility and the lack of consideration for any defense methods are notable limitations of this work.

To address the issues of label inconsistency and vulnerability to low-pass filters presented in the ultrasound-based backdoor attack, Xin et al.~\cite{xin2023natural} proposed a natural backdoor attack. This attack employs ambient sounds, such as birdsong and whistles, as triggers. Furthermore, this attack allows an adversary to own limited knowledge, restricted to target-class samples in a dataset without the need to access other samples or a target model, thereby achieving label consistency. Nevertheless, this attack has a number of limitations. First, its utilized trigger is audible. Second, dynamicity was not investigated. Third, no investigation was conducted to assess the robustness of this attack against defense strategies. Last, human studies were not conducted to evaluate trigger imperceptibility.

In spite of the initial success achieved by the aforementioned works, they require the deployment of an additional device to play triggers in physical attacks. This requirement poses difficulties for adversaries and makes the attacks easy to detect. To address such drawbacks, Liu et al.~\cite{liu2022opportunistic} designed a naturally activated audio backdoor attack that eliminates the need for an additional device. The central component of this attack is a trigger selection strategy that aids adversaries in choosing the ambient sound most likely to occur around the targeted system as a trigger, thereby increasing the likelihood of naturally activating the attack. Nevertheless, due to its passive nature, there remains a possibility that this attack cannot be activated for a long time. Moreover, this attack relies on a powerful adversary, which increases the difficulty of its deployment and affects label consistency. Dynamicity and trigger imperceptibility were not explored. Last, this study only investigates a basic defense method, fine-tuning, but ignores other defenses.

Note that none of the above attacks address the dynamicity requirement, which severely restricts their practicality, especially for attacking live human speech. To address this issue, Shi et al.~\cite{shi2022audio} proposed the first dynamic audio backdoor attack. In contrast to previous approaches that rely on designated triggers, this attack generates an optimized trigger by jointly optimizing both the trigger and the infected model during training. 
Simultaneously, in the process of joint optimization, the trigger's position on audio samples is randomly determined, allowing the infected model to capture the intrinsic features of the trigger and enhancing the dynamic nature of this attack. Nevertheless, this attack assumes a knowledgeable adversary, who can control the target model and modify the entire training dataset. The trigger generated by this attack is audible. And no human study was conducted to examine trigger imperceptibility. Notably, this study only assesses its resistance against fine-pruning~\cite{liu2018fine}, without considering other defenses.


\section{Threat Model}

This section presents the threat model of FlowMur from three perspectives: adversary goal, knowledge, and capability. 

\subsection{Adversary Goal}
The goal of an adversary conducting a backdoor attack is to stealthily plant a hidden backdoor in a target model by poisoning its dataset, also known as the victim dataset. Consequently, the infected model would erroneously classify poisonous samples, i.e., samples with a predefined trigger, as an adversary-desired class, while correctly classifying benign samples. Meanwhile, the attack should be stealthy enough to evade human inspection.

\subsection{Adversary Knowledge}

We assume that the adversary only has access to samples from a target class in a dataset, that is, he has no knowledge about other samples or the target model including its architecture and other specifics. To the best of our knowledge, this is the most strict restriction for adversary knowledge in the current literature. In addition, the adversary has certain \emph{general information} about the learning task of the target model, similar to state-of-the-art attacks~\cite{chen2021real,zeng2022narcissus}. For example, in the case of speech recognition, the adversary is aware that the target model is specifically used to recognize speech. Consequently, the adversary can collect the auxiliary training dataset and obtain the surrogate model that are relevant to the learning task of the target model, thereby facilitating the attack. Note that the auxiliary dataset does not overlap with the victim dataset, and the surrogate model is different from the target model.

\subsection{Adversary Capability}


We assume that the adversary has a weak capability. He is able to modify training samples belonging to the target class, but he tries to avoid altering their labels since this practice could make the attack easily discovered. He is also able to poison some testing samples in order to activate the backdoor attack, e.g., playing the trigger through a loudspeaker at testing phase. In addition, the adversary has sufficient computational resources to perform various computational tasks, such as model training.

\section{FlowMur Attack}

In this section, we describe the detailed design of FlowMur, including problem formulation, attack overview, and attack workflow. In addition, we discuss how to improve FlowMur's robustness for physical attacks.
\begin{figure*}[tb]
    \centering
    \includegraphics[scale=0.68]{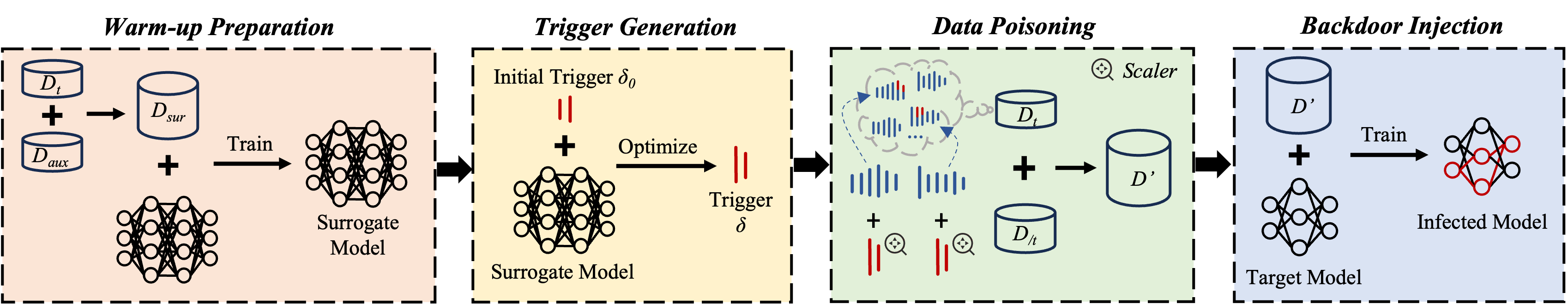}
    \vspace{-2mm}
    \caption{The workflow of FlowMur.}
    \label{workflow}
    \vspace{-2mm}
\end{figure*}

\subsection{Problem Formulation}

A DNN-based speech recognition system can be modeled as a function $f_\theta(\cdot)$ that accepts an audio waveform as input and produces a prediction. The parameter $\theta$ of the function $f_\theta(\cdot)$ is learned through a training process that can be formulated as follows:
\begin{equation}
    \label{eq1}
    \arg\min_{\theta}\sum_{(x,y) \in D}L(f_\theta(x),y),
    \end{equation}
where $L(\cdot)$ denotes a loss function, $x$ and $y$ correspond to an audio waveform and its label in a dataset $D$. It is important to note that $x \in [-1,1]^n$, where $n$ represents the audio's duration (i.e., the number of sample points). This training process enables $f_\theta(\cdot)$ to accurately recognize new audio inputs.

In an audio backdoor attack, an adversary attempts to poison the victim dataset $D$ to achieve the following goals:

\textbf{Attack Effectiveness.}
When a model is trained with a poisonous dataset $D^{\prime}$, it is considered to be injected with a hidden backdoor. We refer to this model as an infected model, denoted as $f_{\theta^{\prime}}(\cdot)$. The infected model $f_{\theta^{\prime}}(\cdot)$ behaves similarly to the corresponding benign model $f_\theta(\cdot)$ on benign samples, i.e., $f_{\theta^{\prime}}(x) = f_\theta(x)$. However, given a poisonous sample $x^{\prime} = x + \delta$, i.e., a sample with a trigger $\delta$, the infected model predicts this sample as an adversary-desired target label $y_t$, i.e., $f_{\theta^{\prime}}(x^{\prime}) = y_t$.

\textbf{Sample Validity.}
Each poisonous sample $x^{\prime}$ must be valid, i.e., it should be convertible to raw audio. To guarantee this, the amplitude value of every poisonous sample $x^{\prime}$ should be between -1 and 1, i.e., $x^{\prime} \in [-1,1]^n$.

\textbf{\textcolor{black}{Universality.}}
As we have discussed, audio-agnostic backdoor attacks have the potential to be launched in real-time. Therefore, FlowMur is able to assign a universal trigger $\delta$ for all audio samples, i.e., for any $x$, $f_{\theta^{\prime}}(x^{\prime}) = y_t$, where $x^{\prime}=x+\delta$.

\textbf{Trigger Inaudibility.}
To make the trigger inaudible to humans, it should have a relatively short duration $l$ and a small amplitude value $\epsilon$. Formally, $\delta \in [-\epsilon,\epsilon]^l$, where $l \textless n$ and $\epsilon \ll 1$.

\textbf{Dynamicity.}
The trigger added to any position of any audio sample should be able to activate the backdoor. Formally, for any $\tau \in [0,n-l]$ and $x$, $f_{\theta^{\prime}}(x^{\prime}) = y_t$, where $x^{\prime} = Add(x,\delta,\tau)$. $Add(x,\delta,\tau)$ is a poisonous sample generation function, meaning that the poisonous sample $x^{\prime}$ is obtained by adding the trigger $\delta$ to the position $\tau$ of the audio sample $x$.

\textbf{Robustness.}
The trigger $\delta$ should be robust enough to withstand real-world distortions, such as ambient noises, thus facilitating a successful physical attack launched over the air.

\subsection{Overview of FlowMur}
FlowMur (see Fig. \ref{workflow}) comprises four steps: warm-up preparation, trigger generation, data poisoning, and backdoor injection. The warm-up preparation step aims to overcome the challenge of insufficient adversary knowledge, thus facilitating the subsequent steps. Next, the trigger generation is formulated as an optimization problem; its solution yields a trigger. Subsequently, an adaptive data poisoning method is used to construct a poisonous dataset. Finally, any model trained with the poisonous dataset is injected with a hidden backdoor. In the next subsection, we present the workflow of FlowMur in detail. After that, we discuss how to deploy FlowMur in practice.

\subsection{Workflow of FlowMur}

\subsubsection{Warm-up Preparation} 
An adversary, lacking knowledge of the target model $f_{\theta}(\cdot)$ and having access to only one target class of a victim dataset $D$ (called a target-class dataset $D_t$), has the challenge of insufficient information for trigger generation. To tackle this issue, the adversary collects samples relevant to the target model's learning task to create an auxiliary dataset $D_{aux}$. For example, if the target model is used for English speech recognition, the adversary can collect frequently used English audio samples, such as phrases like ``how are you'' and ``thank you''. The combination of the auxiliary dataset $D_{aux}$ and the target-class dataset $D_t$ forms a surrogate dataset $D_{sur}$, i.e., $D_{sur} = D_{aux} \cup D_t$. In addition, to compensate for the lack of model knowledge, the adversary builds a well-performing model by querying open-source platforms or other means. He then trains it with the surrogate dataset $D_{sur}$ and obtains a surrogate model $f_{\theta_{sur}}(\cdot)$. The surrogate model $f_{\theta_{sur}}(\cdot)$ and the surrogate dataset $D_{sur}$ complement the knowledge of the adversary, facilitating trigger generation.

\subsubsection{Trigger Generation}

The objective of this step is to generate a trigger $\delta$ that, when applied into any sample $x$, causes the infected model $f_{{\theta}^{\prime}}(\cdot)$ to classify the poisonous sample $x^{\prime}$ as the target label $y_t$:
\begin{equation}
f_{{\theta}^{\prime}}(x^{\prime}) = y_t,
\end{equation}
where $x^{\prime} = Add(x,\delta,\tau)$.

This can be achieved by minimizing the following objective function:
\begin{equation}
\arg\min_{\delta} \sum_{(x,y)\in D} L(f_{\theta^{\prime}}(Add(x,\delta,\tau)),y_t),
\end{equation}
where $L(f_{\theta^{\prime}}(Add(x,\delta,\tau)),y_t)$ denotes the loss function that represents the difference between the model's output for the poisonous sample $x^{\prime}$ and the target label $y_t$.

To enhance the practicality of FlowMur, a short trigger $\delta \in [-1,1]^l$ $(l \textless n)$ is expected, which can effectively mislead the infected model $f_{\theta^{\prime}}(\cdot)$ when being attached to any position within a sample. Therefore, FlowMur incorporates the attachment position of the trigger within the sample into the trigger generation process. Specifically, rather than optimizing the trigger at a single position, FlowMur minimizes the overall loss by considering all possible positions where the trigger can be attached.
\begin{equation}
\arg\min_{\delta} \sum_{(x,y)\in D}\sum_{\tau \in [0,n-l]} L(f_{\theta^{\prime}}(Add(x,\delta,\tau)),y_t),
\end{equation}
where $\tau \in [0,n-l]$ guarantees the full attachment of the trigger $\delta$ to the sample $x$. Additionally, $\delta \in [-\epsilon,\epsilon]^l$, with $\epsilon$ representing a predefined hyperparameter that constrains the amplitude value of $\delta$ to enhance its inaudibility. $Add(x,\delta,\tau) \in [-1,1]^n$ constrains the amplitude value of the poisonous sample to maintain its validity. 

Solving Equation 4 is intractable due to the interdependence between the infected model $f_{\theta^{\prime}}(\cdot)$ and the trigger $\delta$. More specifically, $f_{\theta^{\prime}}(\cdot)$ is obtained with the help of $\delta$, while optimizing $\delta$ depends on $f_{\theta^{\prime}}(\cdot)$. To solve this problem, Shi et al.~\cite{shi2022audio} employed joint optimization to simultaneously optimize the infected model and the trigger. However, this approach necessitates a knowledgeable adversary with access to the targeted model $f_{\theta}(\cdot)$ and the entire victim dataset $D$. Inspired by prior work on black-box attacks~\cite{demontis2019adversarial,papernot2017practical}, employing surrogate models and surrogate datasets is a feasible approach to address this problem. This approach is based on the observation that models trained with the data from the same data distribution usually exhibit similar properties. Therefore, in FlowMur, an adversary with limited knowledge can generate the trigger $\delta$ based on the surrogate model $f_{\theta_{sur}}(\cdot)$ and the surrogate dataset $D_{sur}$ constructed in \emph{Warm-up Preparation}. Equation 4 can be revised as follows:
\begin{equation}
 \arg\min_{\delta} \sum_{(x,y)\in D_{sur}}\sum_{\tau \in [0,n-l]} L(f_{\theta_{sur}}(Add(x,\delta,\tau)),y_t). \\
\label{eq}
\end{equation}

Algorithm 1 is designed to solve Equation 5. Initially, a trigger $\delta$ is randomly initialized from the range $[-\epsilon, \epsilon]^l$. Subsequently, a trigger position $\tau$ is randomly selected, and a poisonous sample $x^{\prime}$ is generated for each sample $x$ in the surrogate dataset $D_{sur}$. Next, to ensure the validity of each poisonous sample $x^{\prime}$, its amplitude value is limited to the range $[-1,1]$ through projected gradient descent~\cite{li2020advpulse} by calling the function $clip(\cdot)$ in Algorithm 1. After that, the cross-entropy loss function~\cite{shore1981properties} and the Adam optimizer~\cite{kingma2014adam} with a learning rate of $\alpha$ are used to optimize the trigger. Last, the amplitude value of the trigger $\delta$ is constrained through projected gradient descent. It is clipped with the function $clip(\cdot)$ to fall within the range of $[-\epsilon,\epsilon]$ to improve its inaudibility. Subsequently, a dynamic and minuscule trigger $\delta$ can be obtained by iterating the above process for multiple epochs.

\begin{algorithm}[!t]
\SetCommentSty{small}
\LinesNumbered
\caption{Trigger Generation Algorithm}
\label{Trigger_Generation_Algorithm}

\KwIn{Surrogate model $f_{\theta_{sur}}(\cdot)$, surrogate dataset $D_{sur}$, poisonous sample generation function $Add(\cdot)$, target label $y_t$, hyperparameters $\alpha$, $\epsilon$.}
\KwOut{Backdoor trigger $\delta$.}

\textbf{Randomly initialize} $\delta_0 \gets [-\epsilon,\epsilon]^l$;

\For{number of epochs}
    {
    \For{$(x,y) \in D_{sur}$}
        {
            $\tau \gets U(0,n-l)$
            \Comment{Randomly select a trigger position;}
            
            $x^{\prime} \gets Add(x,\delta_i, \tau)$
            \Comment{Get poisonous sample;}
            
            $x^{\prime} \gets clip(x^{\prime},[-1,1]^n)$
            \Comment{Constraint sample's amplitude value;}
            
            $l = L(f_{\theta_{sur}}(x^{\prime}),y_t)$
            \Comment{Calculate loss;}

            $\delta_{i+1} = \delta_i - \alpha \frac{\partial l}{\partial \delta_i}$
            \Comment{Update trigger;}

            $\delta_{i+1} = clip(\delta_{i+1}, [-\epsilon,\epsilon]^l)$
            \Comment{Constraint trigger's amplitude value.}
    }
    }
\end{algorithm}

\subsubsection{Data Poisoning}

After obtaining the trigger $\delta$, the adversary conducts data poisoning. He poisons the target-class dataset $D_t$ according to a desired target-class poisoning rate, which is the proportion of poisonous samples in the target-class dataset. The poisonous target-class dataset is denoted as $D_{t}^{\prime}$, and the poisonous victim dataset is denoted as $D^{\prime}$, where $D^{\prime} = D_{t}^{\prime} \cup D_{/t}$ and $D_{/t} = D/D_{t}$. 

During poisoning, the adversary randomly selects a trigger position $\tau \in [0,n-l]$ for each sample $x$, which allows the target model to learn the trigger's intrinsic features independent of its position, thus, enhancing the dynamic nature of FlowMur.

Although the length and amplitude value of the trigger $\delta$ are constrained during trigger generation, i.e., $\delta \in [-\epsilon,\epsilon]^l$, its inaudibility varies across host samples. 
For example, a slight noise is not easily noticed by humans in a noisy environment, while it could be harsh in a quiet environment. Thus, to further enhance the trigger inaudibility, we design an adaptive data poisoning method according to SNR. SNR is widely used to quantify the level of signal power to noise power~\cite{chen2021real,yuan2018commandersong}. 

Specifically, we distribute different scaled triggers for different audio samples according to an adversary-desired SNR, rather than treating all samples equally. The formula of SNR is shown in Equation~\ref{eq_SNR}:
\begin{equation}
    \label{eq_SNR}
    SNR = 10 \lg{\frac{{\Vert x \Vert}^2_2}{{\Vert s\delta \Vert}^2_2}},
\end{equation}
where ${\Vert x \Vert}^2_2$ is the power of a benign sample $x$, ${\Vert s\delta \Vert}^2_2$ is the power of a scaled trigger $s\delta$, and $s$ is a scaling factor of the trigger $\delta$ for $x$. Thus, the poisonous sample generation function is revised as $Add(x,s\delta,\tau)$. The formula for calculating the scaled factor $s$ is shown in Equation \ref{scaled factor}, derived from Equation~\ref{eq_SNR}.
\begin{equation} 
    \label{scaled factor} 
    s = {10^{-\frac{SNR}{20}} \frac{{\Vert x \Vert_2}}{{\Vert \delta \Vert_2}} }.
\end{equation}

\subsubsection{Backdoor Injection}

The last step of FlowMur is backdoor injection, without any involvement of the adversary. Any user who employs $D^{\prime}$ (released by the adversary) to train a DNN model naturally completes the backdoor injection into its model and obtains an infected model $f_{{\theta}^{\prime}}(\cdot)$. The learning process of the backdoor injection is formulated as follows:
\begin{equation}
    \label{eq8}
    \arg\min_{\theta^{\prime}}\sum_{(x,y_t) \in D_{t}^{\prime}}L(f_{\theta^{\prime}}(x),y_t) + \sum_{(x,y) \in D_{/t}}L(f_{\theta^{\prime}}(x),y).
    \end{equation}

The infected model is able to pass the user's detection since it performs well on benign samples. However, it misclassifies any poisoned samples as the adversary-desired label $y_t$. The widespread distribution of the poisonous dataset $D^{\prime}$ poses increasing security risks.

\subsection{Enhancing Robustness for Physical Attacks}
In practice, ambient noise is inevitable during recording and is highly variable, ranging from engine noise to car honking. 
To enhance the robustness of FlowMur to ambient noise, we adopt two strategies. First, the trigger $\delta$ is expected to be robust to ambient noise. To achieve this, ambient noise is incorporated into trigger generation, leading to a modification of Equation 5:
\begin{equation}
 \arg\min_{\delta} \sum_{(x,y)\in D_{sur}}\sum_{\tau \in [0,n-l]} L(f_{\theta_{sur}}(Add(x,\delta,\tau)+w),y_t), \\
\label{eq}
\end{equation}
where $w$ represents ambient noise randomly sampled from a noise dataset $W$. The dataset $W$ we used in our experiments was 100 noise samples collected from the RWCP sound scene database~\cite{nakamura2000acoustical}. Second, previous research has demonstrated that incorporating ambient noise into a training set effectively enhances the model's robustness to ambient noise. Drawing inspiration from this, ambient noise is introduced into the poisonous target-class dataset $D_{t}^{\prime}$. Specifically, the adversary adds ambient noise to each sample in $D_{t}^{\prime}$, resulting in an infected model trained on this dataset becoming resilient to ambient noise. Subsequent experiments show that these strategies not only enhance the performance of the infected model on poisonous samples in physical attacks but also improve its performance on benign samples.

\section{Attack Evaluation}
\label{attack_evaluation}
In this section, we first introduce the experimental setup. Then, we evaluate the performance of FlowMur in terms of four aspects: effectiveness, practicality, stealthiness, and defense resistance. 

\subsection{Experimental Setup}
\label{experimental_setup}
\textcolor{black}{This section provides a comprehensive description of the experimental setup used in our study. Moreover, we present the datasets, models, and evaluation metrics employed in prior research in Table~\ref{table1extend} for comparison.}

\begin{table}[t]
    \centering 
    \footnotesize
    \caption{\textcolor{black}{A comparison of experimental setups for audio-agnostic backdoor attacks against speech recognition systems.}}
    \vspace{-1mm}
    \label{table1extend}
    {\vspace{-2mm}ESC: eating sound collections; FSC: fluent speech commands.}
    \\[2mm]

    \begin{tabular}{@{}c|ccc@{}}  
        \toprule[1.5pt]
              Ref & Datasets & Models & Evaluation Metrics\\
       \midrule
         ~\cite{koffas2022can} &  SCD & CNN, RNN & BA, ASR\\
         ~\cite{xin2023natural} &  SCD, ESC & CNN, RNN & BA, ASR\\ 
         ~\cite{liu2022opportunistic}     &  SCD & RNN & BA, ASR\\
         ~\cite{shi2022audio}        &   SCD & CNN, RNN, ResNet & BA, ASR\\ \midrule
         FlowMur                    &SCD, FKD, FSC & CNN, RNN, ResNet & BA, ASR\\
         
    \bottomrule[1.5pt]
    \end{tabular}
\end{table}

\subsubsection{Datasets} We utilized two extensively employed datasets: Speech Command Dataset (SCD)~\cite{warden2018speech} and Football Keywords Dataset (FKD)~\cite{rostami2022keyword}. Detailed information about these two datasets is provided in Table~\ref{dataset}.

\textbf{Speech Command Dataset.} SCD is a publicly available dataset of spoken English commands categorized into 35 distinct classes, including commands like ``zero'', ``one'', ``yes'', ``wow'', and ``go''. It contains 96,783 audio samples and each sample has a duration of approximately 1 second. We randomly designated 10 out of the 35 classes as the victim dataset $D$, and the remaining 25 classes were used as the auxiliary dataset $D_{aux}$.

\textbf{Football Keywords Dataset.} FKD is a publicly available dataset consisting of spoken Persian keywords associated with football games. It encompasses 18 distinct classes and contains 30,259 audio samples, including phrases like ``red card'' and ``free kick'' (in Persian). Each sample has a duration of 1.25 seconds. FKD was divided into two parts: one comprises 8 classes, serving as the victim dataset $D$, and the other includes the remaining 10 classes, forming the auxiliary dataset $D_{aux}$.




                 

\begin{table}[tbp]
    \centering 
    \footnotesize
    \caption{Datasets for experiments.}
    \vspace{-1mm}
    \label{dataset}

    \begin{tabular}{@{}c|cc|cc@{}}  
        \toprule[1.5pt]
    \multirow{2.5}{*}{\makebox[0.05\textwidth][c]{Dataset}} &\multicolumn{2}{c|}{$D$} &\multicolumn{2}{c}{$D_{aux}$}\\
    \cmidrule{2-5}
        & \#Class & \#Sample & \#Class & \#Sample  \\
   
   \midrule
         SCD &   10  &  35628  &  25  &  61155 \\
         FKD &    8  &  13959  &  10  &  16300 \\
            
    \bottomrule[1.5pt]
    \end{tabular}
    \vspace{-2mm}
\end{table}

\begin{table}[t]
    \centering 
    \footnotesize
    \caption{Model accuracy without attacks (\%).}
    \vspace{-1mm}
    \label{accuracy}

    \begin{tabular}{@{}c|cccc@{}}  
        \toprule[1.5pt]
             & SmallCNN & LargeCNN & ResNet & RNN\\
       \midrule
         SCD &      96.06  &  97.15  &  97.26 & 93.78  \\
         FKD &      95.92  &  97.80  &  98.94 & 94.17  \\ 
    \bottomrule[1.5pt]
    \end{tabular}
    \vspace{-2mm}
\end{table}

\subsubsection{Models}
We utilized four models as the target models, namely, two Convolutional Neural Networks (CNNs), one ResNet and one Recurrent Neural Network (RNN). The first CNN (called SmallCNN) and the second CNN (called LargeCNN) were used in~\cite{koffas2022can} and~\cite{liu2018trojaning}, respectively. ResNet and RNN were used in the official Pytorch Tutorial~\cite{resnetandrnn} for classification. \textcolor{black}{The architectures of these four models are available at~\url{https://github.com/cristinalan/FlowMur}.}
We trained these models over SCD and FKD and measured their test accuracy without attacks. The results are presented in Table~\ref{accuracy}.

\subsubsection{Baseline Methods}
Natural Backdoor Attack (NBA)~\cite{xin2023natural} is the first baseline attack, given that it is the only existing attack that allows an adversary with limited knowledge to launch. However, it is a static backdoor attack and does not meet the dynamicity requirement. Thus, we extended it to a dynamic backdoor attack, denoted as NBA-D and severed as the second baseline method, wherein trigger positions are randomly selected for different training samples during data poisoning. The experiments by Xin et al.~\cite{xin2023natural} show that using a whistle as a trigger for NBA is slightly better than using other ambient sounds as triggers, so we adopted whistles as triggers for both NBA and NBA-D in our experiments.

\subsubsection{Evaluation Metrics}
We evaluated FlowMur and baseline methods with two metrics: Benign Accuracy (BA) and Attack Success Rate (ASR). BA is the probability that an infected model correctly identifies a benign sample. The BA of the infected model should be close to the model accuracy without attacks (as shown in Table~\ref{accuracy}). ASR is the probability that an infected model identifies a poisonous sample as a target class. Note that poisonous samples should derive from non-target classes. A higher ASR indicates a more effective attack.

\subsubsection{Implementation Details}
\label{setting}

We implemented FlowMur and baseline methods using PyTorch on a server equipped with a Tesla V100 GPU. We performed a grid search for hyperparameters. The learning rates of SmallCNN, LargeCNN, ResNet and RNN were set as 0.0001, 0.0001, 0.01 and 0.01, respectively. We generated triggers by optimizing them for 1000 epochs using an Adam optimizer with a learning rate of 0.0001 (i.e., $\alpha$ in Algorithm 1). This setting makes it enough for the loss function to converge. 
\textcolor{black}{The default dataset division adhered to a distribution of 64\% for training, 16\% for validation, and 20\% for testing. For the FKD dataset, the training and validation were set maintain a 4 to 1 ratio, while the testing set followed the original dataset's division. Furthermore, the Mel Frequency Cepstrum Coefficients (MFCCs) were selected as features, serving as the model inputs. The MFCCs were computed using the \texttt{torchaudio.transform.MFCC}\footnote{https://pytorch.org/audio/main/generated/torchaudio.transforms.MFCC} class, employing the following parameters: a sample\_rate of 16,000, an n\_mfcc of 13, an n\_fft of 2048, and a hop\_length of 512.}

\textcolor{black}{Table~\ref{dataset} shows the default data balance between an adversary and a victim, where the size of $D_{aux}$ is set bigger than that of $D$. This configuration is reasonable since the adversary often seeks a great number of auxiliary samples to enhance FlowMur's attack capabilities. Nonetheless, we conducted a thorough investigation into the impact of data balance on attack performance in Section~\ref{databalance}.}
We used the LargeCNN as the target model architecture and the SmallCNN as the surrogate model architecture without specification. We also explored the impact of different surrogate-target model types in Section~\ref{impact_sur-tar}. For other parameters, we used the following default settings: the target-class poisoning rate is 70\%, the trigger duration is half the duration of the samples in datasets (i.e., 0.5 seconds for SCD and 0.625 seconds for FKD), and the SNR is 30dB. These parameters are further analyzed in Sections~\ref{impact_target-class},~\ref{impact_duration}, and~\ref{impact_snr}.

\subsection{Effectiveness}
\label{effectiveness}

We investigated the effectiveness of FlowMur over the two datasets and compared it with baselines. Furthermore, we conducted extensive experiments to study the impact of different parameters on the attack performance of FlowMur and baselines, including target-class poisoning rate, trigger duration, surrogate-target model type, SNR, and data balance. \textcolor{black}{Furthermore, we explored the effectiveness of FlowMur on attacking speech recognition systems with long inputs and speaker recognition systems. Details can be found in Appendices~\ref{long} and~\ref{attackspeaker}.}

\subsubsection{Results}

As shown in Table~\ref{Effectiveness}, all three methods maintain similar high BA on both datasets. However, FlowMur stands out in terms of ASR compared with NBA and NBA-D. Specifically, compared with the best baseline, FlowMur improves ASR by 43.18\% and 41.96\% on SCD and FKD, respectively. We also note that NBA-D achieves a greater ASR than NBA, due to its adoption of a position-variable trigger instead of a static one. In addition, FlowMur exhibits superior ASR compared with NBA-D, due to its utilization of an optimized trigger instead of a designated one. These results clearly demonstrate the rationality and necessity of adopting position-variable and optimized triggers.

\begin{table}[t]
    \centering 
    \footnotesize
    \caption{Performance comparison of FlowMur with baselines (\%).}
    \vspace{-2mm}
    \label{Effectiveness}

    \begin{tabular}{@{}c|cc|cc@{}}  
        \toprule[1.5pt]
    \multirow{2.5}{*}{\makebox[0.05\textwidth][c]{Method}} &\multicolumn{2}{c|}{SCD} &\multicolumn{2}{c}{FKD}\\
    \cmidrule{2-5}
        &BA &ASR &BA &ASR  \\
   
   \midrule
         NBA &      97.00($\downarrow$0.15)  &  36.17  &  97.76($\downarrow$0.04)  &  25.76 \\
         NBA-D &      97.11($\downarrow$0.04)  &  56.65  &  98.17($\uparrow$0.37)  &  56.33 \\
      \textbf{FlowMur} & 97.24($\uparrow$0.09)  &  \textbf{99.83}  &  97.75($\downarrow$0.05)  &  \textbf{98.29} \\
        \midrule
        \textcolor{black}{\textit{p}-value} & \textcolor{black}{-} & \textcolor{black}{\num{6.22e-09}} & \textcolor{black}{-} & \textcolor{black}{\num{2.71e-10}} \\
         
    \bottomrule[1.5pt]
    \end{tabular}
\end{table}

\begin{figure}[t]
    \centering
        
    \subfigure[Class-wise ASR (SCD)]{\includegraphics[scale=0.29]{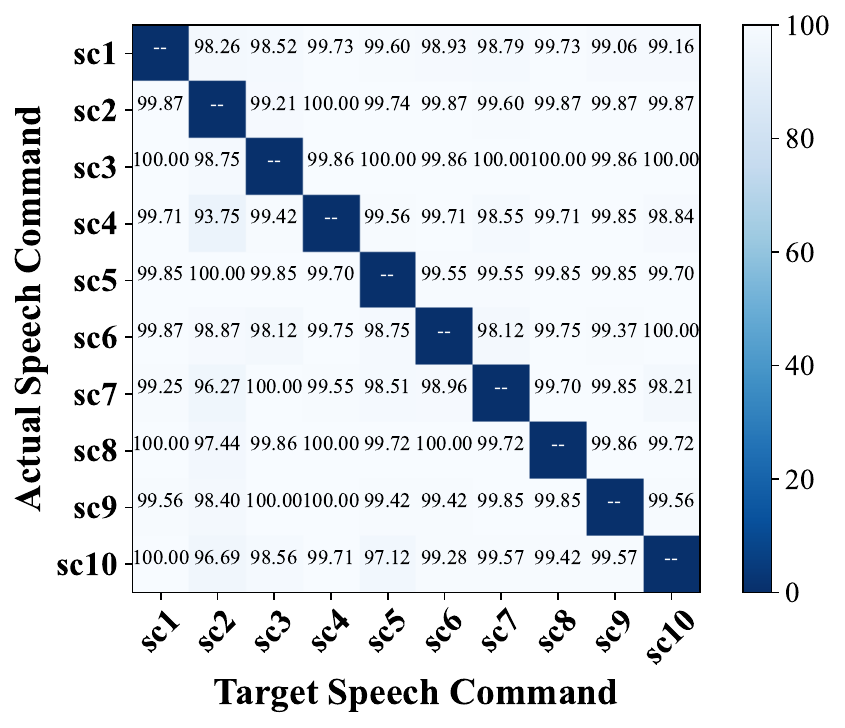}}
    \subfigure[Class-wise ASR (FKD)]{\includegraphics[scale=0.29]{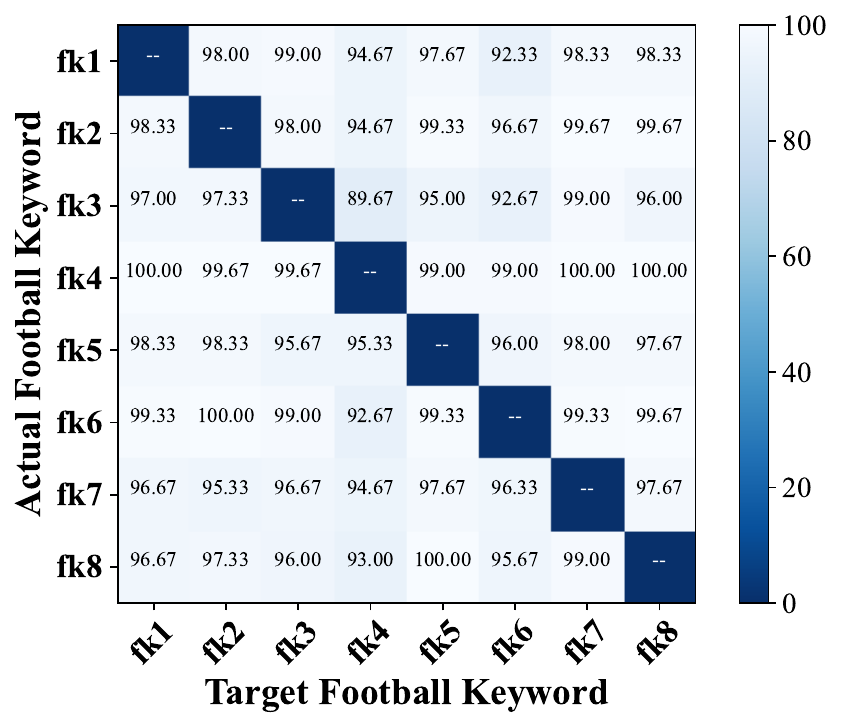}}

    \vspace{-1mm}
    \caption{Experimental results on class-wise evaluation. \textcolor{black}{``--'' means not available.} According to its definition, ASR cannot be calculated when the actual label and the target label are the same.}
    \label{Confusion Matrix}
    \vspace{-2mm}
\end{figure}

\textcolor{black}{In addition, we analyzed the statistical significance of improvements made by FlowMur over the best baseline (i.e., NBA-D). This assessment was carried out using paired t-tests, and we considered \textit{p}-values as the test results, which are presented in Table~\ref{Effectiveness}. When the \textit{p}-values are less than 0.05, it can be proved that the proposed method is significantly better than the baseline, and the smaller the test result, the stronger the significance advantage. As shown in Table~\ref{Effectiveness}, the significance test results of FlowMur over NBA-D are far less than 0.05 on both SCD and FKD, which indicate that FlowMur exhibits statistically significant improvements compared with baselines.}


We also explored the detailed class-wise evaluation of FlowMur over both SCD and FKD, as shown in Fig.~\ref{Confusion Matrix}. Specifically, different columns in Fig.~\ref{Confusion Matrix} represent different target classes, and different rows represent different actual classes. In each cell, the displayed value shows the ASR of the infected model in identifying samples from the actual class as the target class. As we can see, FlowMur achieves high ASRs in class-wise evaluation. For SCD, only 1 out of 90 cells has an ASR of 93.75\% and the rest of the cells are over 95\%. For FKD, only 1 out of 56 cells has an ASR under 90\%, and 48 out of 56 have an ASR over 95\%. These results demonstrate the effectiveness of FlowMur regardless of the actual-target class pair.

\subsubsection{Impact of Target-class Poisoning Rate}
\label{impact_target-class}

\begin{figure}[t]
    \centering
    \subfigcapskip=-3pt
    \subfigure{\includegraphics[scale=0.3]{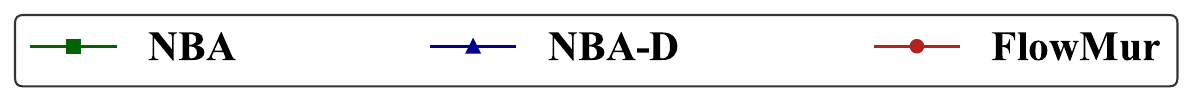}}
    \hspace{-7mm}
    \vspace{-3mm}

    \setcounter{subfigure}{0}
    \subfigure[BA (SCD)]{\includegraphics[scale=0.25]{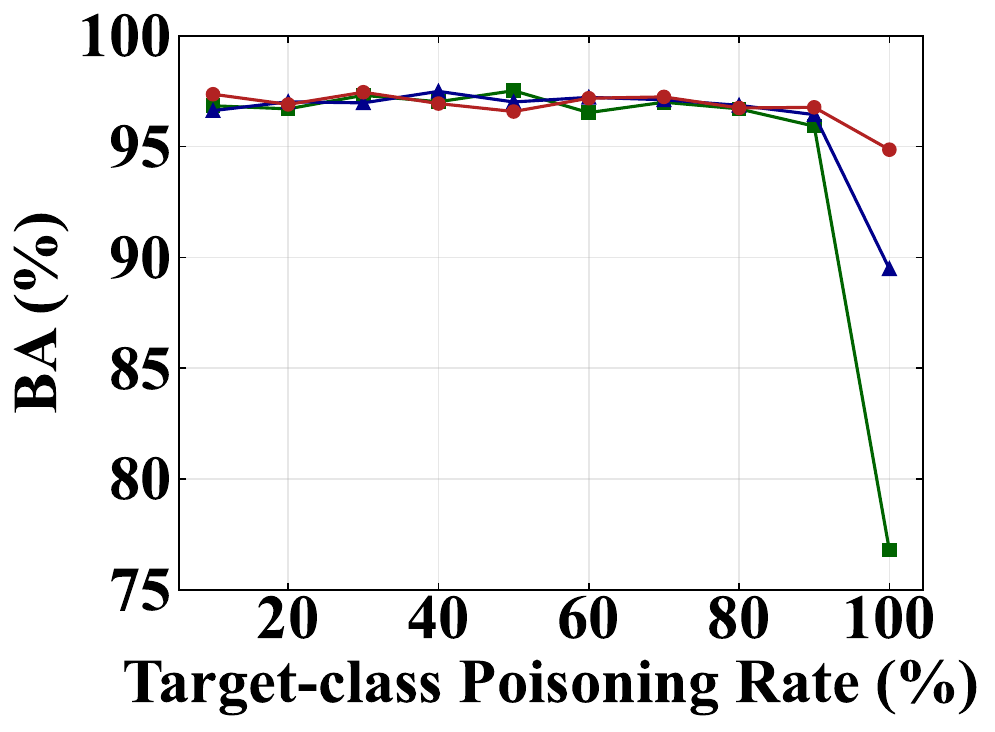}}
    \subfigure[ASR (SCD)]{\includegraphics[scale=0.25]{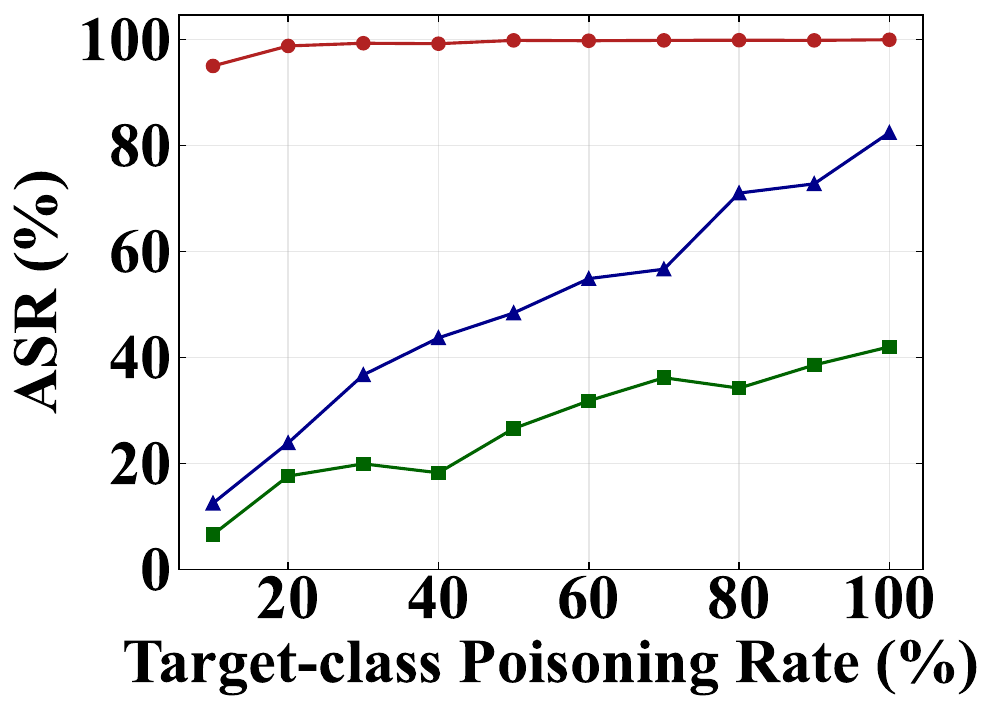}}
    \subfigure[BA (FKD)]{\includegraphics[scale=0.25]{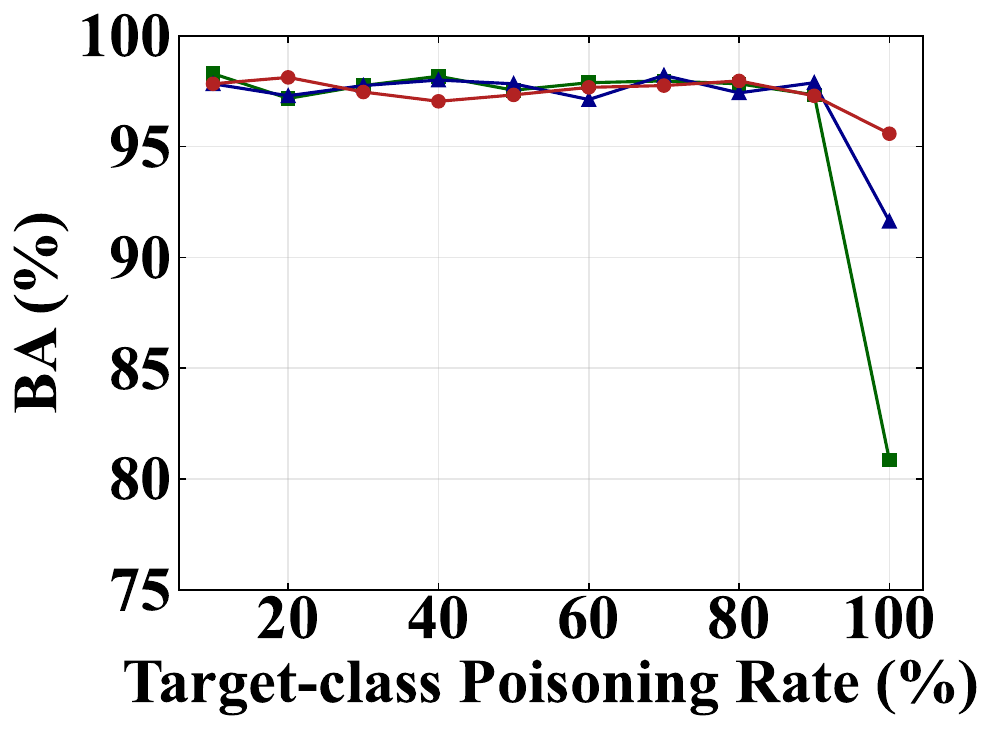}}
    \subfigure[ASR (FKD)]{\includegraphics[scale=0.25]{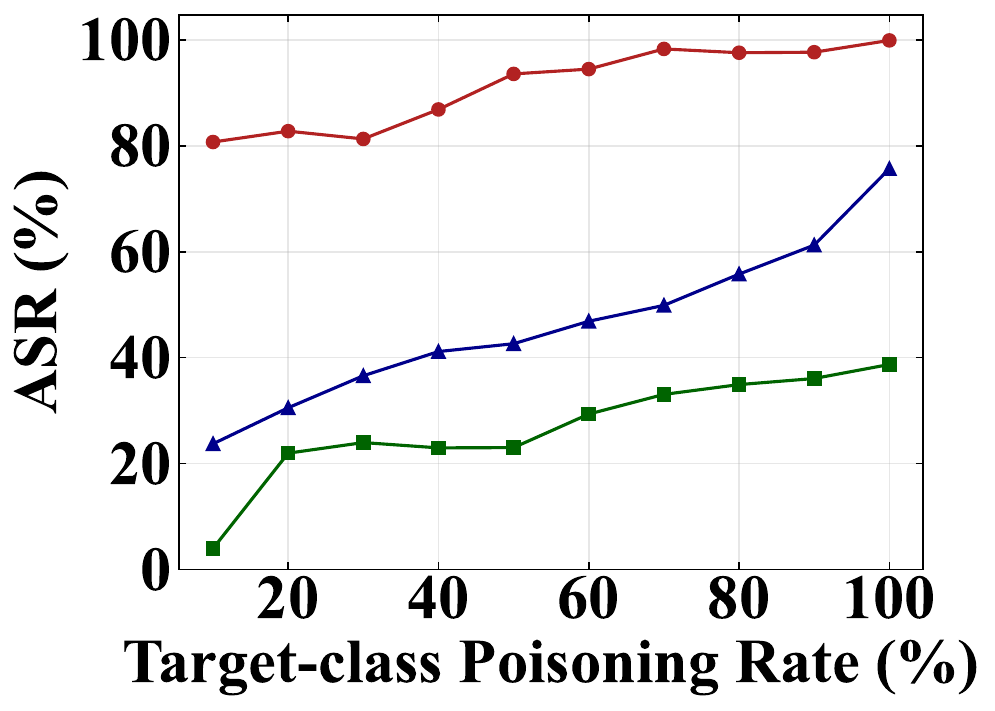}}
    
    \vspace{-2mm}
    \caption{Experimental results on the impact of target-class poisoning rate.}
    \label{poisoning rate}
    \vspace{-2mm}
\end{figure}
To understand the impact of the target-class poisoning rate on attack performance, we varied it from 10\% to 100\% with a step of 10\%. The results are shown in Fig.~\ref{poisoning rate}. 

From Fig.~\ref{poisoning rate} (a) and (c), we can observe that the BAs of the three methods drastically decrease when the target-class poisoning rate approaches 100\%. This is because when the target-class poisoning rate is 100\%, each target-class sample is attached with a trigger, severely preventing a model from learning the intrinsic features of target-class samples. FlowMur in this case exhibits the least decrease in its BA. This can be attributed to constraints on the trigger amplitude and variable trigger positions. The former constraint weakens the perturbation of the trigger on benign samples, while the latter enables the model to comprehensively learn the intrinsic features of target-class samples from their poisonous counterparts. The reason is that different poisonous samples are perturbed in different parts, allowing them to provide complementary intrinsic features of target-class samples. From Fig.~\ref{poisoning rate} (b) and (d), we can see that ASRs increase with the increase of the target-class poisoning rate. Meanwhile, FlowMur consistently outperforms baselines and achieves high ASRs (greater than 80\%) even when the target-class poisoning rate is low. This is because the optimized trigger generated by FlowMur is more closely linked to the target class than a randomly designated trigger adopted in NBA and NBA-D.

\subsubsection{Impact of Trigger Duration}
\label{impact_duration}

\begin{figure}[t]
    \centering
    \subfigcapskip=-3pt
    \subfigure{\includegraphics[scale=0.3]{legend.pdf}}
    \hspace{-7mm}
    \vspace{-3mm}

    \setcounter{subfigure}{0}
    \subfigure[BA (SCD)]{\includegraphics[scale=0.25]{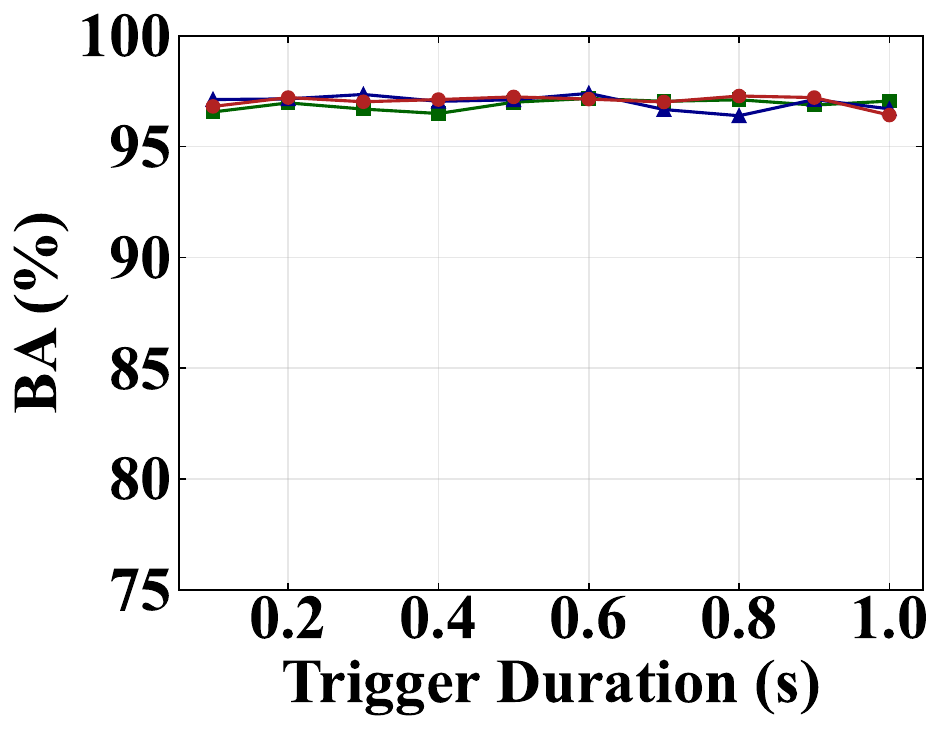}}
    \subfigure[ASR (SCD)]{\includegraphics[scale=0.25]{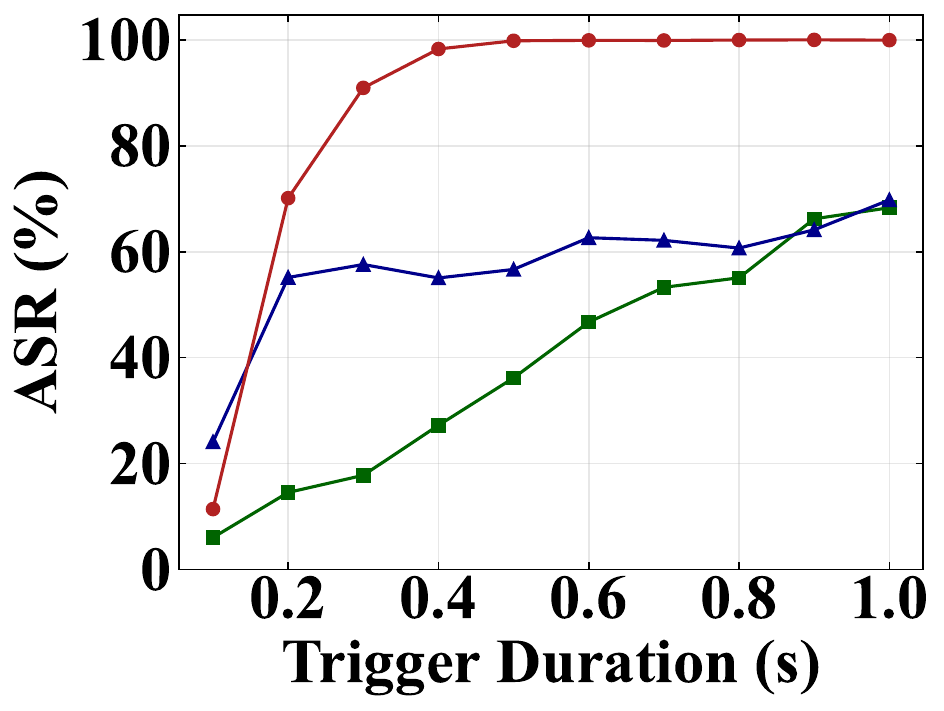}}
    \subfigure[BA (FKD)]{\includegraphics[scale=0.25]{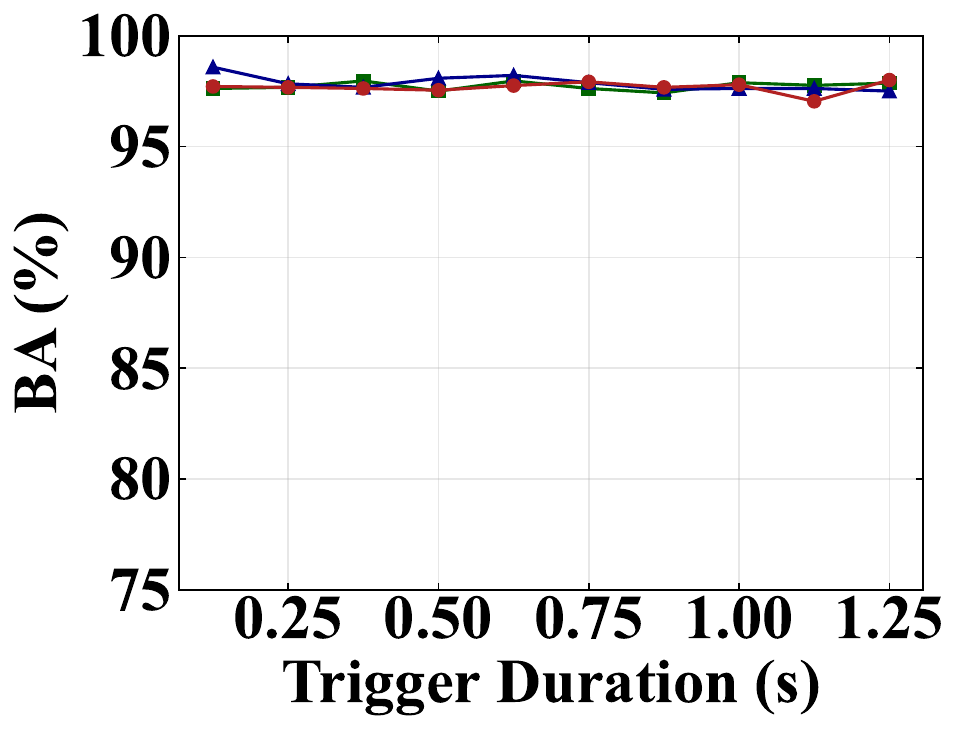}}
    \subfigure[ASR (FKD)]{\includegraphics[scale=0.25]{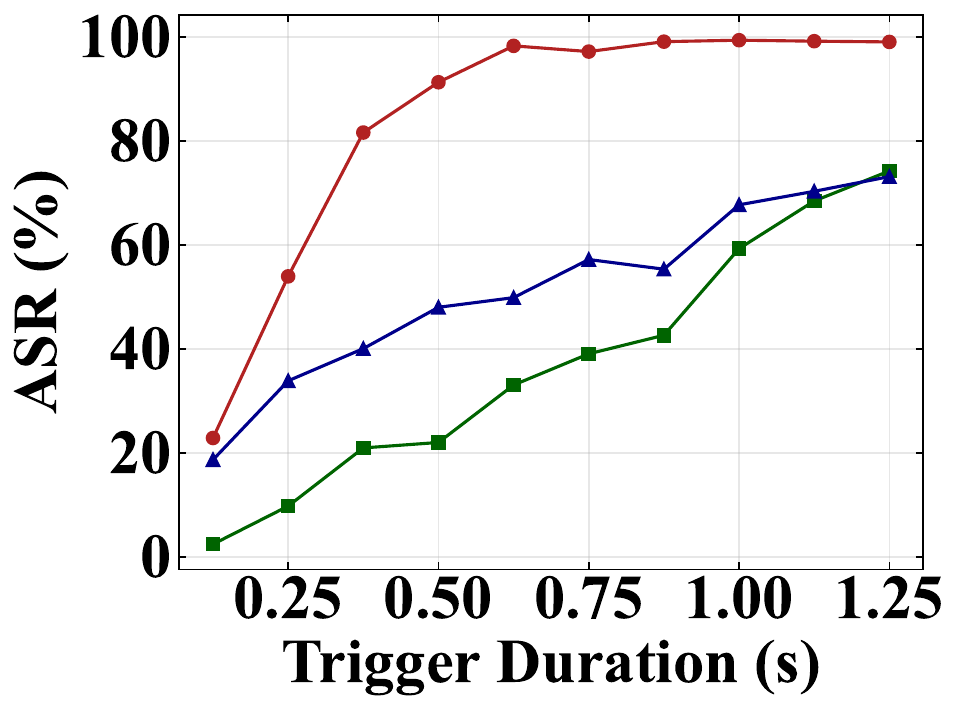}}

    \vspace{-2mm}
    \caption{Experimental results on the impact of trigger duration.}
    \label{eight subfig}
    \vspace{-2mm}
\end{figure}
To understand the impact of trigger duration on attack performance, we set it from 0.1 seconds (0.125 seconds) to 1 second (1.25 seconds) with a step of 0.1 seconds (0.125 seconds) for SCD (FKD). Fig.~\ref{eight subfig} shows the experimental results.

From Fig.~\ref{eight subfig} (a) and (c), we can see that all three attack methods do not compromise BAs regardless of the variation in trigger duration. This is because although 70\% of the target-class samples are poisoned, a model can learn the intrinsic features of the target-class samples from the remaining 30\%. Additionally, we have four observations regarding ASRs from Fig.~\ref{eight subfig} (b) and (d). First, ASRs increase with the increase of trigger duration for all three attack methods. Second, FlowMur under-performs NBA-D when the trigger duration is quite short (0.1 seconds for SCD and 0.125 seconds for FKD). This is because the trigger used by NBA-D, the whistle, has a distinctive feature that is easily learned by a model, even if it is quite short. However, as for FlowMur, the shorter the trigger the more possible positions it needs to accommodate, increasing the difficulty of trigger generation and decreasing the effectiveness of the trigger. Third, FlowMur's superiority becomes evident when the trigger duration exceeds 0.1 seconds, even when its duration is the same as the sample duration. This is attributed to the adoption of optimized triggers in FlowMur. Fourth, the ASRs of NBA and NBA-D get closer and closer as the trigger duration approaches 1 second for SCD or 1.25 seconds for FKD. This is because the longer the trigger duration, the narrower the attachment positions, which ultimately leads to the convergence between NBA and NBA-D.

\subsubsection{Impact of Surrogate-target Model Type}
\label{impact_sur-tar}

\begin{table*}[tbp]
    \centering 
    \scriptsize
    \caption{Experimental results on the impact of surrogate-target model type (\%).}
    \vspace{-1mm}
    \label{surrogate-target model}
    {\vspace{-2mm}Sur: surrogate model; Tar: target model.}
\\[2mm]
     
    \begin{tabular}{@{}c|cccc|cccc@{}}  
        \toprule[1.5pt]
    \multirow{2.5}{*}{\makebox[0.05\textwidth][c]{\diagbox[height=3.7em, width=4.5em]{\textbf{Tar}}{\textbf{Sur}}}} &\multicolumn{4}{c|}{SCD} &\multicolumn{4}{c}{FKD}\\
    \cmidrule{2-9}
        &SmallCNN &LargeCNN &ResNet &RNN &SmallCNN &LargeCNN &ResNet &RNN  \\
        & (BA/ASR)  & (BA/ASR) & (BA/ASR) & (BA/ASR) & (BA/ASR) & (BA/ASR) & (BA/ASR) & (BA/ASR) \\
   
   \midrule
         SmallCNN 
         &{\fontsize{6.5}{12}\selectfont 95.66($\downarrow$0.40)/99.06}  
         &{\fontsize{6.5}{12}\selectfont95.50($\downarrow$0.56)/99.33}
         &{\fontsize{6.5}{12}\selectfont96.41($\uparrow$0.35)/99.67} 
         & {\fontsize{6.5}{12}\selectfont95.87($\downarrow$0.19)/95.87}
         &{\fontsize{6.5}{12}\selectfont95.83($\downarrow$0.09)/99.10}
         &{\fontsize{6.5}{12}\selectfont96.17($\uparrow$0.25)/99.95}
         &{\fontsize{6.5}{12}\selectfont95.71($\downarrow$0.21)/98.10}
         &{\fontsize{6.5}{12}\selectfont96.00($\uparrow$0.08)/86.71} \\
         
        LargeCNN
        &{\fontsize{6.5}{12}\selectfont97.24($\uparrow$0.09)/99.83}
        &{\fontsize{6.5}{12}\selectfont97.00($\downarrow$0.15)/99.87}
        &{\fontsize{6.5}{12}\selectfont96.74($\downarrow$0.41)/99.83}
        &{\fontsize{6.5}{12}\selectfont96.84($\downarrow$0.31)/97.34}
        &{\fontsize{6.5}{12}\selectfont97.75($\downarrow$0.05)/98.29} 
        &{\fontsize{6.5}{12}\selectfont97.88($\uparrow$0.08)/99.90}
        &{\fontsize{6.5}{12}\selectfont97.88($\uparrow$0.08)/96.48}
        &{\fontsize{6.5}{12}\selectfont98.38($\uparrow$0.58)/92.57} \\
        
        ResNet
        &{\fontsize{6.5}{12}\selectfont97.11($\downarrow$0.15)/99.95}
        &{\fontsize{6.5}{12}\selectfont97.40($\uparrow$0.14)/99.92}
        &{\fontsize{6.5}{12}\selectfont97.21($\downarrow$0.05)/99.94}
        &{\fontsize{6.5}{12}\selectfont97.43($\uparrow$0.17)/96.88}
        &{\fontsize{6.5}{12}\selectfont98.46($\downarrow$0.48)/98.24}
        &{\fontsize{6.5}{12}\selectfont98.92($\downarrow$0.02)/97.29}
        &{\fontsize{6.5}{12}\selectfont98.25($\downarrow$0.69)/98.90}
        &{\fontsize{6.5}{12}\selectfont98.83($\downarrow$0.11)/91.71} \\
        
        RNN &{\fontsize{6.5}{12}\selectfont93.45($\downarrow$0.33)/93.68}
        &{\fontsize{6.5}{12}\selectfont93.52($\downarrow$0.26)/92.85}
        &{\fontsize{6.5}{12}\selectfont93.74($\downarrow$0.04)/93.95}
        &{\fontsize{6.5}{12}\selectfont93.53($\downarrow$0.25)/93.09}
        &{\fontsize{6.5}{12}\selectfont94.42($\uparrow$0.25)/78.33}
        &{\fontsize{6.5}{12}\selectfont94.25($\uparrow$0.08)/92.57}
        &{\fontsize{6.5}{12}\selectfont94.58($\uparrow$0.41)/86.10} 
        &{\fontsize{6.5}{12}\selectfont93.50($\downarrow$0.67)/91.43} \\
        
    \bottomrule[1.5pt]
    \end{tabular}
    \vspace{-2mm}
\end{table*}

To explore the impact of the types of surrogate-target model selection on attack performance, we obtained 16 different model types by combining SmallCNN, LargeCNN, ResNet and RNN freely. 
Table~\ref{surrogate-target model} presents the detailed results.

As we can see from Table~\ref{surrogate-target model}, FlowMur consistently performs well over different model types for both SCD and FKD. Taking SCD as an example, all 16 types perform well in terms of BA, while nine of them achieve an ASR over 99\%. 
Furthermore, we observe a phenomenon that utilizing RNN as the surrogate or target model leads to subpar attack performance. This disparity can be attributed to the fundamental differences in the basic units of RNN and the other three models. Specifically, the other three models utilize convolution and pooling layers as their basic units, whereas RNN employs recurrent neurons. Nonetheless, FlowMur still achieves a minimum ASR of 89.72\% (78.33\%) on SCD (FKD), which underscores its effectiveness under different surrogate-target model types. 

\subsubsection{Impact of SNR}
\label{impact_snr}

\begin{figure}[t]
    \centering
    \subfigure[SCD]{\includegraphics[scale=0.26]{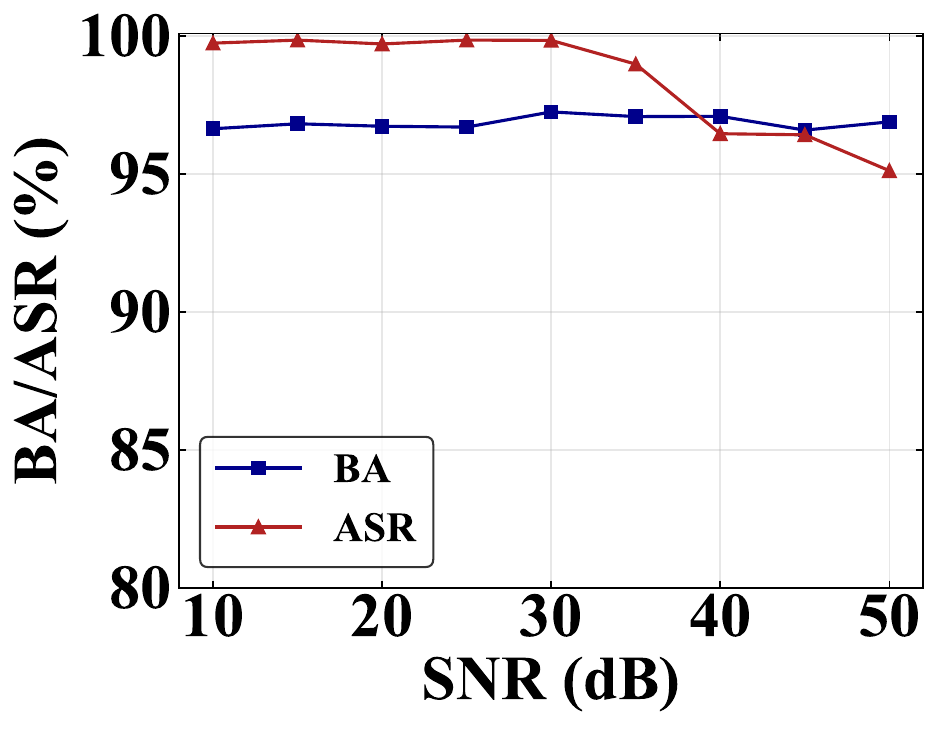}}
    \subfigure[FKD]{\includegraphics[scale=0.26]{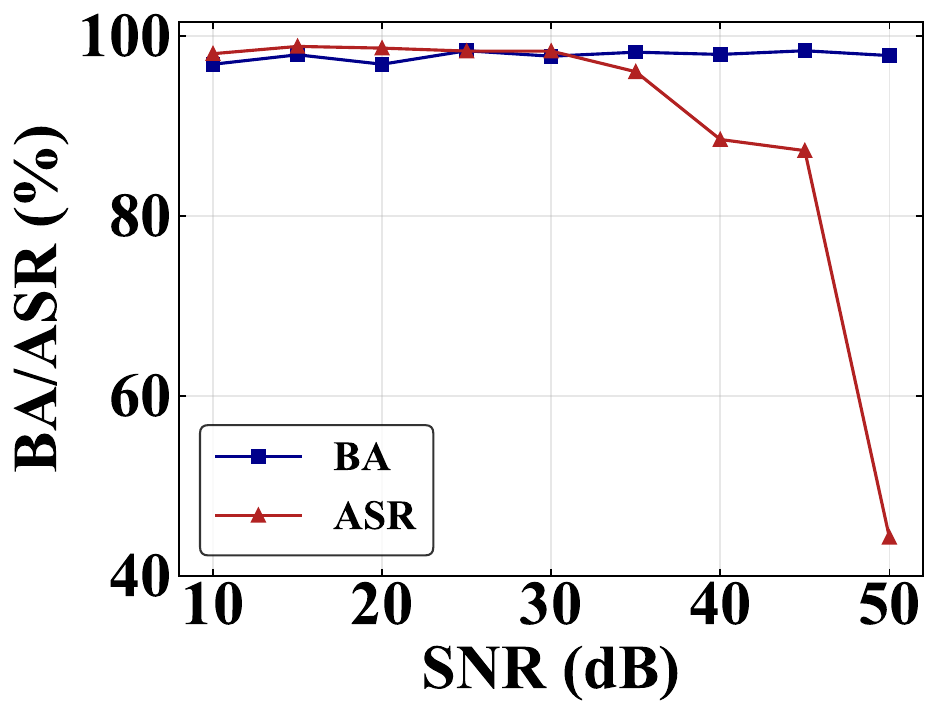}}
    \vspace{-2mm}
    \caption{\textcolor{black}{Experimental results on the impact of SNR.}}
    \label{SNR}
    \vspace{-2mm}
\end{figure}

The SNR between the benign audio and the trigger audio in a poisoned sample is a crucial factor that influences the attack performance of FlowMur. To understand the impact of SNR on attack performance, we set it from 10dB to 50dB with a step of 5dB. The experimental results are presented in Fig.~\ref{SNR}.

\textcolor{black}{From Fig.~\ref{SNR}, we can derive three observations. First, irrespective of SNR fluctuations, FlowMur does not compromise BAs. This resilience stems from the enduring significance of intrinsic benign sample features, which are well captured by the model across varying SNR levels. 
Second, we observe a decline in FlowMur's ASR as SNR exceeds 30dB. This phenomenon arises due to the increasing subtlety of triggers as SNR increases, causing the model to inadequately learn the trigger features. 
Third, a noticeable change occurs at an SNR of 50dB, where FlowMur's ASR exhibits a sudden decrease on FKD, a phenomenon not reflected in SKD. This difference can be attributed to environmental factors, as the samples in FKD are recorded in noisy settings while the samples in SCD are collected in quiet environments. Thus, at the same SNR, the trigger for FKD suffers from greater perturbation compared to the trigger for SCD, causing the subpar ASR of FlowMur on FKD.}

\subsubsection{Impact of Data Balance}
\label{databalance}
\textcolor{black}{To assess the influence of data balance between an adversary and a victim on attack performance, we controlled the size of the victim dataset $D$ and the auxiliary dataset $D_{aux}$ by varying the number of data classes known to the victim and the adversary (i.e., \#class of $D$ and \#class of $D_{aux}$). The experimental results are reported in Table~\ref{data_balance}.}


\begin{table*}[htbp]
    \centering 
    \footnotesize
    \caption{\textcolor{black}{Performance of FlowMur with different data balances between an adversary and a victim.}}
    \vspace{-1mm}
    \label{data_balance}
    {\vspace{-2mm}MA: model accuracy without attacks.}
\\[2mm]

    \begin{tabular}{@{}ccccc|ccccc@{}}  
        \toprule[1.5pt]
    \multicolumn{5}{c|}{SCD}  & \multicolumn{5}{c}{FKD} \\
    \midrule
        \#class of $D$ & \#class of $D_{aux}$ & MA & BA (\%) & ASR (\%) & \#class of $D$ & \#class of $D_{aux}$ & MA & BA (\%) & ASR (\%) \\
   
   \midrule
         5   &  30  &  97.58 & 97.14($\downarrow$0.44)  &  99.75  &  4  &  14  & 99.28 &  98.76($\downarrow$0.52)  &  97.21 \\
         10  &  25  & 97.15 & 97.24($\uparrow$0.09)  &  99.83  &  6  &  12  & 98.57 & 98.29($\downarrow$0.28)  & 98.81\\
         15   &  20  & 96.02 & 96.01($\downarrow$0.01)  &  99.55  &  8  &  10 & 97.80 & 97.75($\downarrow$0.05) & 98.29 \\
         20  &  15  & 94.70 & 94.73($\uparrow$0.03)  &  99.69  &  10  &  8  & 97.67 & 97.77($\uparrow$0.10)  &  98.12 \\
         25   & 10  & 94.07 & 94.17($\uparrow$0.10)  &  99.02  &  12  &  6  & 97.32 & 97.45($\uparrow$0.13)  &  97.88 \\
         30  &  5  & 93.32 & 93.43($\uparrow$0.11)  &  99.74  &  14  &  4  & 97.52 & 97.57($\uparrow$0.05)  &  95.83 \\
    \bottomrule[1.5pt]
    \end{tabular}
    \vspace{-2mm}
\end{table*}

\textcolor{black}{As Table~\ref{data_balance} shows, FlowMur consistently performs well over different data balances for both FKD and SCD. Specifically, FlowMur achieves an ASR exceeding 99\% (95\%) on SCD (FKD) without compromising BA, even when the adversary knows 5 (4) classes, while the victim knows other 30 (14) classes.}


\subsection{Practicality}

To evaluate the practicality of FlowMur, we conducted two physical attacks in the real world, against recorded audio and live human speech, respectively. For the attack on recorded audio, we played and recorded poisonous samples via a loudspeaker and a microphone respectively, and then fed the recorded audio to the infected model. As for the attack on live human speech, we played the trigger while participants were interacting with the infected model, and recorded the mixture of the trigger and human speech. Then, the mixed audio was fed to the infected model. We validated these physical attacks in two different places: a lab room and a restaurant. The size of the lab room is approximately $4m \times 5m \times 3m $, and the main noise is the sound of keyboards, footsteps, and conversations. The restaurant size is approximately $8m \times 6m \times 3.5m$, and the main noise is conversations, machine noises, and clinking of cutlery. The volumes of background noise in the lab room and the restaurant were approximately 38dB and 50dB, respectively.

In our experiments, we utilized an iPhone 13 as the loudspeaker and a Macbook as the microphone without specification. The default distance between the loudspeaker and the microphone is 1 meter, and the default volume of sounds, including triggers, benign samples, and human speech, is 55dB, which is close to the volume of a normal conversation. As for other settings, unless specially stated, we used the same default settings as stated in Section \ref{experimental_setup}.

\subsubsection{Physical Attack on Recorded Audio}

In this study, we first analyzed whether the incorporation of ambient noise can enhance the robustness of FlowMur in physical attacks. We then performed extensive experiments by varying the device (loudspeaker and microphone) and the distance between the loudspeaker and microphone to thoroughly evaluate FlowMur with consideration of ambient noise.

\begin{figure}[t]
    \centering
    \subfigcapskip=-3pt
    \subfigure{\includegraphics[scale=0.3]{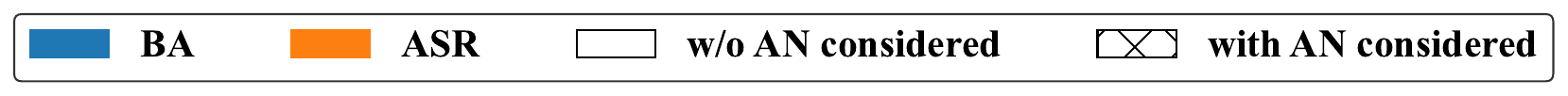}}
    \vspace{-3mm}
    \hspace{-1mm}

    \setcounter{subfigure}{0}
    \subfigure[Lab room (SCD)]{\includegraphics[scale=0.26]{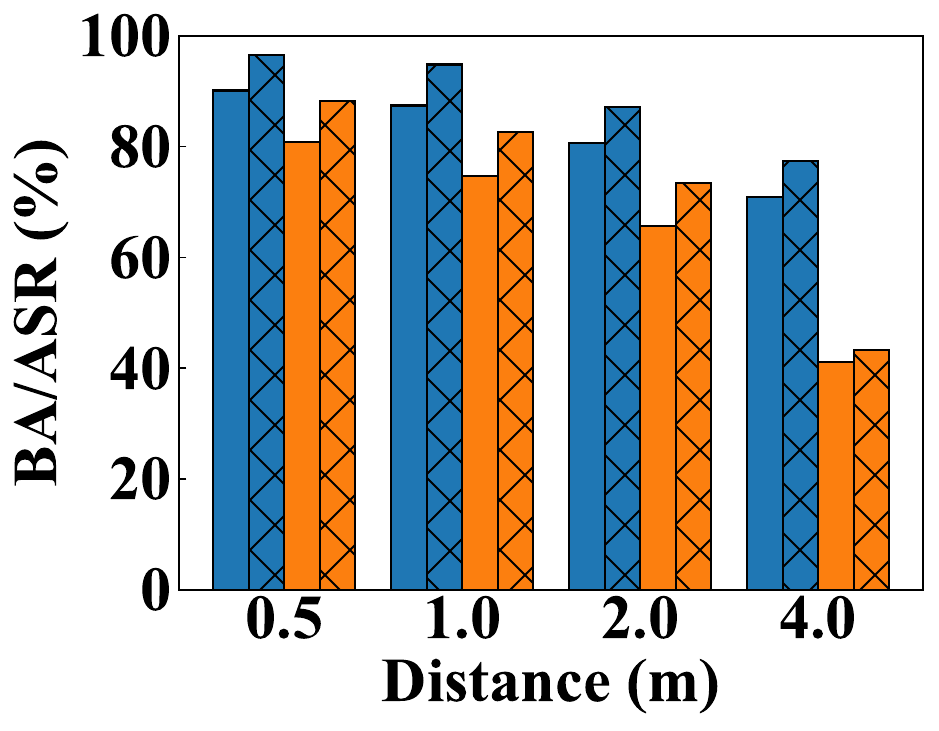}}
    \subfigure[Lab room (FKD)]{\includegraphics[scale=0.26]{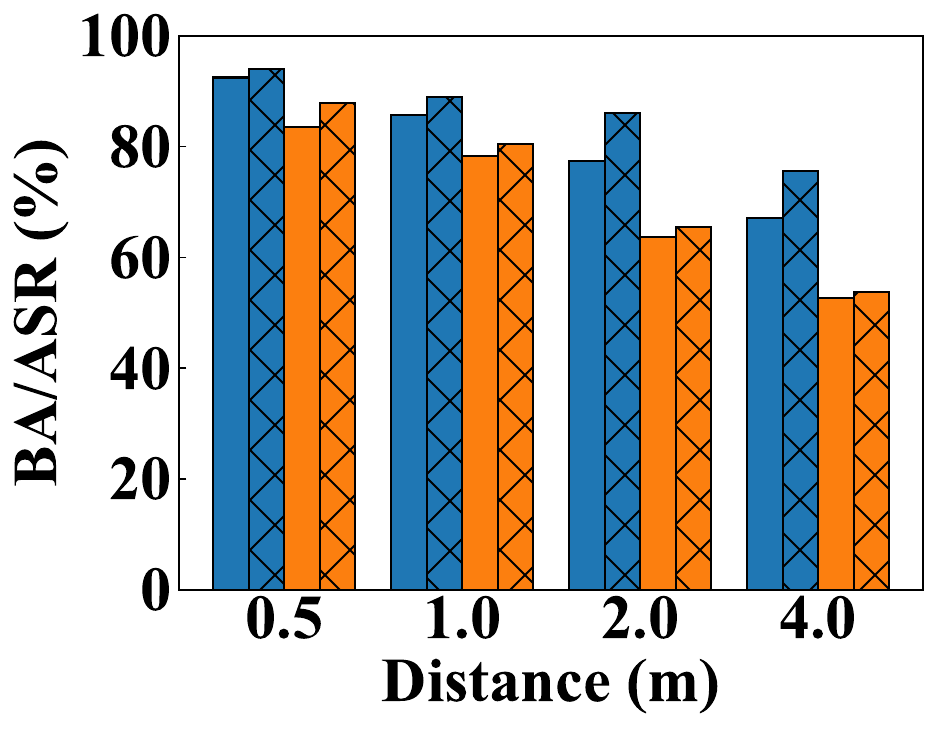}}
    \subfigure[Restaurant (SCD)]{\includegraphics[scale=0.26]{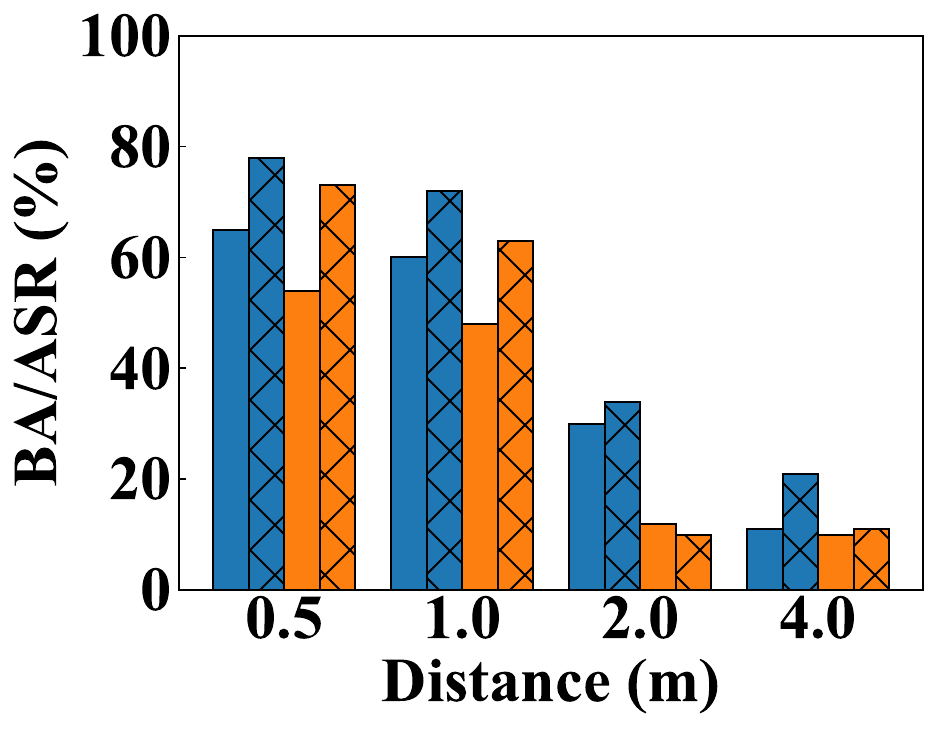}}
    \subfigure[Restaurant (FKD)]{\includegraphics[scale=0.26]{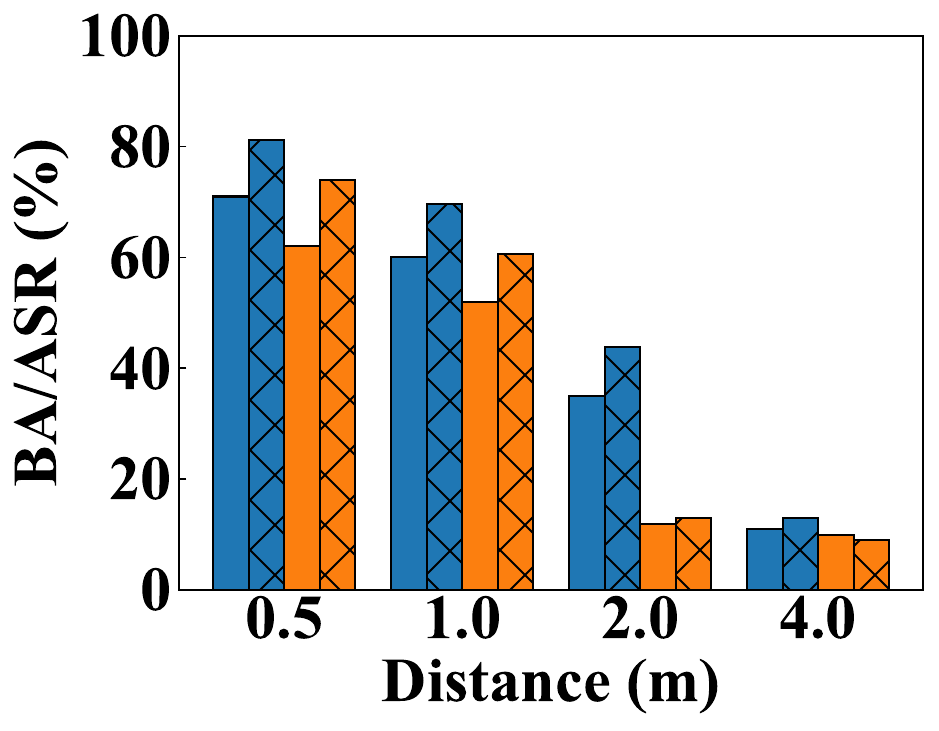}}
   
    \vspace{-2mm}
    \caption{Experimental results on physical attacks by playing recorded audio over the air. \textcolor{black}{AN means ambient noise. ``w/o'' means without. The shaded portion shows FlowMur’s attack performance with ambient noise considered, and the unshaded portion shows its attack performance without ambient noise considered.}}
    \label{physical attack distance}
    \vspace{-2mm}
\end{figure}

\textbf{Results.}
Fig.~\ref{physical attack distance} presents the experimental results of FlowMur on recorded audio with and without considering ambient noise in various settings, including different places, datasets and distances. Take Fig.~\ref{physical attack distance} (a) as an example, when the distance between the loudspeaker and the microphone is 1 meter, FlowMur's BA and ASR increase by 7.41\% and 8.06\%, respectively, when ambient noise is considered compared to when it is not. Similar trends are observed in other settings. These results demonstrate that incorporating ambient noise into the process of trigger generation and data poisoning not only enhances FlowMur's robustness against ambient noise, but also mitigates its impact on the BA of an infected model.

\textbf{Impact of different distances.}
To examine the impact of distance between microphones and loudspeakers on attack performance, we conducted experiments at varying distances of 0.5, 1, 2, and 4 meters, referring to the experimental results shown in Fig.~\ref{physical attack distance} with ambient noise consideration.

From Fig.~\ref{physical attack distance}, we can observe that as the distance between the loudspeaker and the microphone increases, attacks exhibit reduced performance, especially in a noisy context (the restaurant herein). This phenomenon can be attributed to the attenuation and disturbance of sound as it propagates through the air. On the one hand, the longer the distance, the more the sound is weakened and disrupted. On the other hand, the more noisy the context, the more the sound is disrupted. Despite this, at distances of up to 1 meter, the ASR of FlowMur exceeds 80\% in a quiet context (i.e., the lab room) and 60\% in a noisy context (i.e., the restaurant).

\begin{table}[tbp]
    \centering 
    \footnotesize
    \caption{Experimental results on different devices (\%).} 
    \vspace{-1mm}
    \label{physical attack device}
    {\vspace{-2mm}Mic: microphone; Lou: loudspeaker.}
\\[2mm]

    \begin{tabular}{@{}c|cc|cc@{}}  
        \toprule[1.5pt]
    \multirow{2.5}{*}{\makebox[0.05\textwidth][c]{\diagbox[height=3.7em, width=4.3em]{\textbf{Lou}}{\textbf{Mic}}}} &\multicolumn{2}{c|}{SCD} &\multicolumn{2}{c}{FKD}\\
    \cmidrule{2-5}
     &Macbook &Dell &Macbook &Dell\\
        & (BA/ASR)  & (BA/ASR) & (BA/ASR) & (BA/ASR) \\
   
   \midrule
   \multicolumn{5}{c}{Setting 1: Lab Room}\\
   \midrule
         iPhone 13 & 94.81/82.67  & 93.21/81.78  & 88.04/80.48 & 87.66/75.24  \\
        Dell &94.96/83.77 & -- & 89.45/78.86& --   \\
    \midrule
     \multicolumn{5}{c}{Setting 2: Restaurant}\\
   \midrule
        iPhone 13 & 72.69/63.56  & 70.95/61.39  & 69.86/60.69 & 69.02/59.40  \\
        Dell & 71.90/64.11 & -- & 69.54/60.24 & --   \\
        \bottomrule[1.5pt]
    \end{tabular}
    \vspace{-2mm}
\end{table}

\textbf{Impact of different devices.}
In order to assess the impact of different devices on attack performance, we used two different devices as the microphone, i.e., a Macbook and a Dell laptop, and two devices as the loudspeaker, i.e., an iPhone 13 and a Dell laptop. Experimental results are presented in Table \ref{physical attack device}. 

From Table~\ref{physical attack device}, we can observe that for each combination of the loudspeaker and the microphone, FlowMur achieves a minimum of 87.66\% BA and 75.24\% ASR in the lab room, and a minimum of 69.02\% BA and 59.40\% ASR in the restaurant. Thus, we can conclude that FlowMur remains effective across various devices. Furthermore, physical attacks utilizing the MacBook as a microphone consistently outperform attacks using the Dell laptop as a microphone (at least 0.38\% increase for BA and 0.89\% increase for ASR). One possible explanation for this performance disparity is that the recording quality of the MacBook is slightly superior to that of the Dell laptop. The attack performance using different loudspeakers, i.e., iPhone 13 and Dell laptop, shows similarities. This may arise from the similarity in sound quality of the two devices.

\subsubsection{Physical Attack on Live Human Speech}
\label{physcalattack}

\begin{figure}[t]
    \centering
    \subfigure[Experimental setup]{\includegraphics[scale=0.042]{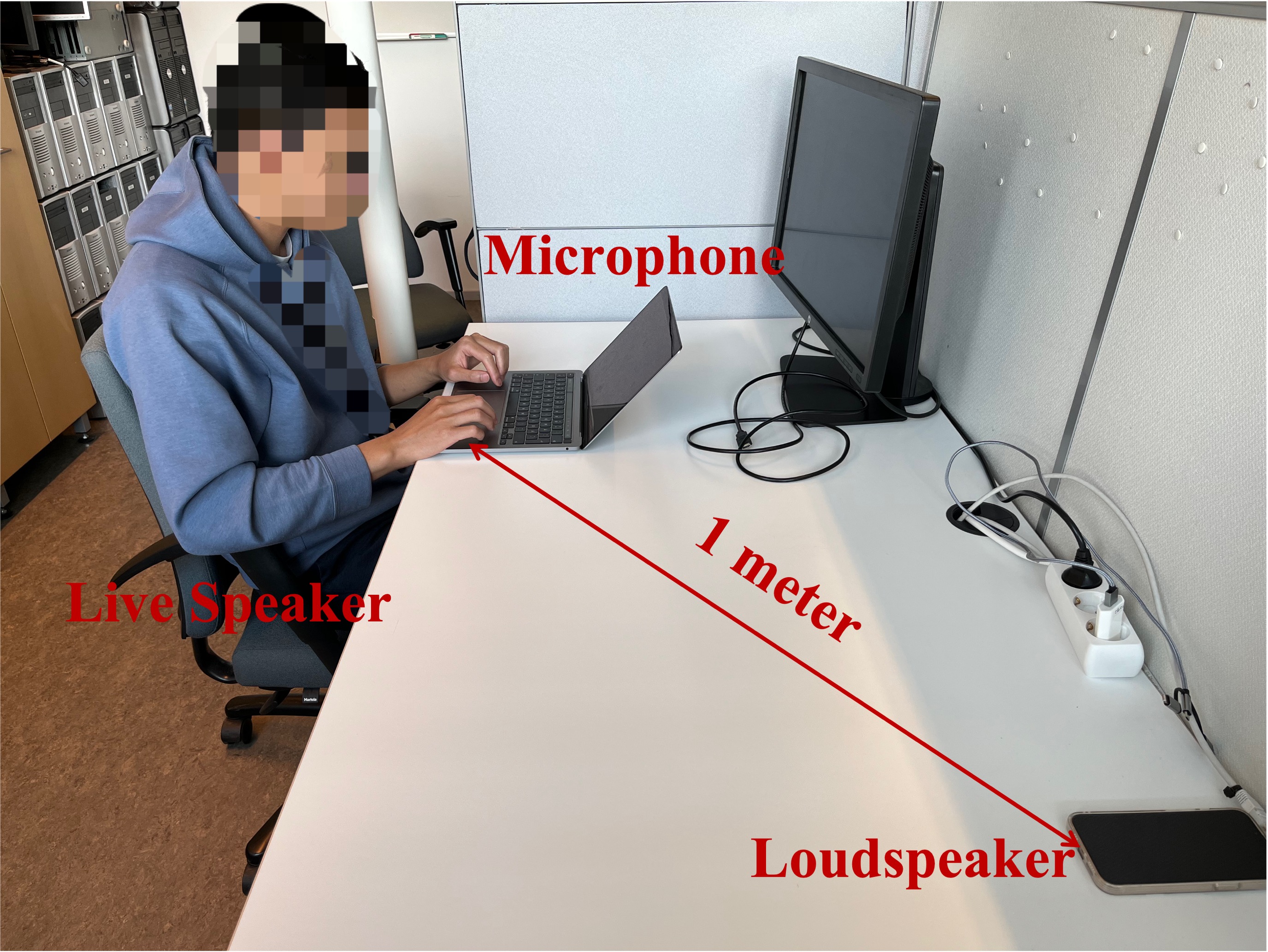}}
    \subfigure[Experimental results]{\includegraphics[scale=0.25]{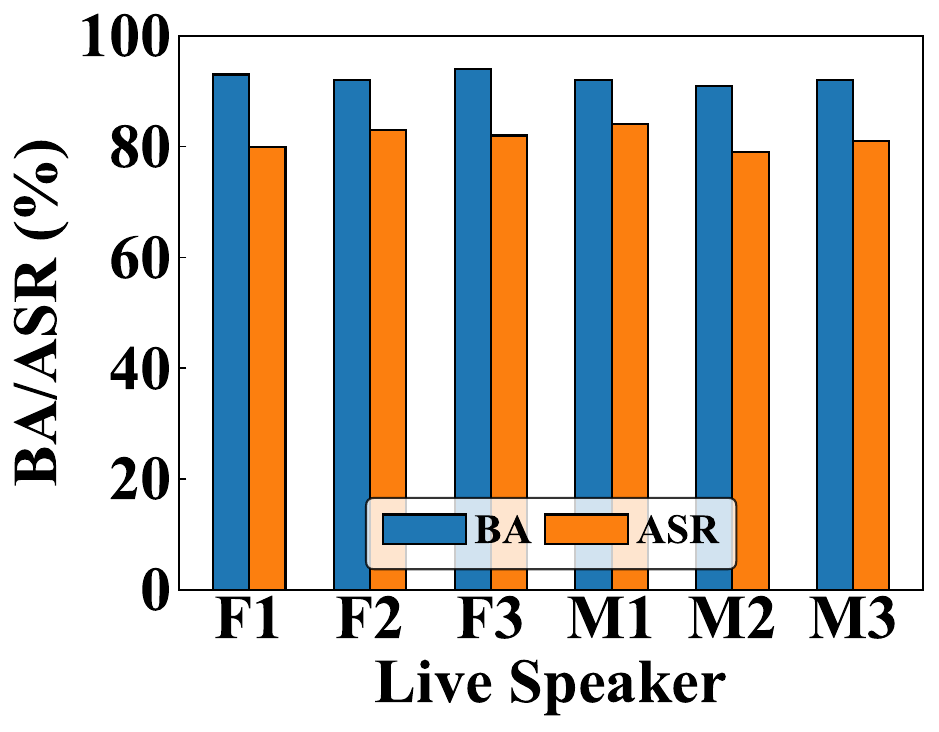}}
    \vspace{-2mm}
    \caption{Experimental setup and results on physical attacks against live human speech, where F and M denote female and male, respectively.}
    \label{humanstudy_live}
    
    \vspace{-2mm}
\end{figure}

We recruited six participants (three males and three females) to assess the effectiveness of FlowMur in attacking live human speech. As shown in Fig.~\ref{humanstudy_live} (a), we asked each participant to sit in front of a microphone (a Macbook) with a loudspeaker (an iPhone 13) placed a meter away to play triggers. Initially, participants were instructed to speak each speech command 10 times without playing triggers to measure BA. Subsequently, they were asked to speak each command 10 times while playing triggers to measure ASR. The live human speech experiments were conducted in the lab room, utilizing SCD only, since FKD is in Persian and poses difficulty for the participants to speak.

\textbf{Results.}
Fig.~\ref{humanstudy_live} (b) shows the attack performance of FlowMur on live human speech. \textcolor{black}{The results indicate that FlowMur achieves a BA of over 90\% and an ASR of over 79\% for all six participants.} These findings indicate the effectiveness of FlowMur in attacking live human speech, showcasing its capability to conduct successful backdoor attacks on real-world human-computer speech interaction systems.
We also studied the impact of trigger volume on FlowMur's ASR, please see the results in Appendix \ref{trigger_volume}.
 
\subsection{Stealthiness}
As for stealthiness, we primarily focused on trigger imperceptibility. Human studies are a feasible way to investigate trigger imperceptibility. However, human studies conducted by previous work primarily focus on detecting similarities between benign and poisonous samples, such as ABXTest~\cite{kreuk2018fooling} and similarity test~\cite{Gong2022attention}. They may not be suitable for evaluating the imperceptibility of audio triggers, given their flexibility and variability. For instance, a poisonous sample containing a whistle trigger, as used in NBA and NBA-D attacks, may lack similarity to its benign counterpart; however, this does not imply that it is suspicious to humans. Therefore, we designed and conducted a new human study applicable to audio backdoor attacks, named \emph{clean or suspicious}.

\begin{figure}[t]
    \centering
    \subfigure[Task 1]{\includegraphics[scale=0.26]{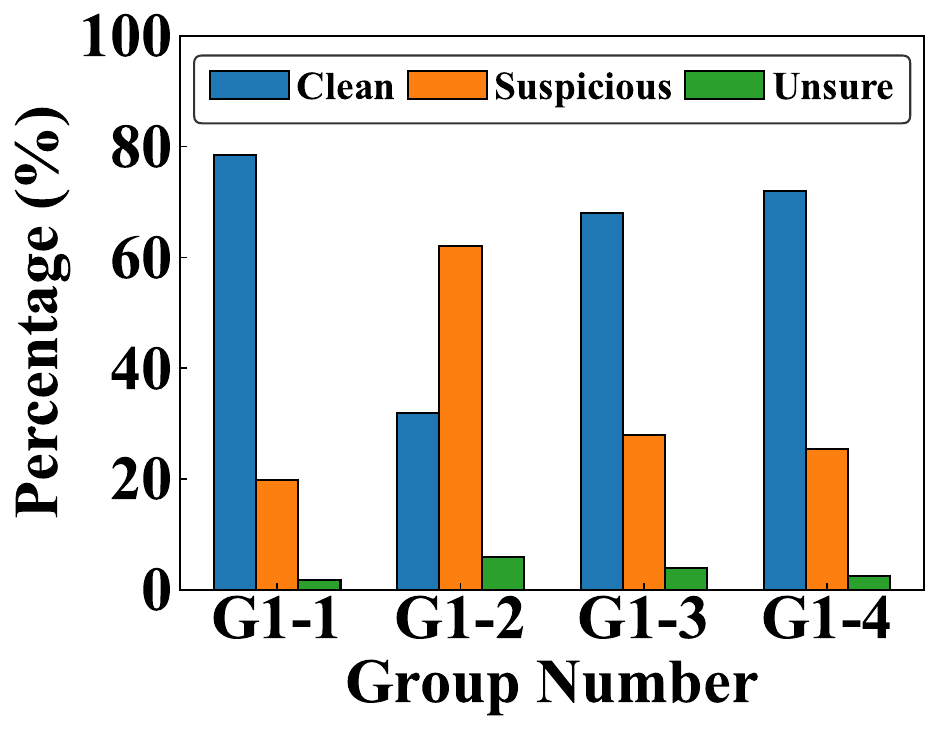}}
    \subfigure[Task 2]{\includegraphics[scale=0.26]{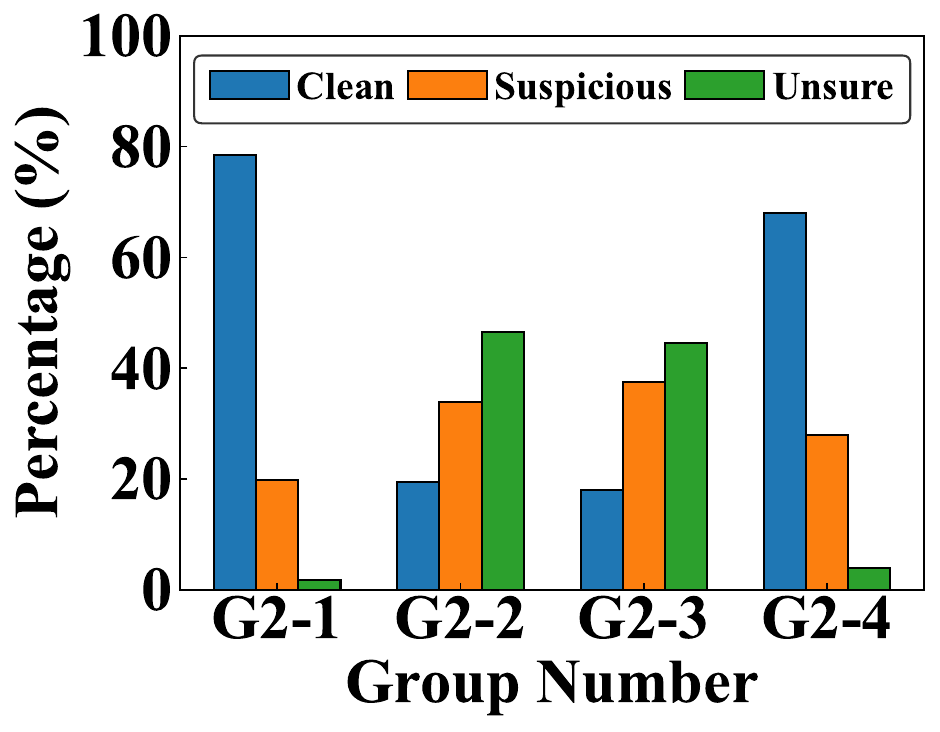}}
    \vspace{-2mm}
    \caption{Experimental results of human studies on imperceptibility. \textcolor{black}{(a): imperceptibility of FlowMur with different SNRs: 20dB (G1-2), 30dB (G1-3), and 40dB (G1-4). (b): imperceptibility of different attack methods: NBA (G2-2), NBA-D (G2-3), and FlowMur (G2-4). G1-1 and G2-1 are benign samples for comparison.}}
    \label{humanstudy}
    
    \vspace{-2mm}
\end{figure}

\textbf{Clean or suspicious.} This survey asked the participants whether a played audio sample is clean or suspicious. The test set consists of a combination of benign samples and poisonous samples. To simulate a real poisonous training dataset, the benign samples vastly outnumber the poisonous samples. It is essential to note that the samples in the test set are selected randomly and do not overlap. Finally, the participants were asked to identify whether the samples in the test set are clean or suspicious, with three response options: \emph{clean, suspicious, or unsure}. We conducted two tasks using \emph{clean or suspicious}, described in detail as follows.

\emph{Task 1:} Task 1 aims to measure the trigger imperceptibility of FlowMur with different SNRs, namely 20dB, 30dB, and 40dB. The test set for this task comprises four audio sample groups: the first group (G1-1) consists of 70 benign samples, while the remaining three groups each have 10 poisonous samples generated by FlowMur with SNRs of 20dB (G1-2), 30dB (G1-3), and 40dB (G1-4), respectively.

\emph{Task 2:} Task 2 evaluates the trigger imperceptibility of different attack methods, including NBA, NBA-D, and FlowMur. The test set of this task is also composed of four groups. The first group (G2-1) consists of 70 benign samples, and the remaining three groups consist of 10 poisonous samples generated by NBA (G2-2), NBA-D (G2-3), and FlowMur (with SNR of 30dB) (G2-4), respectively.

\textbf{Results of the human study.} We recruited 20 volunteers, 10 males and 10 females, from the campus forum. Before the tests, we briefly introduced the basics of backdoor attacks to the volunteers. We collected 20 * 100 = 2000 answers for each task. The results of the human study are shown in Fig.~\ref{humanstudy}. 

For task 1, as shown in Fig.~\ref{humanstudy} (a), 32\% participants believed the samples in G1-2 to be clean, while 62\% considered them to be suspicious. In contrast, 68\% and 72\% participants believed that samples in G1-3 and G1-4, respectively, were clean, close to the baseline 78.4\% of G1-1. These findings suggest that the triggers generated by FlowMur with 20dB are easily perceptible by humans, while FlowMur with 30dB and 40dB exhibits good trigger imperceptibility.

For task 2, as shown in Fig.~\ref{humanstudy} (b), 68\% participants considered the samples in G2-4 to be clean, close to the baseline 78.4\% of G2-1, indicating that the triggers generated by FlowMur are imperceptible to humans. In contrast, only 19.5\% and 18\% participants believed the samples in G2-2 and G2-3 to be clean, and 46.5\% and 44.5\% of participants were unsure. This is likely because the samples in G2-2 and G2-3 used whistles as triggers, which, despite being a common ambient sound, can still be perceptible when frequently played. Therefore, the trigger imperceptibility of NBA and NBA-D, which use ambient sound as triggers, could be also influenced by their poisoning rate (30\% herein). \textcolor{black}{We provided audio waveforms, spectrograms, and MFCCs for different cases in Appendix~\ref{appendix_samplevisualization}.}

\subsection{Defense Resistance}
\label{defense_resistance}
\begin{figure}[t]
    \centering
    \subfigure[SCD (NBA)]{\includegraphics[scale=0.18]{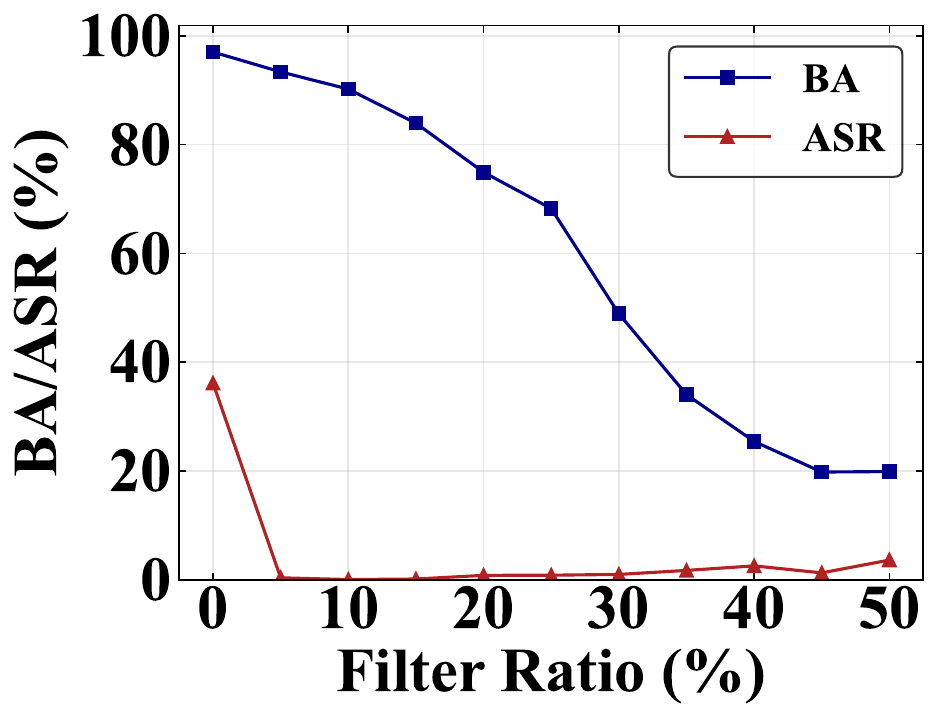}}
    \subfigure[SCD (NBA-D)]{\includegraphics[scale=0.18]{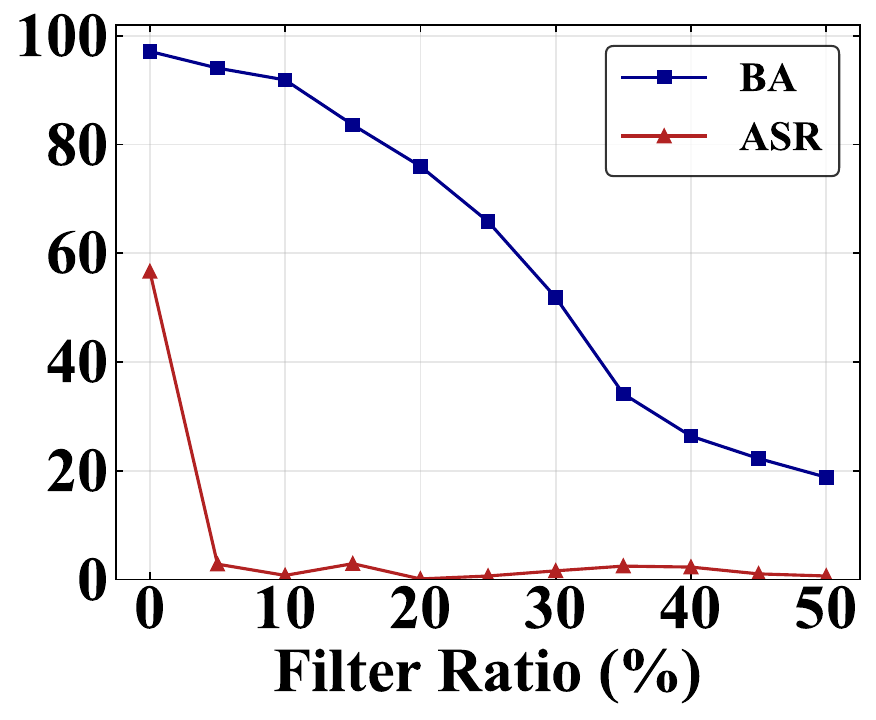}}
    \subfigure[SCD (FlowMur)]{\includegraphics[scale=0.18]{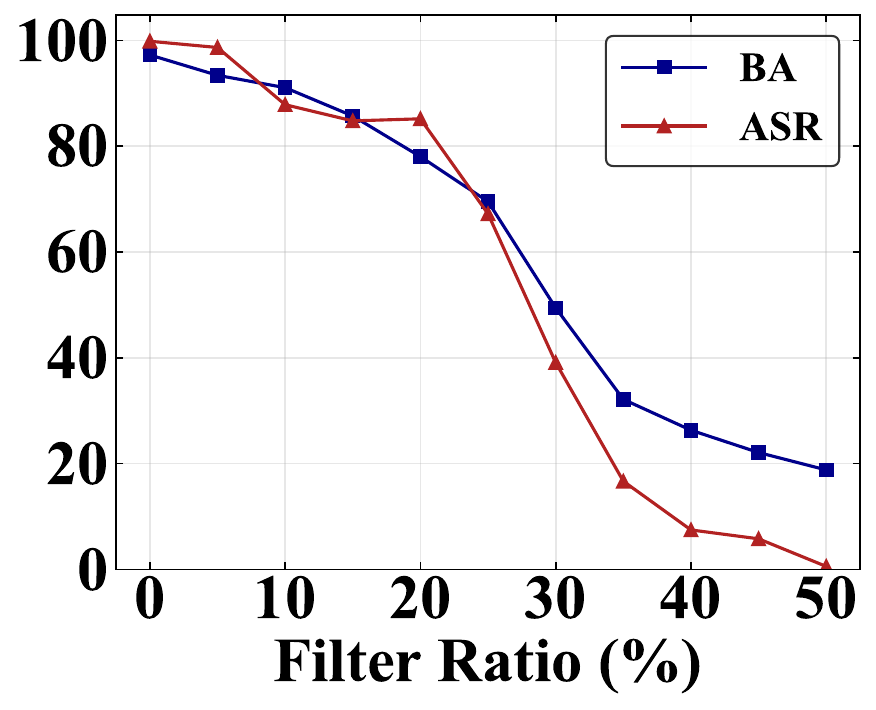}}
    \subfigure[FKD (NBA)]{\includegraphics[scale=0.18]{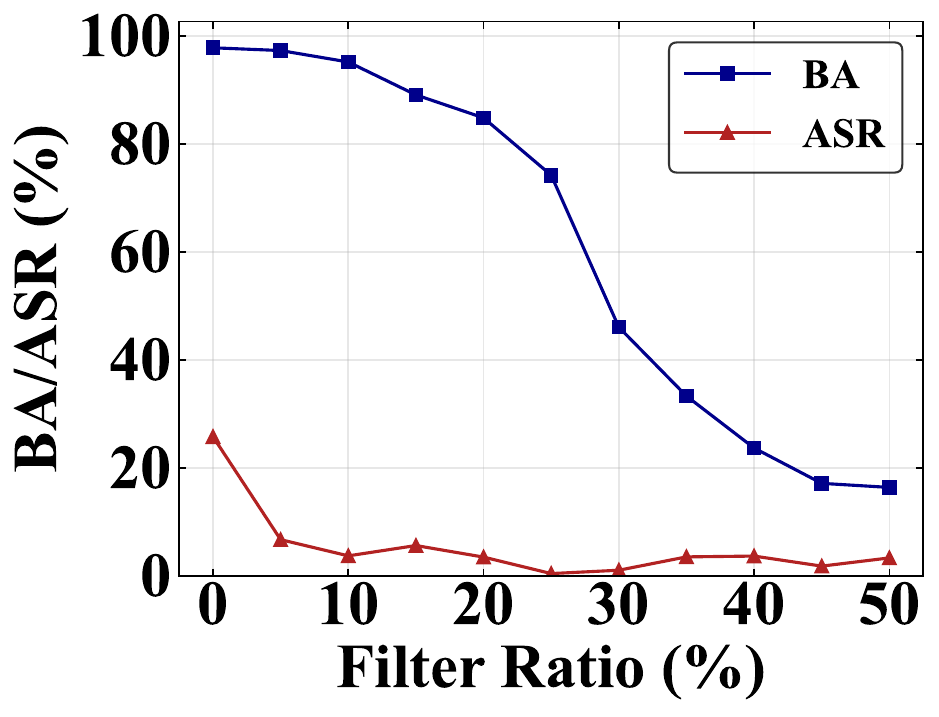}}
    \subfigure[FKD (NBA-D)]{\includegraphics[scale=0.18]{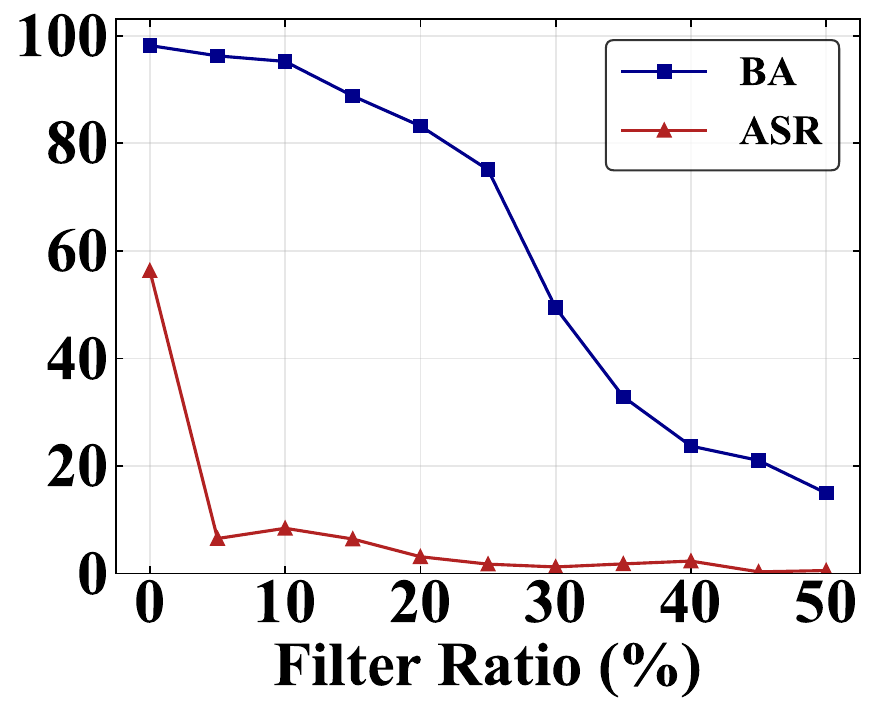}}
    \subfigure[FKD (FlowMur)]{\includegraphics[scale=0.18]{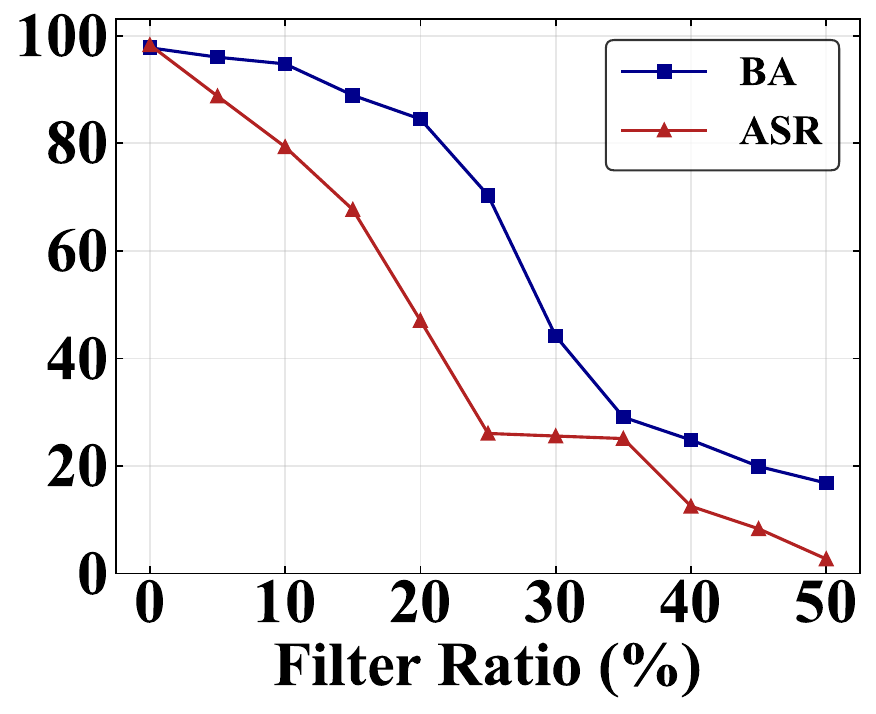}}
    \vspace{-2mm}
    \caption{\textcolor{black}{Defense performance of filters on FlowMur and baselines.}}
    \label{filter}
    \vspace{-2mm}
\end{figure}
Although numerous backdoor defense methods have been proposed, the majority of them are centered around the image domain and only fine-tuning~\cite{tajbakhsh2016convolutional}, fine-pruning~\cite{liu2018fine}, STRIP~\cite{gao2021design,gao2019strip} and Beatrix~\cite{ma2023beatrix} have been proven effective in the audio domain. Meanwhile, reference~\cite{liu2018fine} reports that fine-pruning significantly outperforms fine-tuning. \textcolor{black}{Additionally, filters~\cite{carlini2016hidden} serve as a commonly used defense strategy in the audio domain. Therefore, we assessed the defense resistance of FlowMur and baselines against filters and three advanced defense methods: fine-pruning~\cite{liu2018fine}, STRIP~\cite{gao2021design,gao2019strip} and Beatrix~\cite{ma2023beatrix}}.

\subsubsection{Filters}
\label{filters_defense}

\textcolor{black}{The filter serves as a common defense method within the realm of audio~\cite{carlini2016hidden}. This method is rooted in an observation that marginally degrading the quality of a benign sample does not impact the model's prediction of it, whereas diminishing the quality of a poisonous sample significantly alters the model's prediction. Thus, reducing the quality of audio samples emerges as a feasible defense method against audio backdoor attacks. Specifically, given a filter ratio $f$ and an audio sample with $s$ sample points, we remove $f \cdot s$ sample points chosen uniformly and preserve the remaining $(1-f) \cdot s$ sample points.}



\textcolor{black}{Fig.~\ref{filter} shows the attack performance of NBA, NBA-D, and FlowMur over SCD and FKD after applying filters. We can notice that with the filter ratio increasing, the ASRs of all attack methods exhibit a decreasing trend. However, FlowMur shows the slowest decline. While the filter effectively defends against FlowMur, it significantly affects BA. Thus, it is challenging for defenders to balance between ASR and BA when applying this defense.}


\subsubsection{Fine-pruning}

Fine-pruning~\cite{liu2018fine} is a backdoor defense method based on an observation that neurons activated by triggers are often different from those activated by benign samples. Thus, pruning neurons that are not easily activated by benign samples can defend against backdoor attacks. Specifically, the defender first ranks neurons in ascending order according to the activation of benign samples. Then, the defender prunes away the neurons in order.

\begin{figure}[t]
    \centering
    \subfigure[SCD (NBA)]{\includegraphics[scale=0.18]{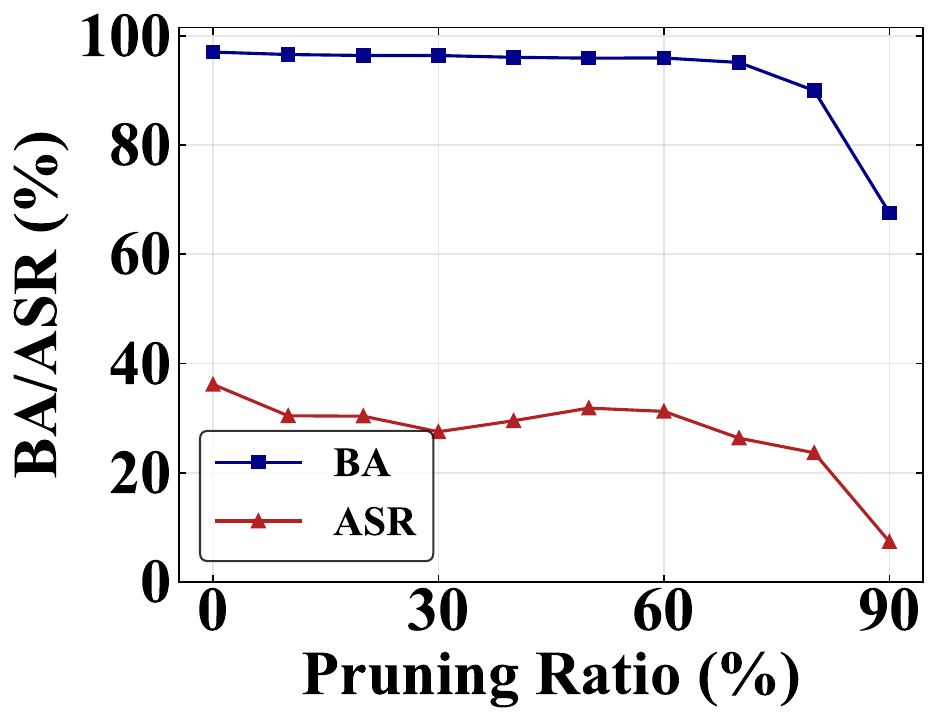}}
    \subfigure[SCD (NBA-D)]{\includegraphics[scale=0.18]{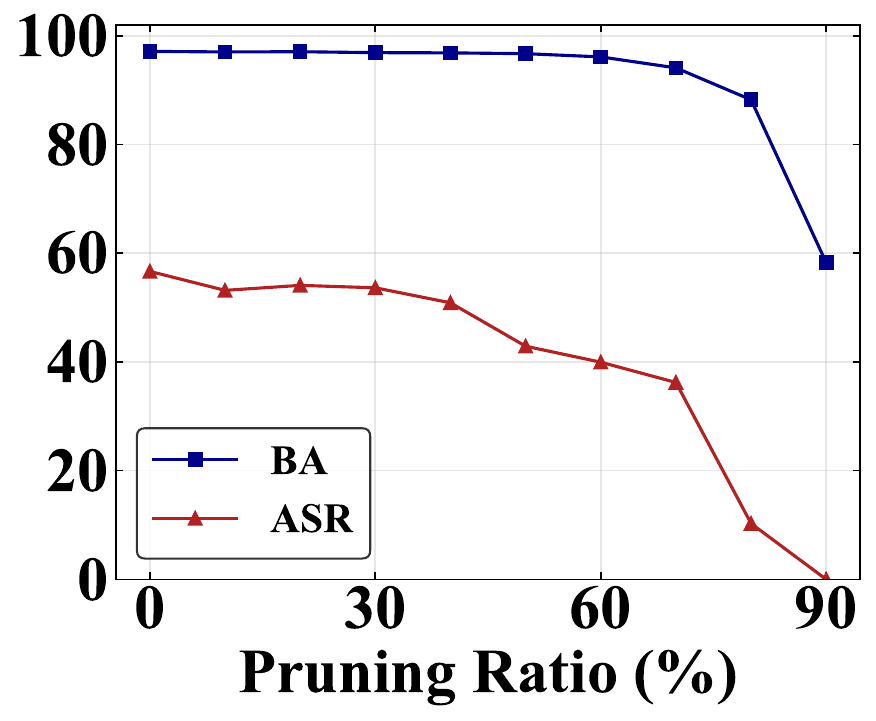}}
    \subfigure[SCD (FlowMur)]{\includegraphics[scale=0.18]{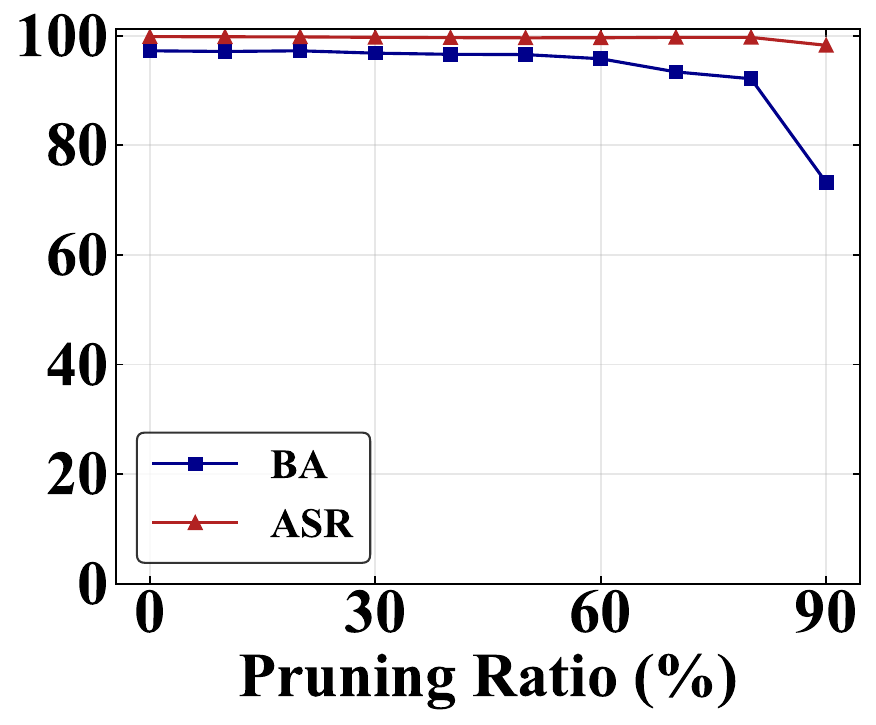}}
    \subfigure[FKD (NBA)]{\includegraphics[scale=0.18]{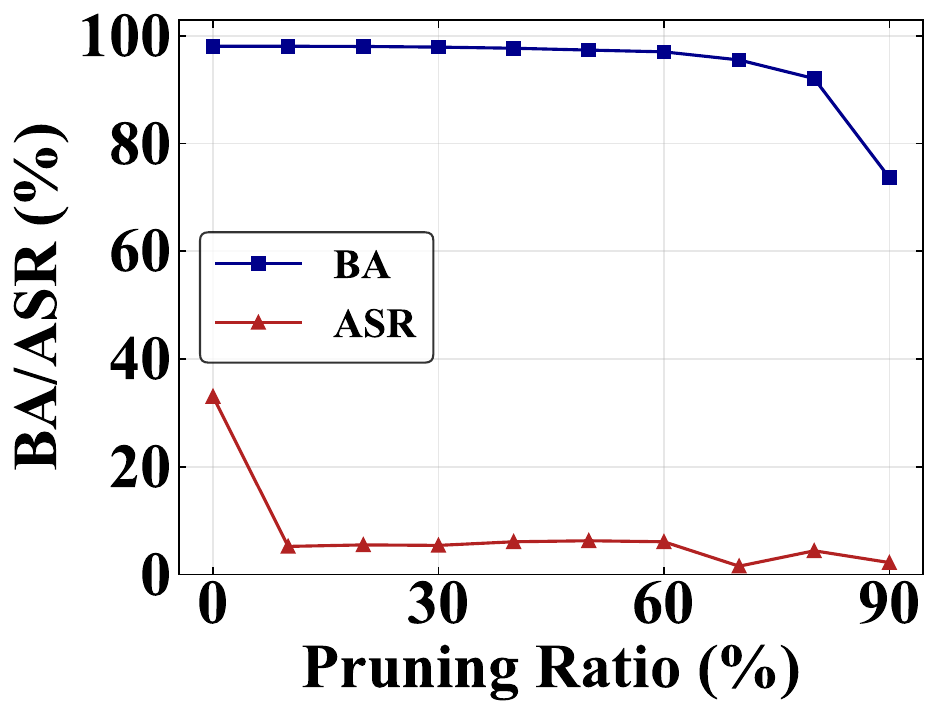}}
    \subfigure[FKD (NBA-D)]{\includegraphics[scale=0.18]{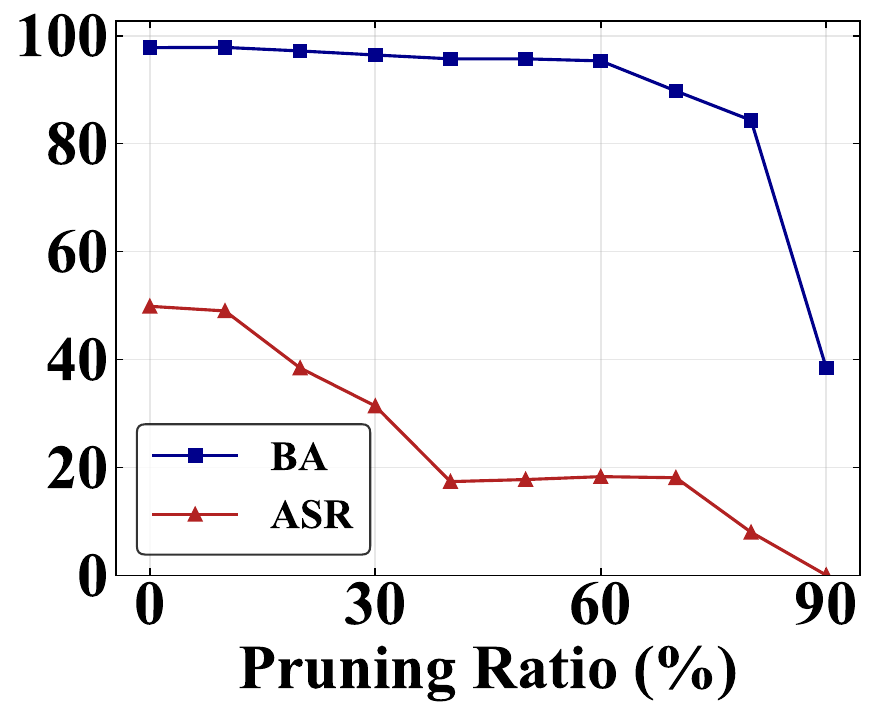}}
    \subfigure[FKD (FlowMur)]{\includegraphics[scale=0.18]{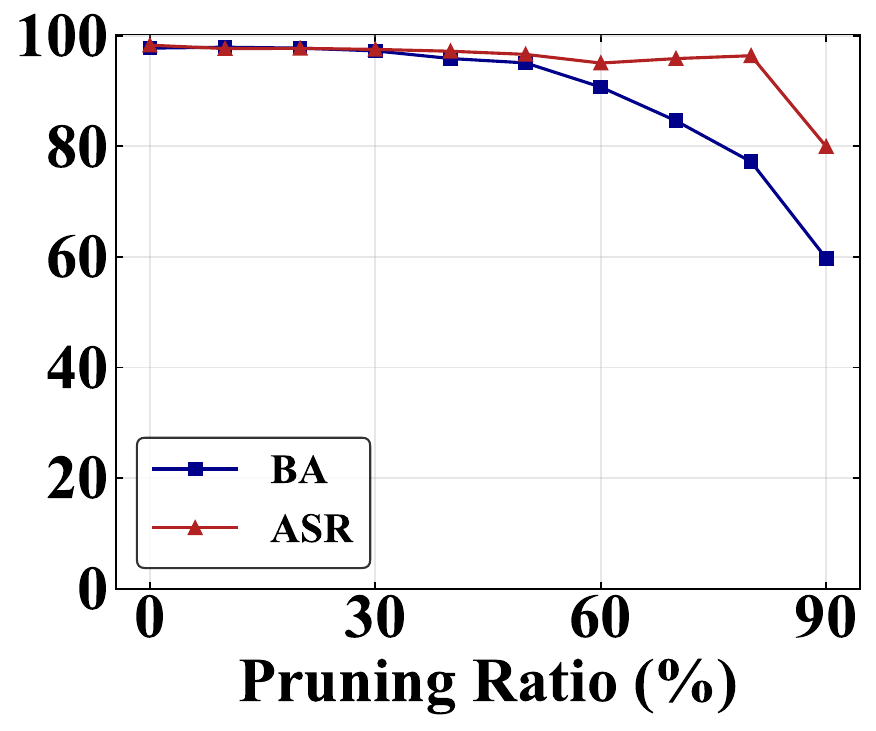}}
    \vspace{-2mm}
    \caption{Defense performance of fine-pruning on FlowMur and baselines.}
    \label{fine-pruning}
    
    \vspace{-2.8mm}
\end{figure}

Fig.~\ref{fine-pruning} shows the attack performance of NBA, NBA-D, and FlowMur over SCD and FKD after applying fine-pruning. From Fig.~\ref{fine-pruning} (a), (b), (d) and (e), we can observe that as the pruning ratio increases, the ASRs of NBA and NBA-D decrease first, followed by a decrease of BAs, implying that the neurons activated by the trigger are pruned earlier than those activated by benign samples. Thus, fine-pruning is effective for defending against NBA and NBA-D. In contrast, from Fig.~\ref{fine-pruning} (c) and (f), we can see that when the pruning ratio exceeds 50\%, the BAs of FlowMur decrease first, followed by a decrease of ASRs. This observation suggests that fine-pruning is unable to differentiate between backdoor neurons and uninfected neurons within an infected model generated by FlowMur. Thus, FlowMur is able to bypass fine-pruning.

\subsubsection{STRIP}

STRIP~\cite{gao2021design,gao2019strip} is a defense method against sample-agnostic backdoor attacks. The key insight of STRIP is that, for an infected model and a poisonous input, regardless of the strength of perturbation on the input, the perturbed input is still predicted as an adversary-desired target class with high confidence by the infected model. Thus, for each input, STRIP first perturbs it to generate a set of perturbed inputs by superimposing other different samples on the input. Then, the predictions of these perturbed inputs are used to determine the nature of the input. Specifically, based on the key insight, the entropy of predictions on a poisonous input is expected to be lower than that of a benign sample. Thus, the defender can indicate an entropy threshold to discern the nature of the input.

\begin{figure}[t]
    \centering
    \subfigure[SCD (NBA)]{\includegraphics[scale=0.18]{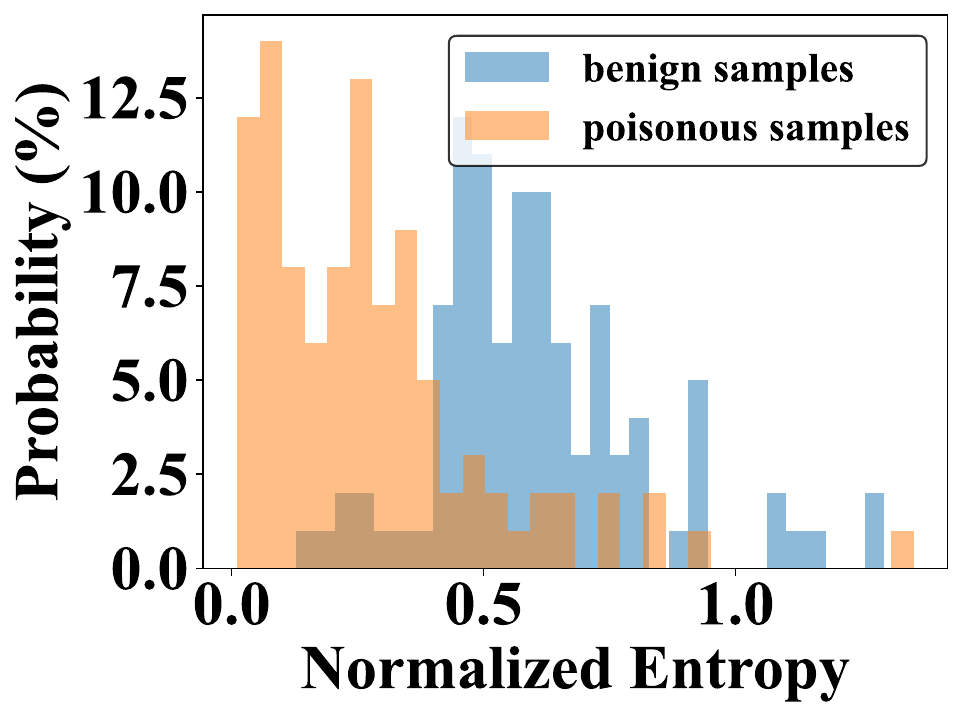}}
    \subfigure[SCD (NBA-D)]{\includegraphics[scale=0.18]{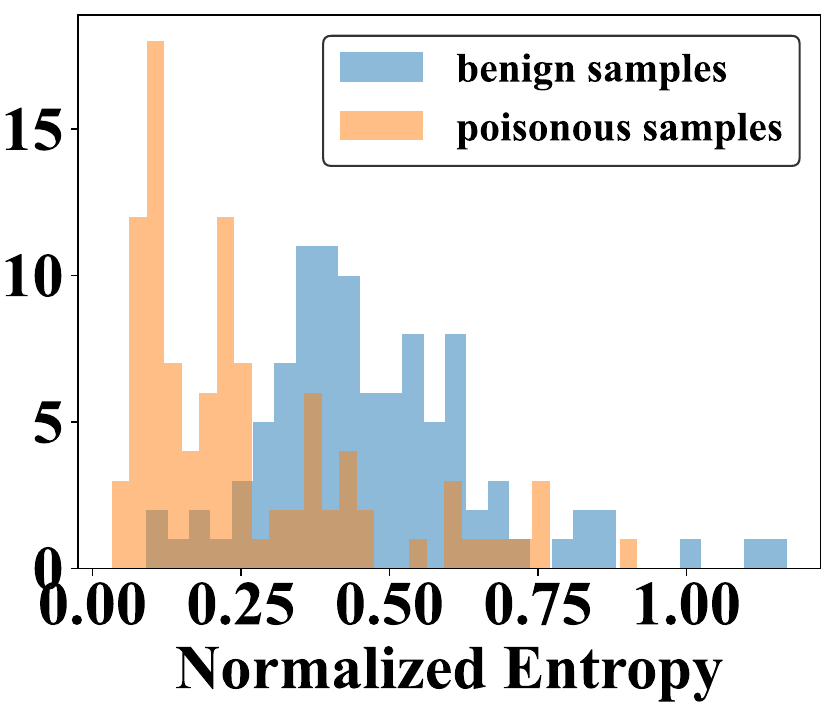}}
    \subfigure[SCD (FlowMur)]{\includegraphics[scale=0.18]{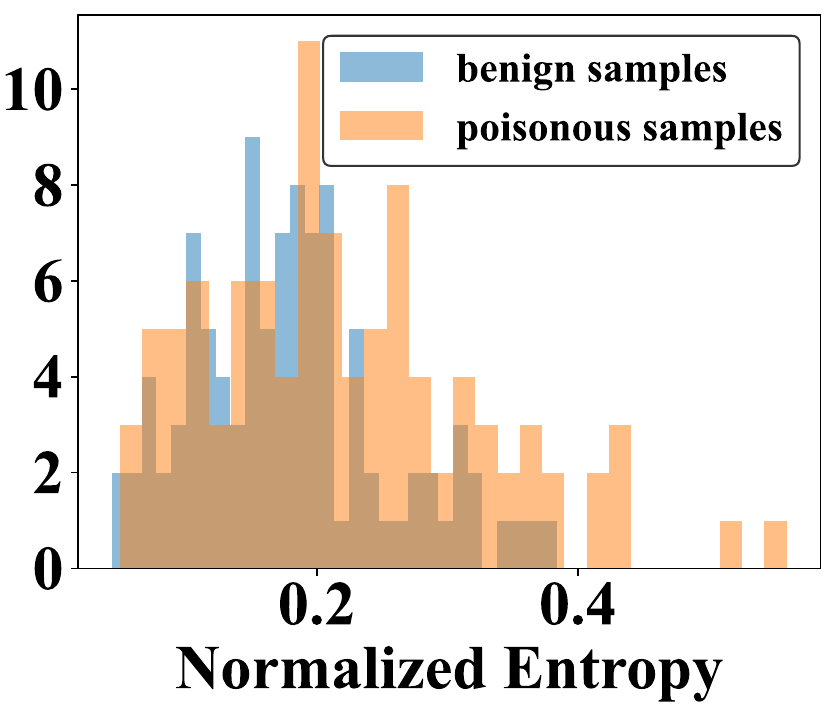}}
    \subfigure[FKD (NBA)]{\includegraphics[scale=0.18]{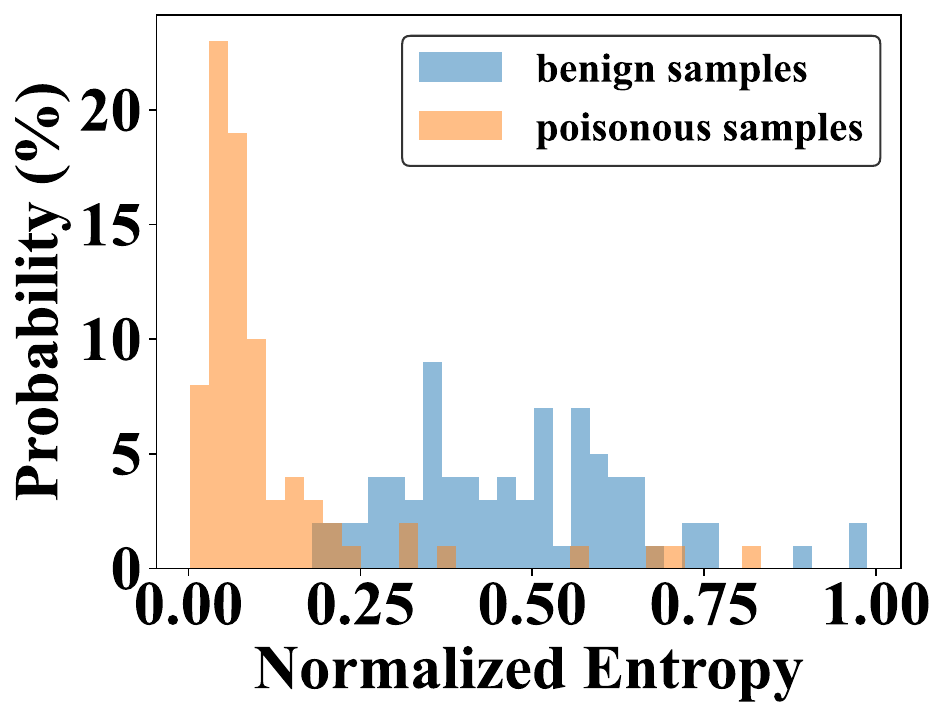}}
    \subfigure[FKD (NBA-D)]{\includegraphics[scale=0.18]{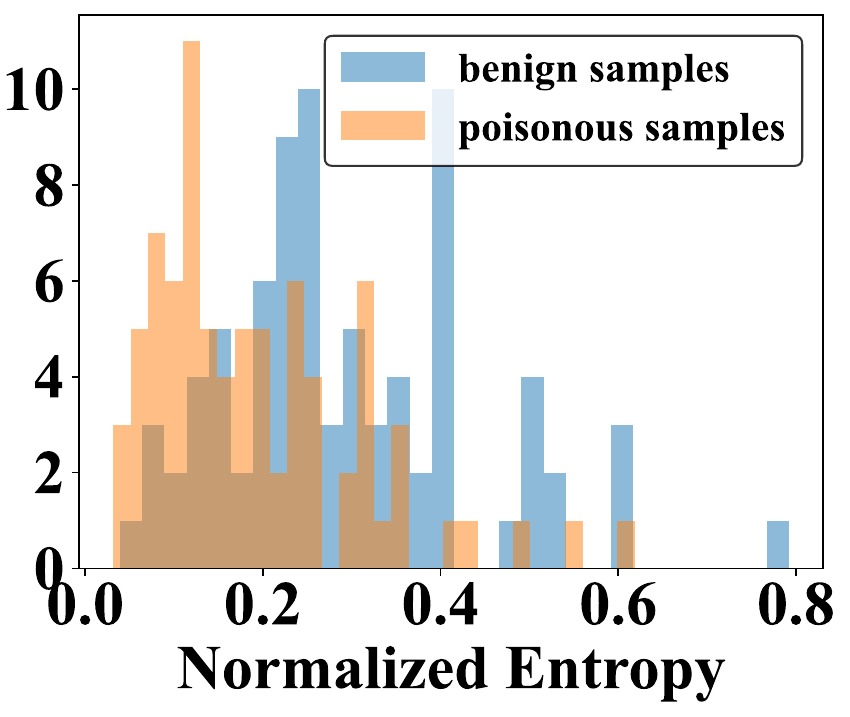}}
    \subfigure[FKD (FlowMur)]{\includegraphics[scale=0.18]{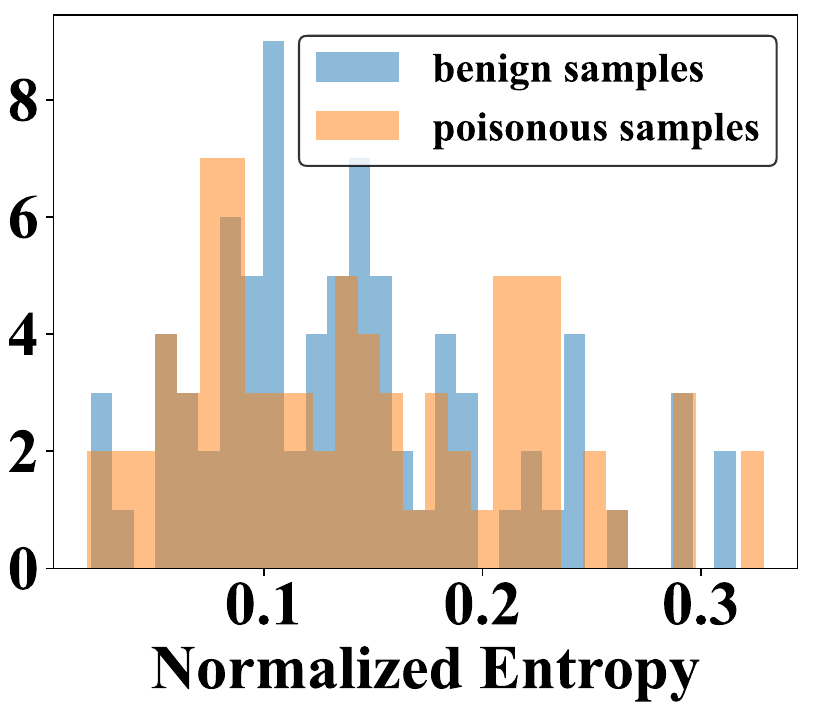}}
    \vspace{-2mm}
    \caption{Defense performance of STRIP on FlowMur and baselines.}
    \label{strip}
    
    \vspace{-2mm}
\end{figure}

Fig.~\ref{strip} presents the defense performance of STRIP against NBA, NBA-D, and FlowMur over SCD and FKD. The results indicate that STRIP is effective for defending against NBA and NBA-D, as the defender can find a proper entropy threshold to effectively distinguish benign and poisonous samples, as shown in Fig.~\ref{strip} (a), (b), (d) and (e). However, for FlowMur, the entropy distributions of benign and poisonous samples have a large overlap, suggesting that STRIP fails to differentiate between benign and poisonous samples. This is because variable trigger positions increase the likelihood of the trigger overlap with the vocalized segments in a sample, thus perturbing the trigger's characteristics. In addition, the subtle nature of the trigger generated by FlowMur makes it susceptible to perturbations. Therefore, FlowMur is able to bypass STRIP.

\subsubsection{Beatrix}
\label{beatrix_defense}

\begin{figure}[t]
    \centering
    \subfigure[SCD]{\includegraphics[scale=0.26]{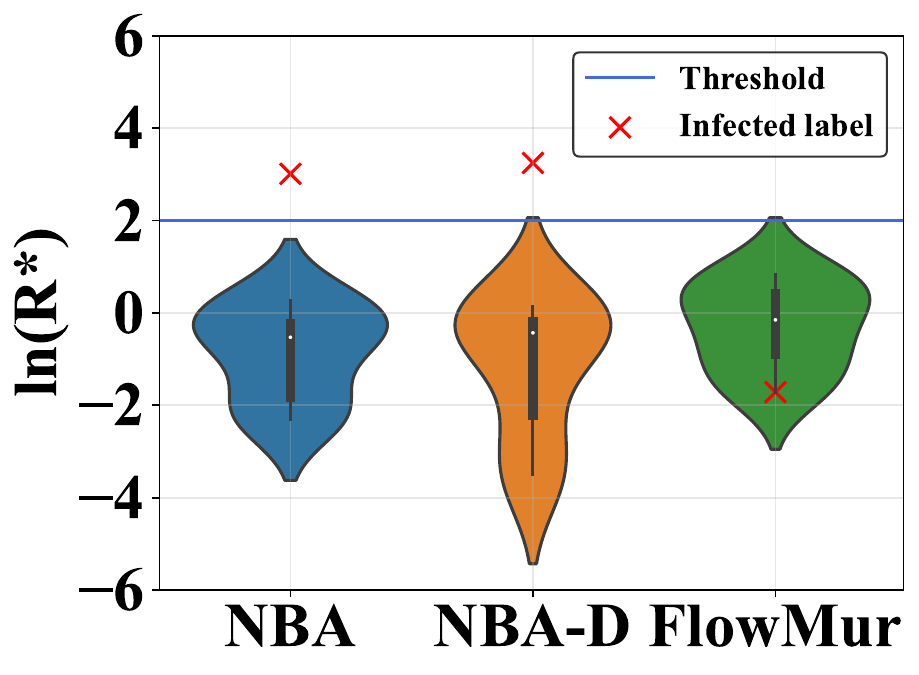}}
    \subfigure[FKD]{\includegraphics[scale=0.26]{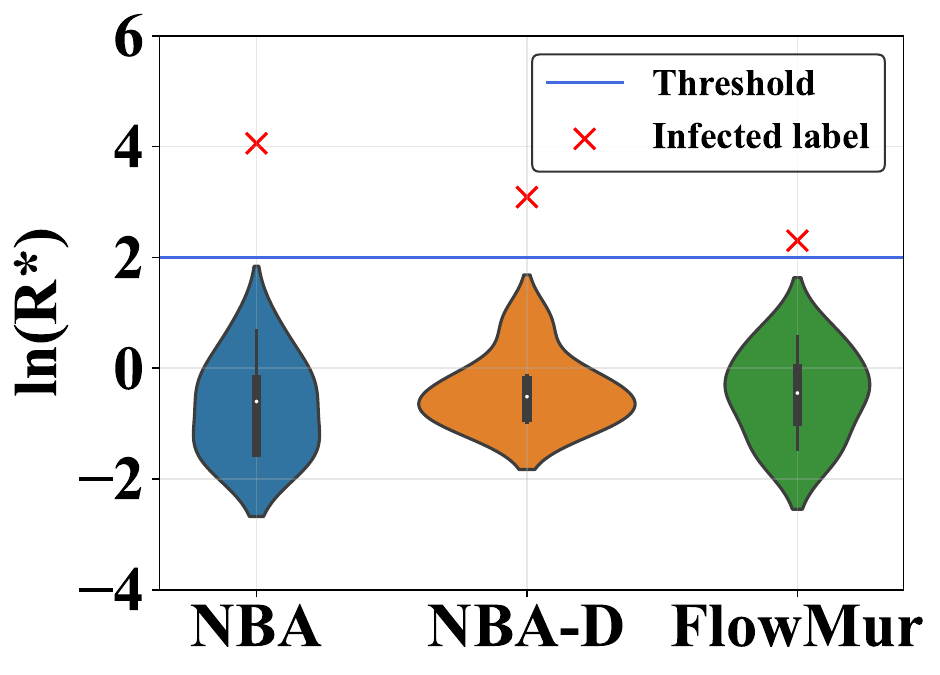}}
    \vspace{-2mm}
    \caption{\textcolor{black}{Defense performance of Beatrix on FlowMur and baselines.}}
    \label{beatrix}
    \vspace{-2mm}
\end{figure}

\textcolor{black}{Beatrix~\cite{ma2023beatrix} is a backdoor defense method based on an observation that although an infected model identifies both clean samples of the target class and poisonous samples as the target class, these two sets of samples are disjoint in the pixel space. Consequently, the intermediate representations of the poisonous samples differ from those of the clean samples. Built upon this, Beatrix leverages Gram Matrices to enlarge the discrimination between benign and poisonous samples. Additionally, it employs kernel-based testing to identify the infected label (i.e., the target class). Figure~\ref{beatrix} presents the defense performance of Beatrix on FlowMur and baselines.}

\textcolor{black}{In Fig.\ref{beatrix}, $R^*$ denotes the anomaly index. In alignment with Ma et al.\cite{ma2023beatrix}, we set the threshold for the anomaly index as $e^2$. Fig.~\ref{beatrix} reveals that the anomaly index of the infected class exceeds the threshold for NBA and NBA-D on both SCD and FKD, while the anomaly indexes for non-infected classes remain below the threshold. This indicates that Beatrix is effective in detecting NBA and NBA-D. Furthermore, Beatrix successfully detects FlowMur on FKD. The reason is that Beatrix breaks the assumption that all poisonous samples share the same trigger, enabling it to defend against dynamic backdoor attacks. However, Beatrix does not perform well in detecting FlowMur on SCD. This finding suggests that Beatrix's defense performance against FlowMur exhibits a degree of instability.}

\section{Discussion}
This section discusses the strategies to mitigate FlowMur and its limitations. We also explore its positive applications, refer to Appendix~\ref{positive_purposes}.

\subsection{Mitigation of FlowMur}
\label{mitigation of flowmur}
\textcolor{black}{In Section~\ref{defense_resistance}, we assessed the resistance of FlowMur against four defense methods. Filters are commonly used in the audio domain as a general defense method. Although they have an impact on FlowMur’s ASR, they also compromise the BA of the infected model. Thus, it is essential to consider a trade-off when using filters. Fine-pruning and STRIP rely on specific observations that may not be applicable to all backdoor attacks, particularly trigger-optimized attacks and dynamic attacks. Consequently, FlowMur is able to bypass them. Beatrix, on the other hand, is built on an observation (or a fact) inherent in all backdoor attacks. In theory, it can defend against all backdoor attacks, from trigger-designated attacks, static attacks to trigger-optimized attacks and dynamic attacks. For example, it can effectively defend against FlowMur on FKD, as shown in Section~\ref{beatrix_defense}. However, it requires a knowledgeable defender who can access the infected model and some benign samples, which hinders its practicality. Therefore, there is still a need to explore backdoor defenses that operate with limited defender knowledge.}

\textcolor{black}{Neural vocoders~\cite{yamamoto2020parallel,kong2020hifi} employ neural networks to generate speech waveforms from acoustic features, such as MFCCs and spectrograms. As neural vocoders are trained based on real speech and designed to synthesize real speech, they can produce real speech even when distorted or attacked acoustic features are provided. Therefore, we argue that neural vocoders can be utilized to purify poisonous samples and defend against FlowMur since the trigger generated by FlowMur is a meaningless noise segment. However, it is essential to note that neural vocoders may fail to defend against backdoor attacks that utilize real speech as triggers, such as whistles and birdsongs, as these triggers encompass semantic information. Moreover, as a DNN, the neural vocoder is also vulnerable to various threats, including evasion attacks and backdoor attacks.}




\subsection{Limitations of FlowMur}

Speech recognition systems can be classified into two categories: speech command recognition systems and automatic speech recognition systems. The former employs a sequence-to-vector model, producing a command label as output, whereas the latter transcribes speech signals into text using a sequence-to-sequence model, resulting in a sequence of words as output. FlowMur is a backdoor attack specifically targeting speech command recognition systems. Due to limitations imposed by existing publicly available datasets, we evaluated FlowMur's effectiveness over two short speech command datasets, i.e., SCD and FKD. Nevertheless, we argue that FlowMur is effective for any speech command recognition system, regardless of input speech length.

As for the automatic speech recognition systems, they are typically more complicated than the speech command recognition systems. This heightened complexity renders a challenge for adversaries to plant backdoors into them. The reason is that this practice requires significant computational resources. To this end, the utility of FlowMur in attacking automatic speech recognition systems becomes limited. We consider modifying the model's architecture~\cite{zong2023trojanmodel} or loss function~\cite{blind} as potential methods for conducting backdoor attacks on the automatic speech recognition systems, which we leave as our future work.

\section{Conclusion}
In this paper, we proposed FlowMur, a stealthy and practical audio backdoor attack that only requires target-class samples to be launched. FlowMur addresses the drawbacks of current audio backdoor attacks, including full knowledge requirement, limited stealthiness, poor practicality, and weak defense resistance. Specifically, we constructed a surrogate dataset to train a surrogate model for augmenting adversary knowledge in FlowMur. The trigger generation of FlowMur was formulated as an optimization problem and the optimized trigger can be attached to various positions of a sample, allowing FlowMur to achieve dynamicity. We also designed an adaptive data poisoning method to enhance the stealthiness of FlowMur. Additionally, we incorporated ambient noise into the process of trigger generation and data poisoning, enhancing its robustness to ambient noise and improving its practicality. We demonstrated the superiority of FlowMur with extensive experiments in terms of effectiveness, practicality, stealthiness and defense resistance, by comparing with two baselines. Our results show the vulnerability of speech command recognition systems to FlowMur, emphasizing the need for advanced defense methods to enhance their security.

\section{Ethical Considerations}
All human-related studies and experiments conducted in this paper have been approved by the Institutional Review Board of our institutes. Personal information, such as native language, age and region, was not recorded, except for the participants' answers in the human study and their gender. Sensitive data, including gender, were encrypted and protected with access control.

\section*{Acknowledgement}
This work is supported in part by the National Natural Science Foundation of China under Grant 62072351; in part by the Key Research Project of Shaanxi Natural Science Foundation under Grant 2023-JC-ZD-35; in part by the open research project of ZheJiang Lab under grant 2021PD0AB01; in part by the 111 Project under Grant B16037; and in part by the Fundamental Research Funds for the Central Universities, No. YJSJ23007.

\bibliographystyle{IEEEtran}
\bibliography{reference}

\appendices

\section{Attacking Speech Recognition Systems with Long Inputs}
\label{long}

\textcolor{black}{In Section~\ref{attack_evaluation}, we evaluated FlowMur's performance in attacking two short speech command datasets, namely SCD and FKD. In this subsection, we extend our investigation to assess FlowMur's effectiveness on a long speech command dataset. Our dataset comprises 18 distinct speech commands, including phrases like ``put on the music'' and ``bedroom heat up'', sourced from the Fluent Speech Commands (FSC) dataset~\cite{FSC}. However, each of these speech commands is accompanied by roughly 100 audio samples in FSC, which is inadequate for training an effective speech recognition model. Consequently, we utilized data augmentation techniques to enhance our dataset. Specifically, we applied variations in volume (amplified, unchanged, reduced), pitch (raised, unchanged, reduced), and speed (sped up, unchanged, slowed down) to the original samples, resulting in an augmented dataset containing a total of 51,624 samples. Within this augmented dataset, we designated 8 classes as the victim dataset, while the remaining 10 classes formed the auxiliary dataset. The trigger duration was established at 1.5 seconds, considering that the sample duration in this dataset typically falls within the range of 2 to 4 seconds. All other experimental configurations, including feature engineering, target and surrogate models, and the target-class poisoning rate, adhered to the default settings outlined in Section~\ref{setting}. Table~\ref{fsc} displays the attack performance of FlowMur and baselines on FSC.}

\begin{table}[b]
    
    \centering 
    \footnotesize
    \caption{\textcolor{black}{Performance comparison between FlowMur and baselines on FSC (\%).}}
    \vspace{-1mm}
    \label{fsc}
    {\vspace{-2mm}MA: model accuracy without attacks.}
\\[2mm]
    
      \begin{tabular}{@{}c|ccc@{}}  
        \toprule[1.5pt]
    \multirow{2.5}{*}{\makebox[0.05\textwidth][c]{Method}} &\multicolumn{3}{c}{FSC} \\
    \cmidrule{2-4}
        & MA & BA & ASR \\
   
   \midrule
         NBA &      94.50 & 95.05($\uparrow$0.55)  &9.19   \\
         NBA-D &   94.50 & 94.77($\uparrow$0.27)  & 43.59 \\
      \textbf{FlowMur} & 94.50 &94.06($\downarrow$0.44)  &  \textbf{89.05} \\
    \midrule
        \textit{p}-value & - & - & \num{1.50e-06} \\
        
    \bottomrule[1.5pt]
    \end{tabular}
\end{table}


\textcolor{black}{Three observations can be derived from Table~\ref{fsc}. First, none of the three attack methods compromise their BAs. Second, FlowMur achieves an ASR of 89.05\%, significantly outperforming the baselines. Third, the \textit{p}-value $1.50 \times 10^{-6}$ emphasizes the statistically significant improvements introduced by FlowMur in comparison to the baselines.}

\section{Attacking Speaker Recognition Systems}
\label{attackspeaker}


\textcolor{black}{We argue that the effectiveness of FlowMur extends not only to speech command recognition systems but also to a variety of classification models within the audio domain. To support this claim, we applied FlowMur and baselines to launch attacks on a speaker recognition model. We built a dataset by randomly selecting 30 distinct classes, corresponding to 30 speakers, from the CSTR VCTK corpus~\cite{vctk}. Within this dataset, 10 classes were chosen as the victim dataset, while the remaining 20 constituted the auxiliary dataset. Since the samples in this dataset have varying durations, ranging from 3 seconds to over 10 seconds, we set the trigger duration as 2.5 seconds. All other experimental parameters, including feature engineering, target and surrogate models, and the target-class poisoning rate, remained consistent with the default settings described in Section~\ref{setting}. The attack performance of FlowMur and baselines for speaker recognition are presented in Table~\ref{speaker}.}

\begin{table}[t]
    
    \centering 
    \footnotesize
    \caption{\textcolor{black}{Performance comparison between FlowMur and baselines for speaker recognition (\%).}}
    \vspace{-1mm}
    \label{speaker}
    {\vspace{-2mm}MA: model accuracy without attacks.}
\\[2mm]
    
      \begin{tabular}{@{}c|ccc@{}}  
        \toprule[1.5pt]
    \multirow{2.5}{*}{\makebox[0.05\textwidth][c]{Method}} &\multicolumn{3}{c}{VCTK} \\
    \cmidrule{2-4}
        & MA & BA & ASR \\
   
   \midrule
         NBA &      98.44 & 98.82($\uparrow$0.38)  &14.43   \\
         NBA-D &   98.44 & 98.77($\uparrow$0.33)  & 59.49 \\
      \textbf{FlowMur} & 98.44 &98.83($\uparrow$0.39)  &  \textbf{98.20} \\
    \midrule
        \textit{p}-value & - & - & \num{1.62e-04} \\
        
    \bottomrule[1.5pt]
    \end{tabular}
\end{table}


\textcolor{black}{Table~\ref{speaker} offers three observations. Firstly, none of the three attack methods compromise their BAs. Secondly, FlowMur achieves an ASR of 98.20\%, substantially surpassing the baselines. Thirdly, the results of the significance test demonstrate the statistically significant advancements introduced by FlowMur in comparison to the baselines.}

\section{Impact of Trigger Volume}
\label{trigger_volume}

In Section~\ref{physcalattack}, we demonstrated the effectiveness of FlowMur on attacking live human speech with a trigger volume of 55dB. In this section, in order to assess the impact of different trigger volumes on attack performance. We set the trigger volumes from 45dB to 65dB with a step of 5dB. Experimental results are presented in Table \ref{triggervolume}.

\begin{table}[t]
    \centering 
    \small
    \caption{ASR of FlowMur under different trigger volumes.}
    \vspace{-1mm}
    \label{triggervolume}

    \begin{tabular}{@{}c|ccccc@{}}  
        \toprule[1.5pt]
             &45dB & 50dB & 55dB & 60dB & 65dB\\
       \midrule
         ASR (\%) & 57.96 & 79.63 & 81.48 & 92.96 & 94.07  \\
    \bottomrule[1.5pt]
    \end{tabular}
    \vspace{-2mm}
\end{table}
From Table \ref{triggervolume}, we can observe a positive correlation between the ASR of FlowMur and the trigger volume. In particular, when the trigger volume is greater than or equal to 60dB, the ASR of FlowMur exceeds 90\%. However, excessively increasing the trigger volume would render it easily detectable by humans.

\begin{figure*}[t]

	\centering
	\subfigure[A benign sample]{
	\begin{minipage}[c]{0.22\textwidth}
	    \includegraphics[scale=0.18]{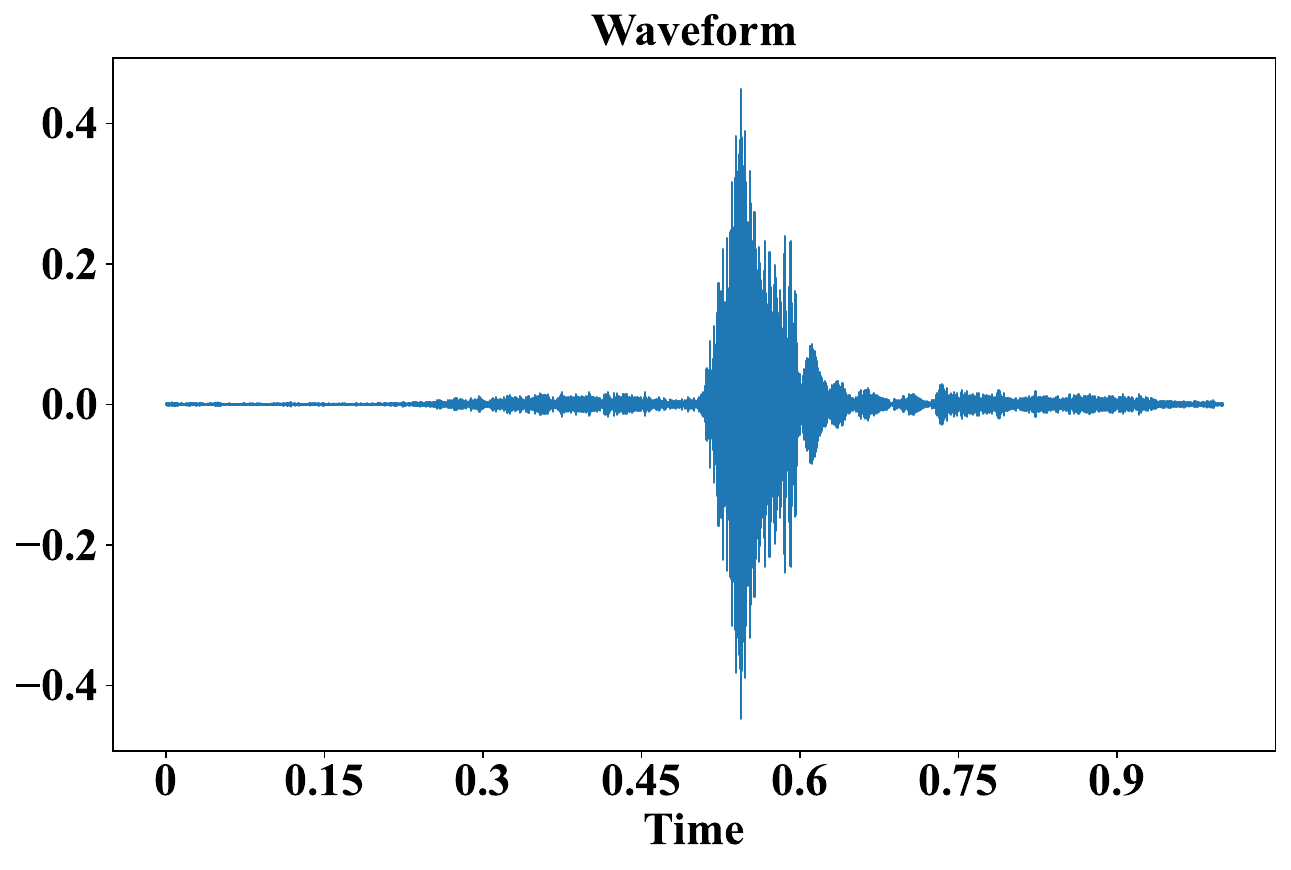}\\
		\includegraphics[scale=0.18]{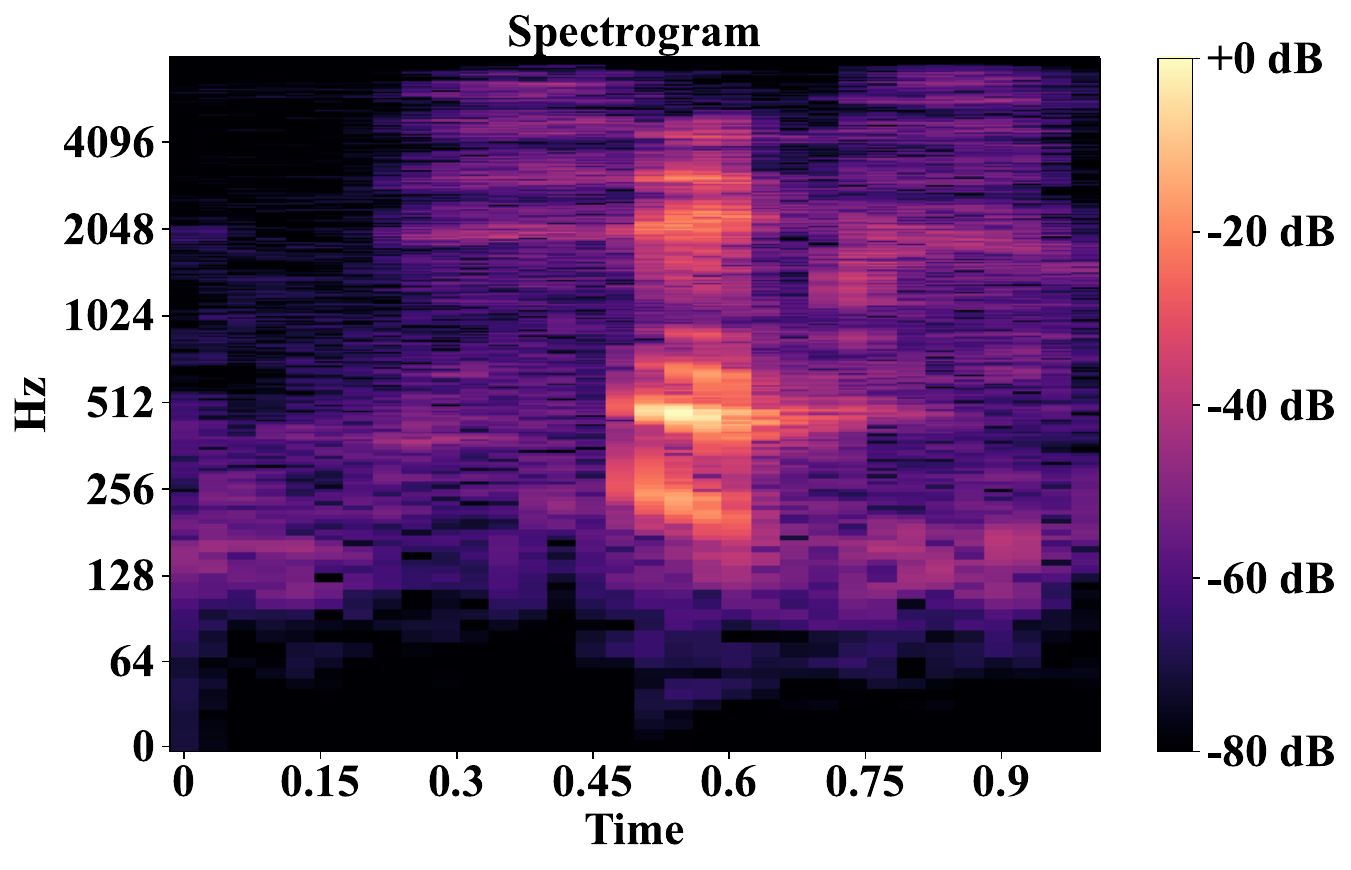}\\
            \hspace*{\fill}\hspace*{0.116\textwidth}\includegraphics[scale=0.18]{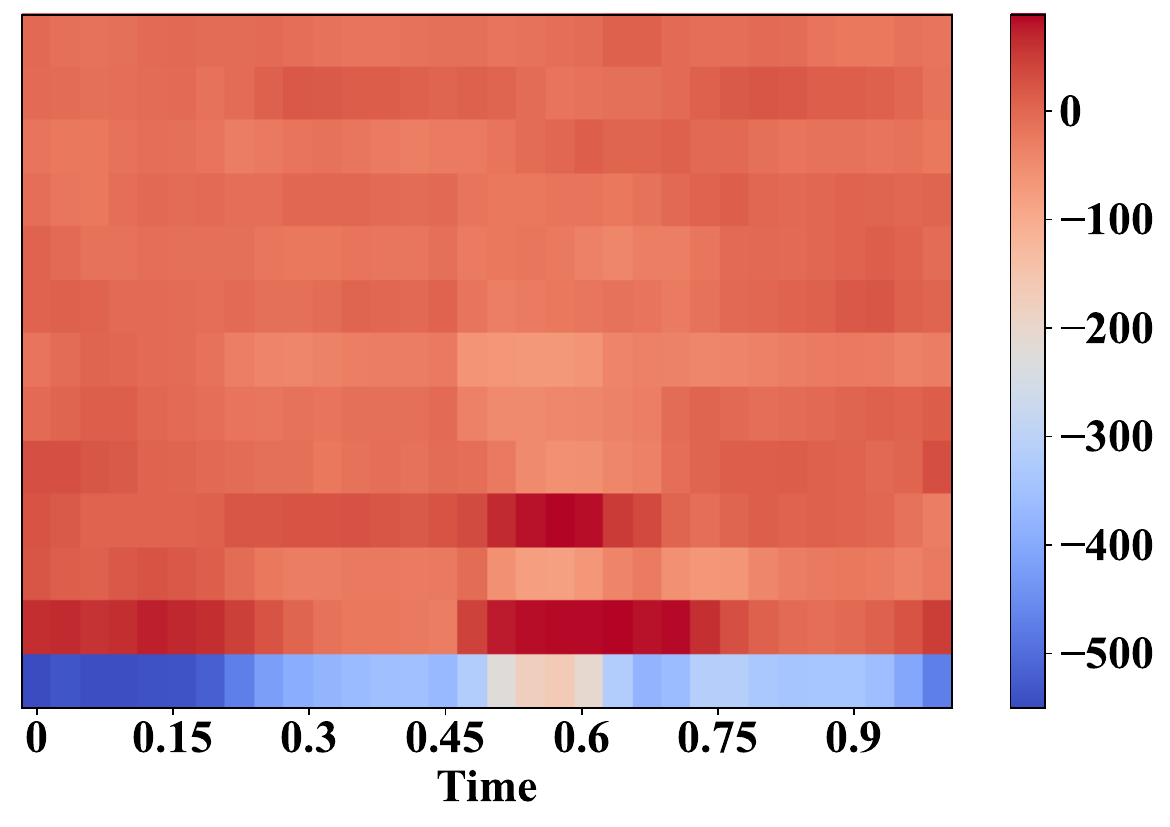}
	\end{minipage}
}
	\subfigure[A sample with ambient noise]{
	\begin{minipage}[c]{0.22\textwidth}
		\includegraphics[scale=0.18]{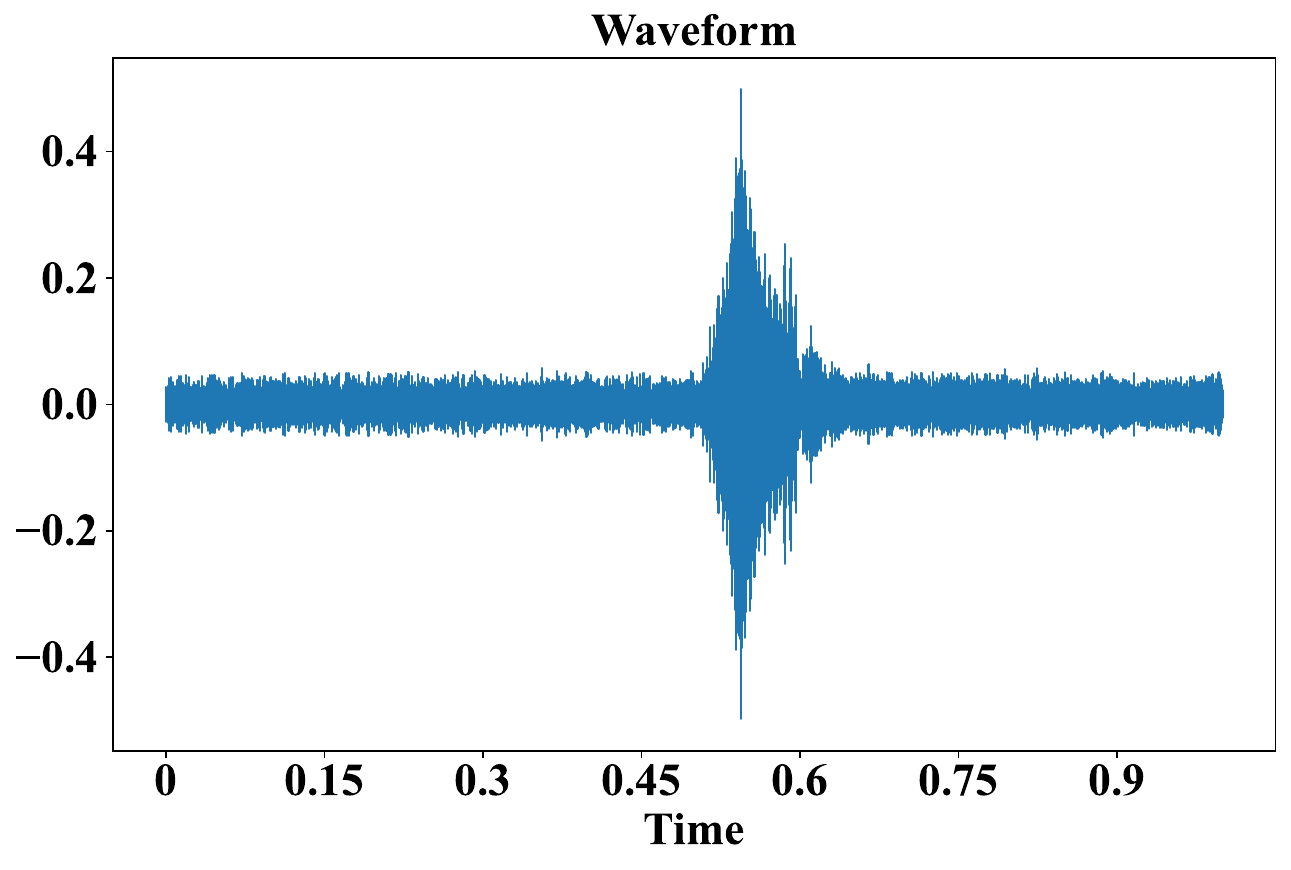} \\
	    \includegraphics[scale=0.18]{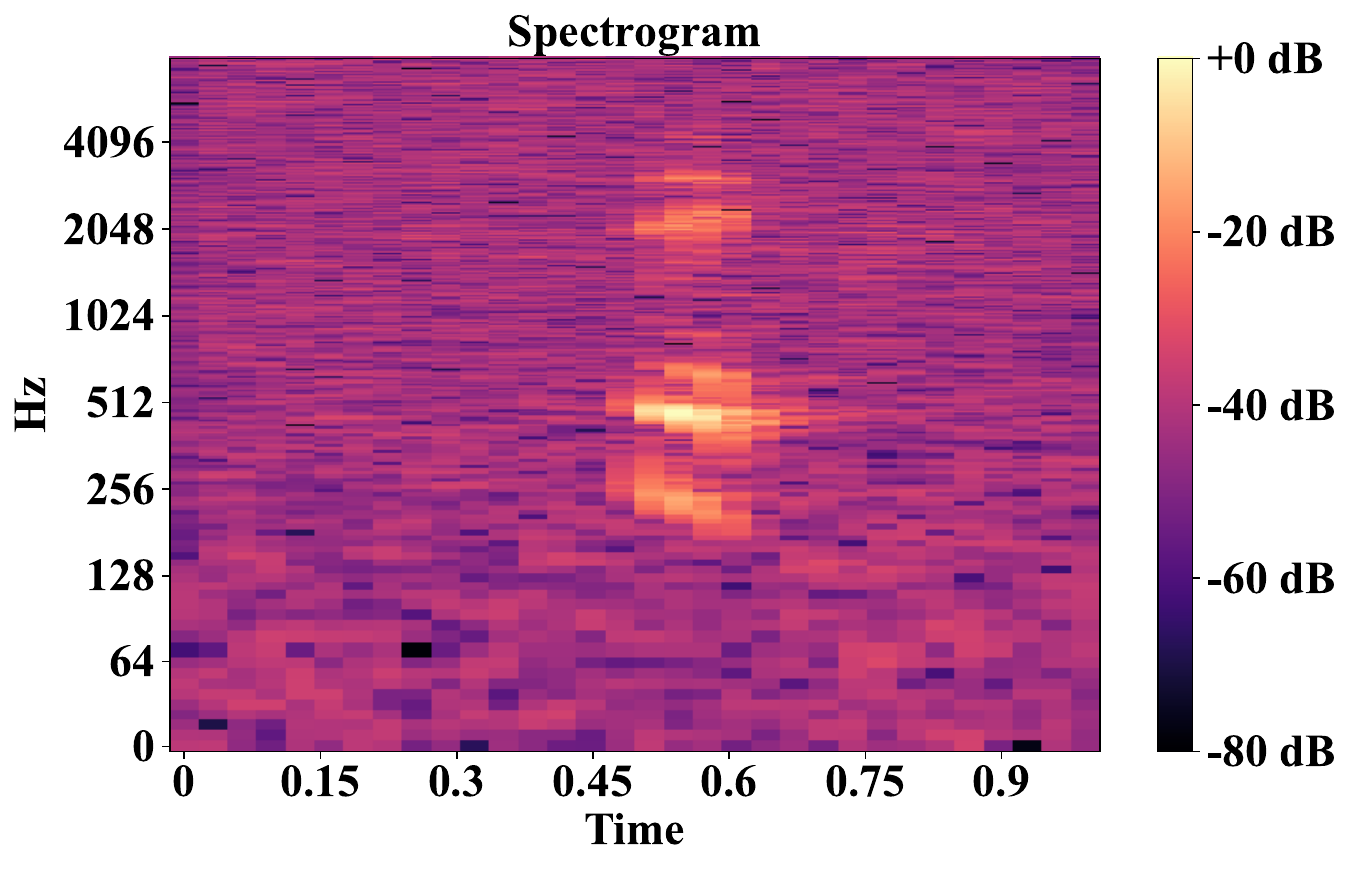}\\
            \hspace*{\fill}\hspace*{0.116\textwidth}\includegraphics[scale=0.18]{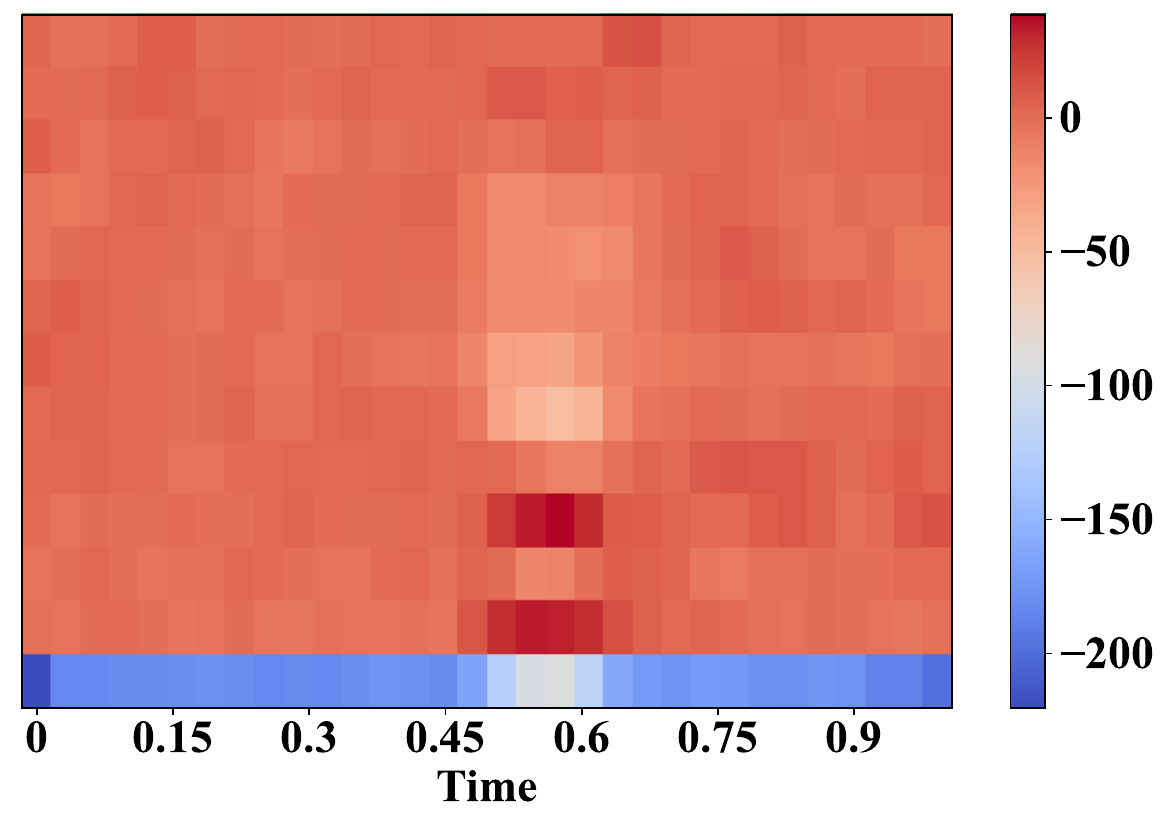}
	\end{minipage}
}
	\subfigure[A poisonous sample of NBA(-D)]{
	\begin{minipage}[c]{0.22\textwidth}
		\includegraphics[scale=0.18]{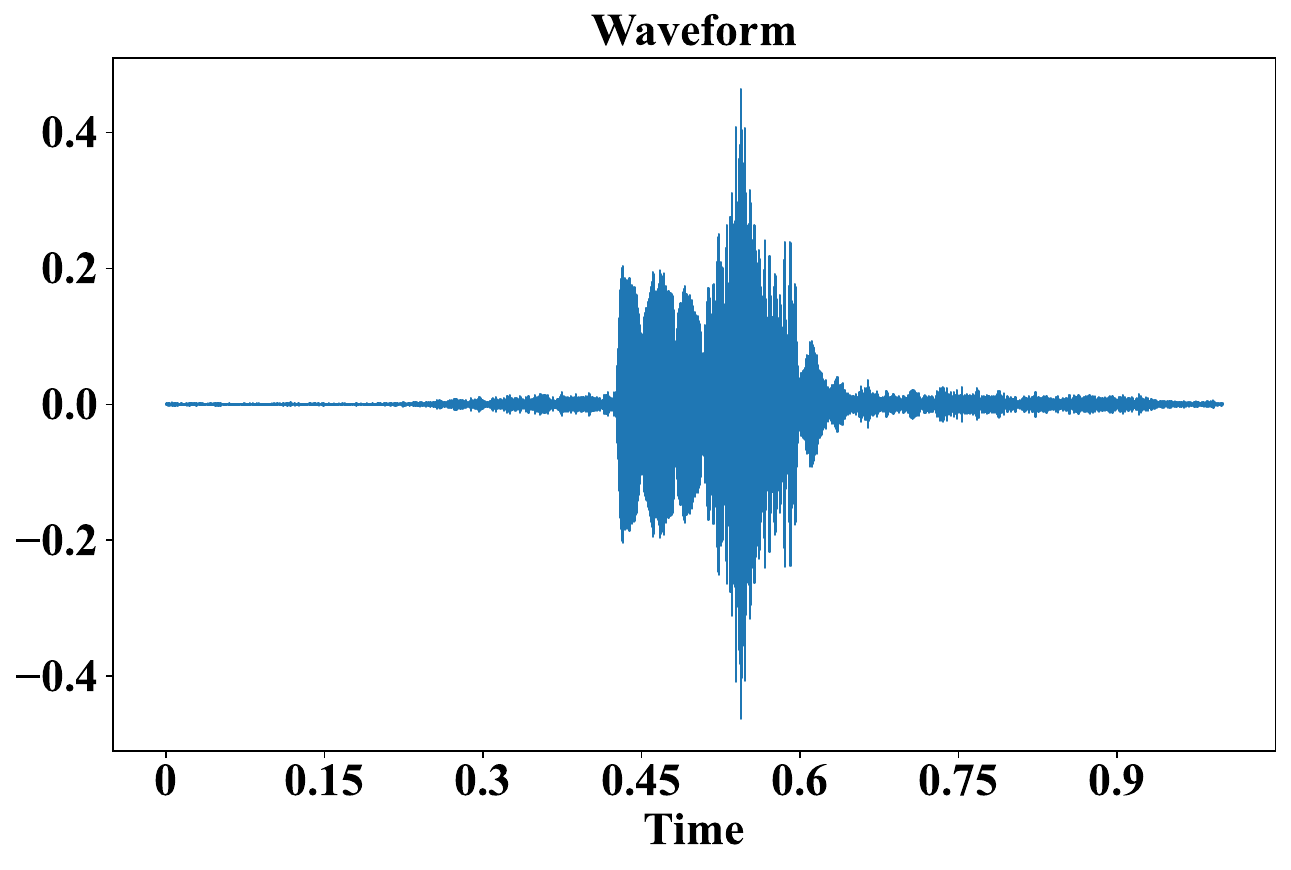} \\
		\includegraphics[scale=0.18]{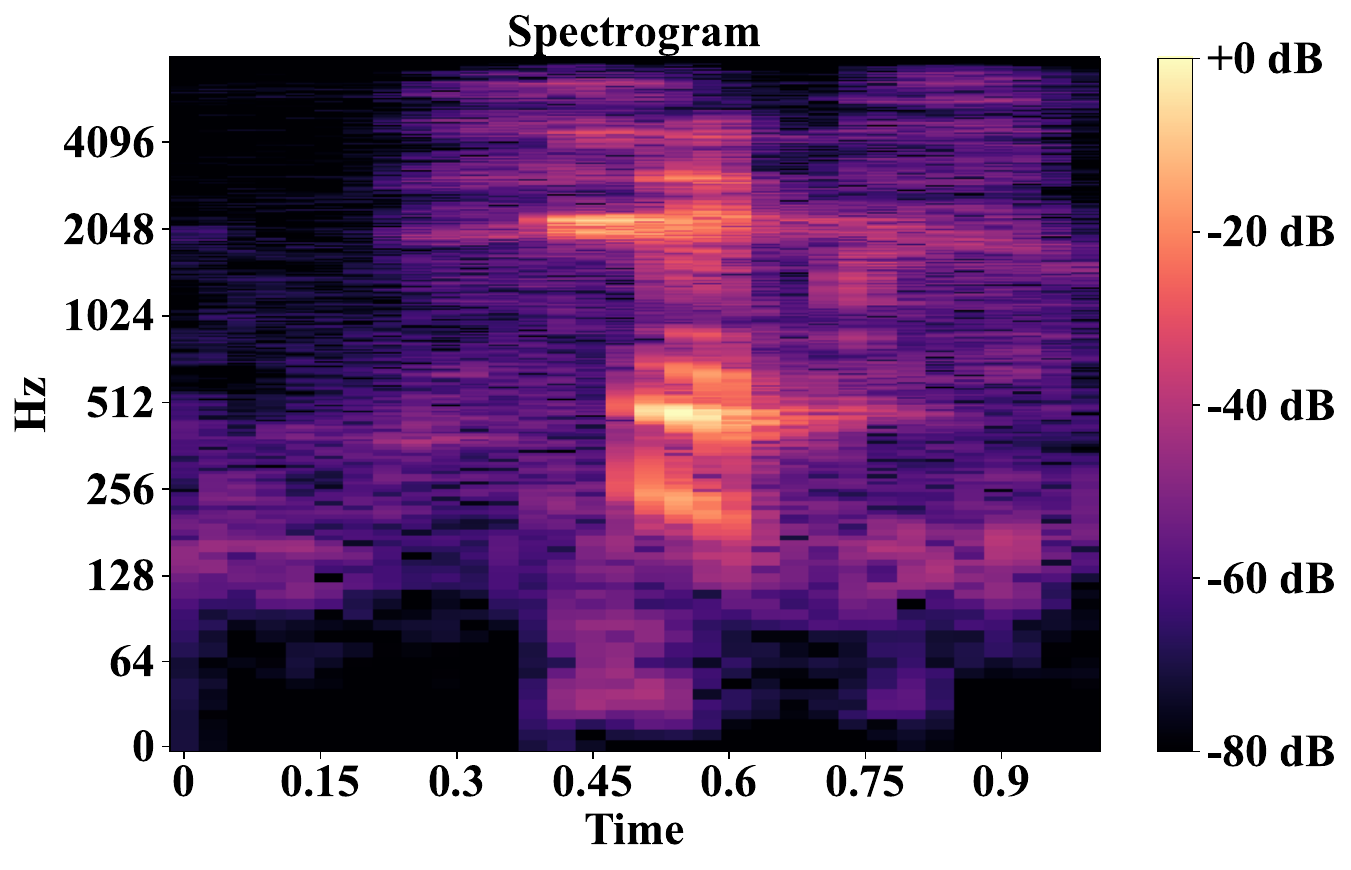}\\
            \hspace*{\fill}\hspace*{0.116\textwidth}\includegraphics[scale=0.18]{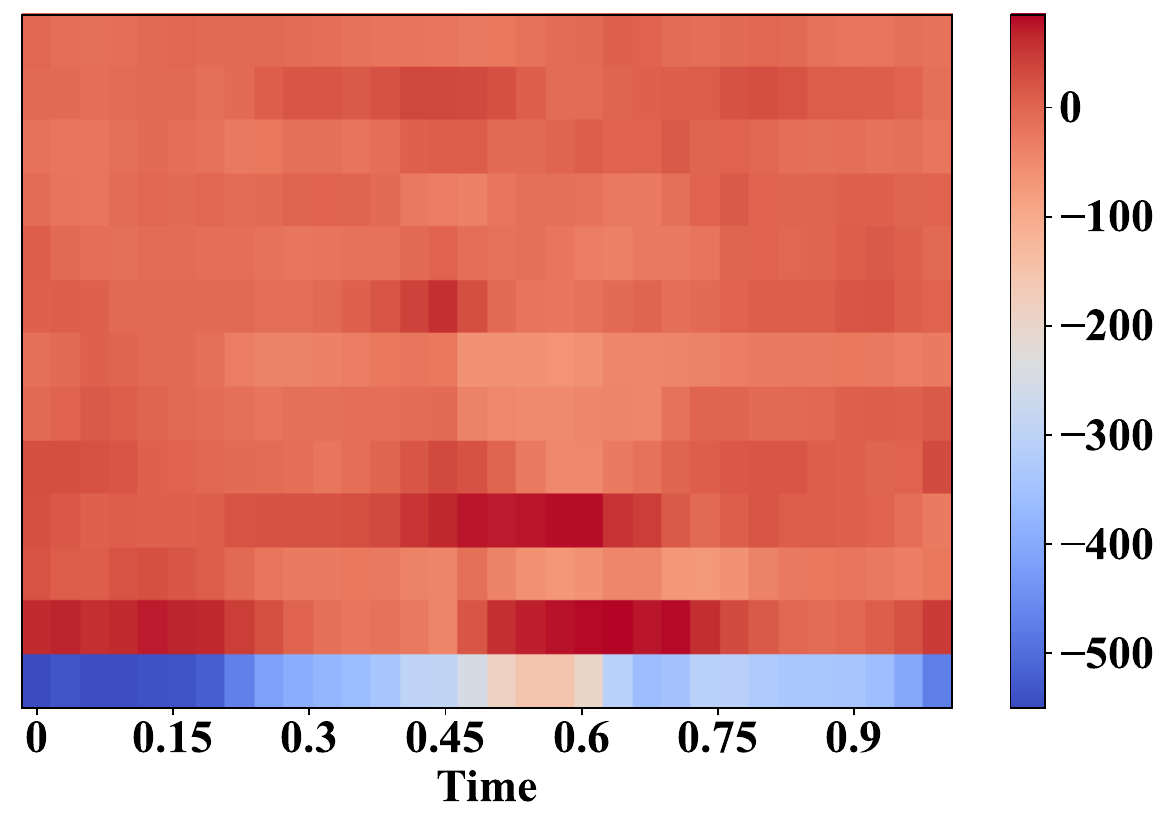}
	\end{minipage}
}
	\subfigure[A poisonous sample of FlowMur]{
	\begin{minipage}[c]{0.22\textwidth}
		\includegraphics[scale=0.18]{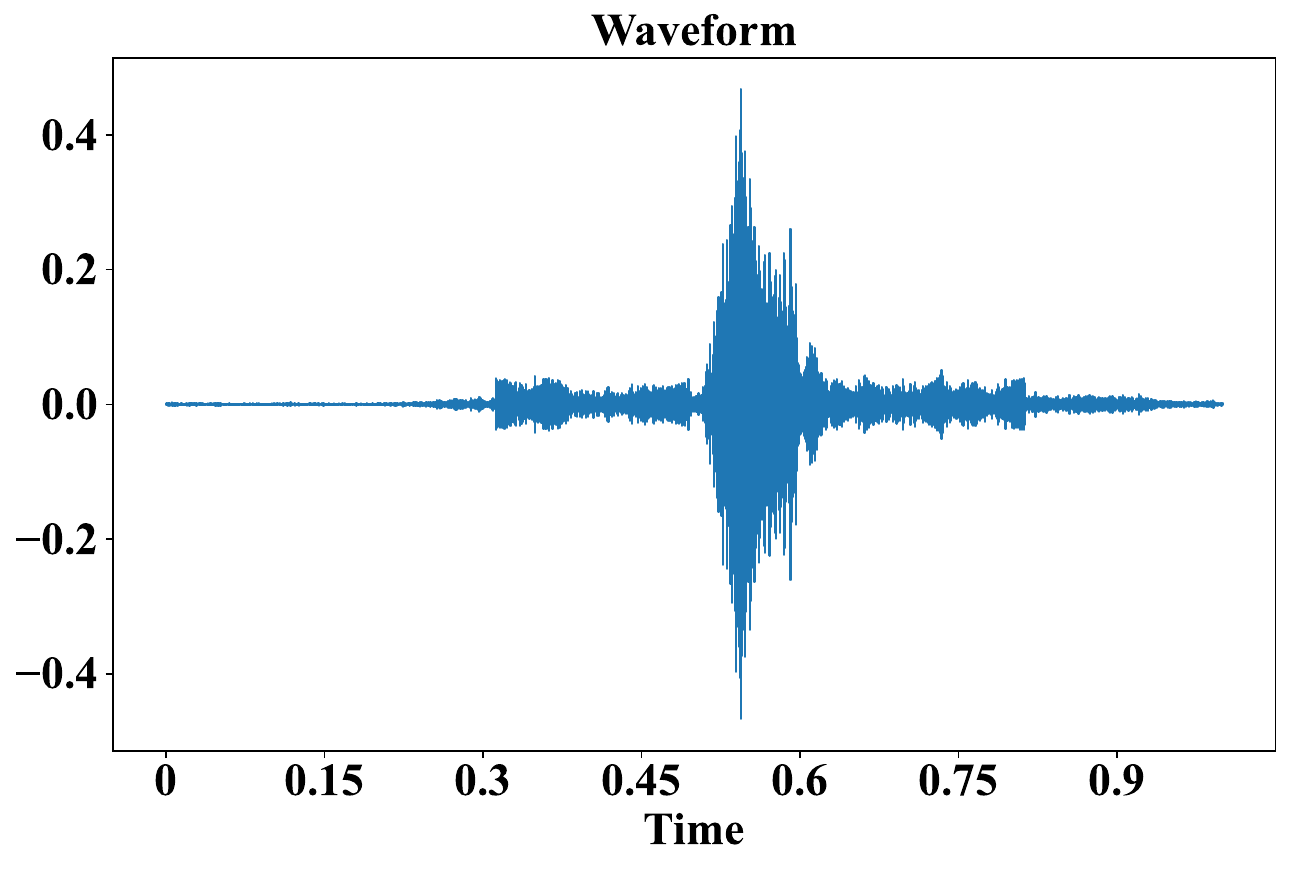} \\
		\includegraphics[scale=0.18]{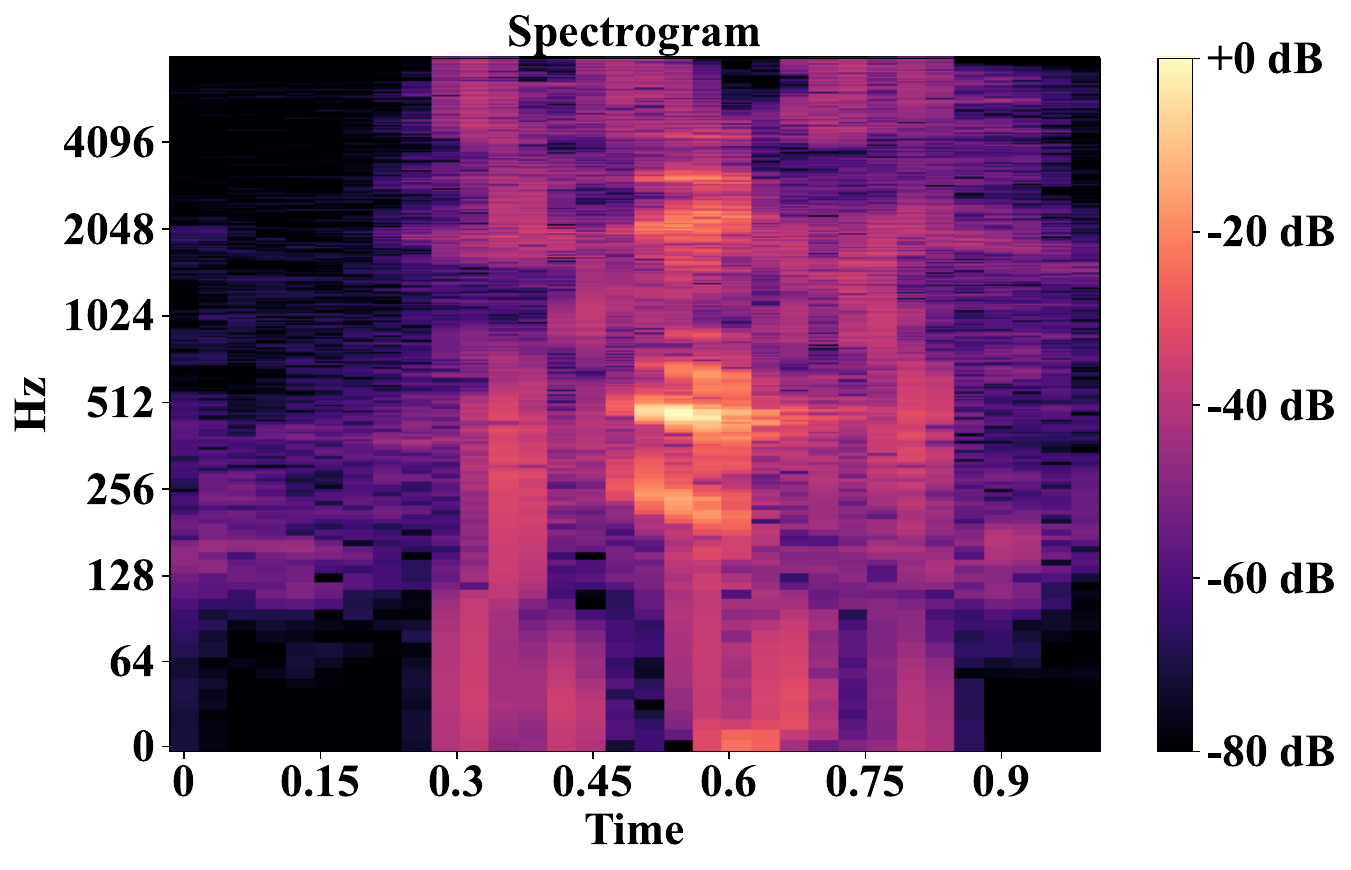}\\
            \hspace*{\fill}\hspace*{0.116\textwidth}\includegraphics[scale=0.18]{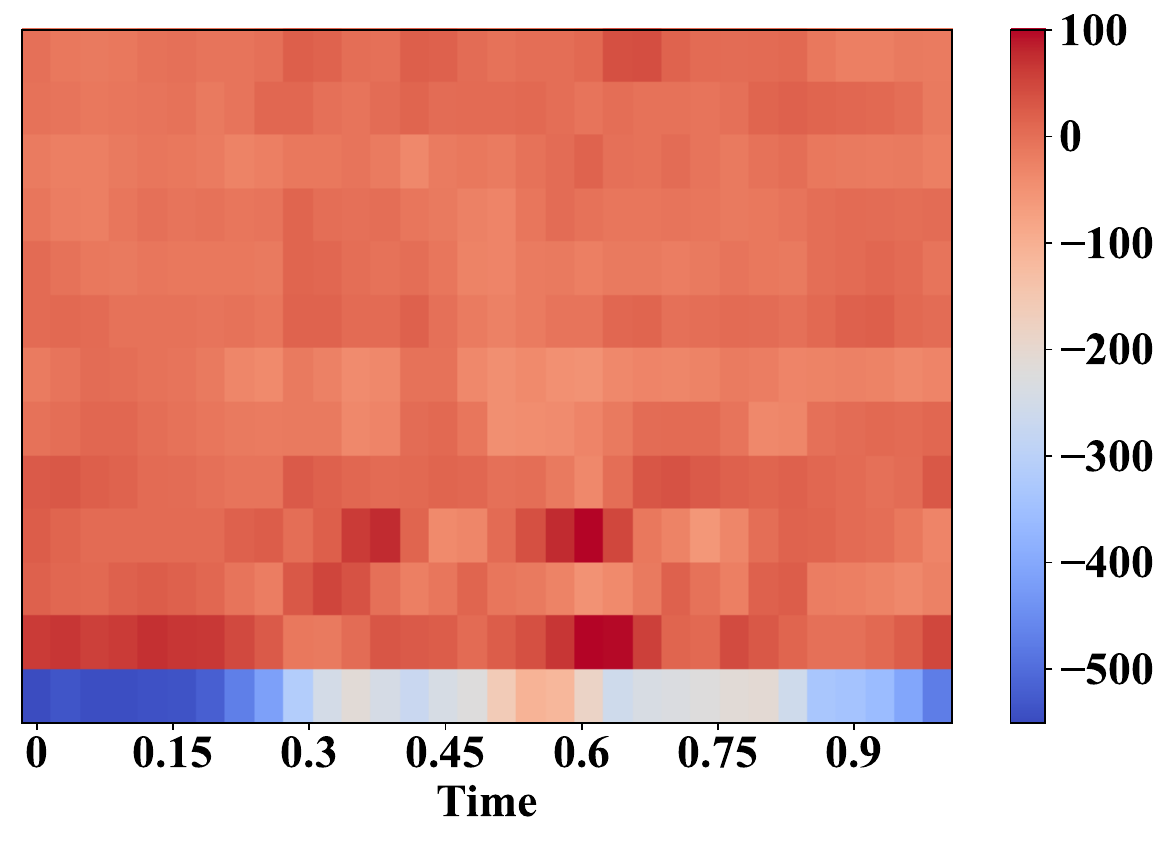}
	\end{minipage}
}
\caption{Audio waveforms, spectrograms and MFCCs for different cases.}
\label{samplevisualization}
\end{figure*}

\section{Positive Applications of FlowMur}
\label{positive_purposes}
While the primary purpose of designing FlowMur is to explore the vulnerabilities of speech recognition systems, it can be employed for positive purposes, specifically protecting intellectual property rights or digital copyrights. 

The process of collecting a dataset or training a DNN model is indeed complex and time-consuming, requiring significant effort. As a result, there is a growing demand for effective methods to safeguard the intellectual property rights associated with these models and datasets. Previous work has demonstrated that backdoor attacks can serve as a viable way of dataset or model ownership verification~\cite{adi2018turning,li2023black}. To elaborate, for a dataset, its owner can introduce poisonous samples, i.e., samples with triggers, into it. Consequently, any model trained with this dataset would contain hidden backdoors. Thus, by examining the output of unknown models for samples with triggers, the dataset owner can identify whether his/her dataset has been misused without authorization. Similarly, a model owner can also insert hidden backdoors into his/her model. By inspecting the output of unknown models for samples with triggers, it is easy to determine whether the model is used unlawfully.

Therefore, FlowMur can be used for protecting the intellectual property rights of speech datasets and speech recognition models by being applied to verify their ownership. Meanwhile, due to its excellent attack performance and stealthiness, FlowMur allows efficient ownership verification of speech datasets and speech recognition models in a senseless manner.

\section{Sample Visualization for Different Cases}
\label{appendix_samplevisualization}
Audio waveforms, spectrograms, and MFCCs for different cases are presented in Fig~\ref{samplevisualization}.

\end{document}